# The Nu Class of Low-Degree-Truncated, Rational, Generalized Functions. I. IMSPE in Design of Computer Experiments: Integrals and Very-Low-N, Single-Factor, Free-Ranging Designs

Selden Crary

*Palo Alto, CA, USA*

**Abstract**

We provide detailed algebra for determining the integrated mean-squared prediction error ($IMSPE$) of designs of computer experiments, with one factor and one or two points, under the exponential, Gaussian, or either of two Matérn correlation functions. This algebra shall provide the basis for the identification of the $IMSPE$ as a member of a special class of low-degree-truncated rational functions, which we name, here, the Nu class. We shall detail this function class in a series of papers, of which this is the first.

## 1.    Introduction to the series

In a study of exact, statistical, optimal design of computer experiments, we have discovered a class of pole-free, special-range-of-parameter, low-degree-truncated rational functions, with essential discontinuities at and only at movable clusters of zero-separation design points. We name this class the "Nu class," which we write with the upper-case, sans-serif, Greek-letter nu, thusly: "N class," or simply "N." In direct contrast to members of N, Padé approximants are high-degree-truncated rational functions. Members of N are formally not functions, due to multivaluedness at their discontinuities.

The algebraic form of each member of N depends upon the number and geometry of its clusters. For example, the algebraic form associated with a cluster of three points depends on whether the clustered points are oriented syzygetically. An example is given, three paragraphs below. Details shall appear in Part IV of this series, including almost-never exceptions to the concise statements in this Section.

In Part II of this series, we shall present examples of designs with clusters of $n$ highly proximal points in two or more factors, what we previously dubbed, in the case of optimal designs, "$n$-uplets" [1]. The integrated mean-square prediction error ($IMSPE$) of a given design depends on the orientations of its proximal points, even in their zero-separation limits, where directional derivatives are assumed to exist that replace directional differences [2]. This dependence leads to essential discontinuities and to the identification of $IMSPE$ functions as members of N.

Example: The $IMSPE$, as a function of the loci of four design points in two factors and assumed Gaussian correlation with $\boldsymbol{\theta} = (0.064, 0.00016)$ was plotted as a hue plot in Fig. 7 of Ref. 2, which provided details. For the sake of simplicity, the plot was a projection which had two points, say $x_1$ and $x_2$, fixed on the abscissa at $[\pm 0.767117, 0.0]$, with the other two points forced to have inversion symmetry with respect to one another, i.e. $x_4 = -x_3$, but otherwise free to move. Thus, the plot had two independent variables, and hue represented the dependent variable. A 3D plot for this example is given below. In each plot, the $IMSPE$ is $C^\infty$ continuous, except when either of the movable points coincide, i.e. when $x_3 =$ $x_4 = 0$, or when each of the movable points coincides with one of the fixed points, i.e. either when $x_3 = x_1$ and $x_4 = x_2$ or when $x_3 = x_2$ and $x_4 = x_1$. At these three exceptional points in the plots, and nowhere else, there are essential discontinuities.

In Part III, we shall prove a theorem, outlined previously [1], relating zero-separation-cluster-design $IMSPE's$ with corresponding members of N. It shall be shown these correspondences arise via "miraculous cancellations," to borrow an informal term from mathematics, physics, and computer science.

As mentioned above, Part IV shall provide details of N.

In Part V, we shall present extensions of our earlier research showing that each finite-$n$ $IMSPE$-optimal design can be assigned one of a finite number of phases, with each design in a given phase sharing common symmetry properties with all others in the same phase [1]. We shall identify an order parameter for such designs, and shall demonstrate this order parameter is discontinuous at transitions between phases, thus identifying the $IMSPE$-optimal-design transitions as similar to quantum phase transitions, despite the fact that neither mass, dynamics, temperature, energy, thermodynamics, nor quantum mechanics is invoked in the former. We shall posit that quantum phase transitions are a subset of a larger class of discontinuous phase transitions, which also includes those arising in regard to N. We shall discuss possible irreducible attributes of this larger class in finite systems, such as coexisting centering effects, other topological effects due to finite size, and inter-point repulsion.

Beyond this series: Given the key role played historically by new function classes, e.g. solitons as Painlevé-transcendent-class solutions of Painlevé equations, we plan to investigate potential applications of N transcendents in science and engineering. The following are a few applications that come to an indulgent, speculative mind: resolution of the "notorious node problem" in computation of trajectories in the de Broglie-Bohm formulation of quantum mechanics [3], resolution of the "black-hole firewall paradox" in cosmology [4], and the classification of qubits in quantum-information theory [5].

## 2. Introduction to this paper

In this Part I of the five-part series, we provide background material for computing – symbolically, when reasonable – examples of $IMSPE$-optimal designs from the field of statistical design of computer experiments. Our focus is on designs with one factor and just one or two points, under the exponential, Gaussian, or either of two Matérn correlation functions. These simple examples provide the background for our subsequent development of the concept of the N class and for showing $IMSPE$ is a member of this class. An extensive set of appendices is provided as a repository of symbolic algebra and techniques potentially useful in mathematical proofs related to the development of the concepts in this series of papers. Sub-section R.3 provides the simplest example of the "miraculous cancellations," ahead.

## 3. Outline

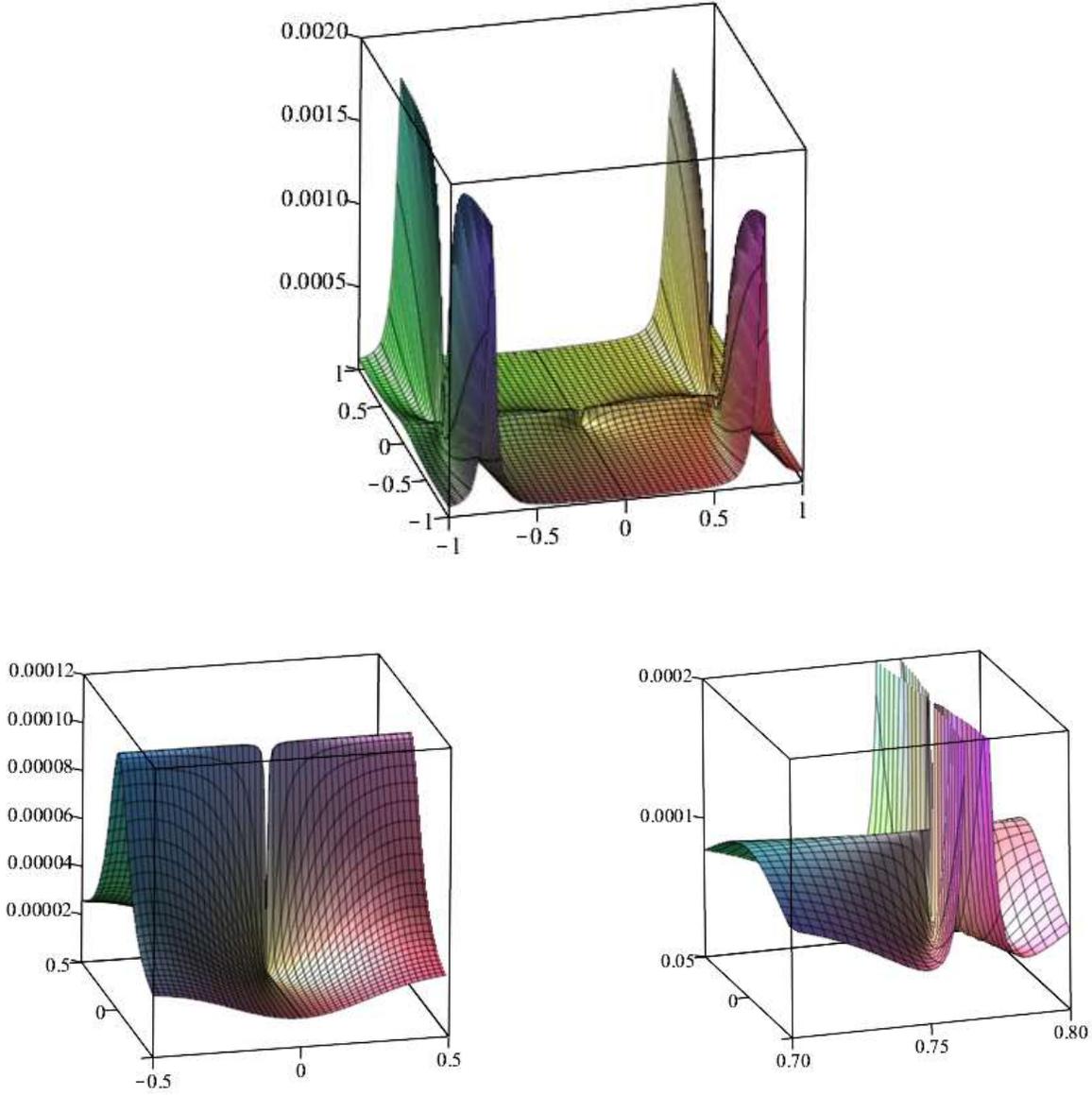

Fig. 1. The upper 3D plot of the example $IMSPE$ of Section 1 is shown as a function of the coordinates of $\boldsymbol{x_1}$, with the abscissa running from left to right, along the plot's obvious base. The region of the origin's essential discontinuity, which is magnified in the lower-left-hand plot, is comprised of the global minimum at the origin, along a rift valley running along the ordinate; as well as a mild local maximum at the origin, along a high-lying ridge running along the abscissa. Moving along the abscissa and away from the origin, the $IMSPE$ has mild minima at $x_{3,1} = \pm 0.3675\cdots$. The two other essential discontinuities are evident in the upper plot, where the locus of the movable point coincides with either of the two fixed design points, $[\pm 0.767117, 0.0]$. The essential discontinuity on the upper plot's right-hand side, which is magnified in the lower-right-hand plot, is comprised of the following: (i) a highest point, $IMSPE = 0.0001068\cdots$, which lies along both a mildly rising ridge along the abscissa and a very strong minimum parallel with the ordinate; (ii) a lowest point that is the common maximum of two paths crossing at an acute angle; and (iii) other paths passing through the singularity at intermediate heights. The essential discontinuity on the upper plot's left-hand side is symmetric to the one on that plot's right-hand side.

# 4. Notation

Our starting point is Eq. 2.9 of the Y1989 Sacks, Schiller, and Welch paper [6], Eq. 4.1, below. A brief summary of the notation is provided in this section.

Symbols for vectors and matrices are bolded, as are headings of tables.

Response and covariance functions: The response, $Y(\pmb{x})$, is a assumed to be a Gaussian process over a $d$-dimensional design domain, via $Y(\pmb{x}) = \beta_0 + Z(\pmb{x})$, where $\beta_0$ is a constant, and the covariance of $Z(\pmb{x})$ depends upon only pairwise distances between points, via a real, symmetric matrix $\pmb{V}$ chosen from one of the four covariance functions given in Table 4.1, below, where we have drawn from Table 4.1 of [7] and defined $Q_k \equiv \theta_k (x_{i,k} - x_{j,k})^2$ and $\tilde{Q}_k \equiv \theta_k (x_{i,k} - x_k)^2$. $\pmb{\theta}$ is a $dx1$ vector of positive covariance hyper-parameters.

Detail on covariance functions is available in Rasmussen and Williams [8], and emphasis is placed on the Matérn class in Stein [9].

| **Correlation-function class** | **Class param.** | **Continuity class** | $V_{i,j}/\sigma_z^2$ | $v_{i,k}/\sigma_z^2$ |
|---|---|---|---|---|
| Exponential power | $p = 1$ | $C^0$ | $e^{-\sum_{k=1}^d \theta_k |x_{i,k} - x_{j,k}|}$ | $e^{-\sum_{k=1}^d \theta_k |x_{i,k} - x_k|}$ |
| Matérn | $v = 3/2$ | $C^1$ | $\prod_{k=1}^d (1 + \sqrt{3Q_k}) e^{-\sqrt{3Q_k}}$ | $\prod_{k=1}^d (1 + \sqrt{3\tilde{Q}_k}) e^{-\sqrt{3\tilde{Q}_k}}$ |
| Matérn | $v = 5/2$ | $C^2$ | $\prod_{k=1}^d \left(1 + \sqrt{5Q_k} + \frac{5Q_k}{3}\right) e^{-\sqrt{5Q_k}}$ | $\prod_{k=1}^d \left(1 + \sqrt{5\tilde{Q}_k} + \frac{5\tilde{Q}_k}{3}\right) e^{-\sqrt{5\tilde{Q}_k}}$ |
| Matérn | ⋮ | ⋮ | ⋮ | ⋮ |
| Matérn | $v$ | $C^{v-1/2}$ | *long expression* | *long expression* |
| Matérn | ⋮ | ⋮ | ⋮ | ⋮ |
| Exponential power | $p = 2$ (Gaussian) | $C^\infty$ | $e^{-\sum_{k=1}^d \theta_k (x_{i,k} - x_{j,k})^2}$ | $e^{-\sum_{k=1}^d \theta_k (x_{i,k} - x_k)^2}$ |

Table 4.1. For the set of correlation-function classes [exponential power with parameter $p = 1$; Matérn with parameter $v = 3/2$; Matérn with parameter $v = 5/2$; exponential power with parameter $p = 2$, also known as Gaussian] this table gives the continuity class for each, as well as expressions for the normalized elements of the correlation matrix $V_{i,j}/\sigma_z^2$ and the quantities $v_{i,k}/\sigma_z^2$ defined in Appendix C. Nota bene: In the right-most column, the $\tilde{Q}_k$ have accent marks.

Ref. 6 prescribes a search for the design that minimizes the integrated mean-squared prediction error, $IMSPE$, over all possible $N$-point designs, as

$$IMSPE_0 = \sigma_Z^2 \min_{\omega_N}[1 - tr(\boldsymbol{L}^{-1}\boldsymbol{R})], \qquad (4.1)$$

where real, symmetric matrices $\boldsymbol{L}$ and $\boldsymbol{R}$ are defined as follows:

$$\boldsymbol{L} \equiv \begin{pmatrix} 0 & | & 1 & \cdots & 1 \\ -- & | & -- & -- & -- \\ \cdot & | & & & \\ \vdots & | & & \boldsymbol{V} & \\ \cdot & | & & & \end{pmatrix}, \text{ and}$$

$$\boldsymbol{R} \equiv \frac{1}{2^d} \int_{-1}^{1} \int_{-1}^{1} \cdots \int_{-1}^{1} \begin{pmatrix} 1 & | & v_1 & v_2 & \cdots & v_n \\ --- & | & --- & --- & --- & --- \\ \cdot & | & v_1^2 & v_1 v_2 & \cdots & v_1 v_n \\ \cdot & | & v_1 v_2 & v_2^2 & \cdots & v_2 v_n \\ \vdots & | & \vdots & \vdots & \ddots & \vdots \\ \cdot & | & v_1 v_n & v_2 v_n & \cdots & v_n^2 \end{pmatrix} dx_1 dx_2 \cdots dx_d, \text{ where}$$

$x_k$, $k = 1, 2, \cdots, d$ are the coordinates of the design domain; $\boldsymbol{x}_i = (x_{i,1}, x_{i,2}, \cdots, x_{i,d})$, $i = 1, 2 \cdots n$ are the design points; and the indices $i$ and $j$ of the twin points, if a pair is present, are assigned to the largest values, viz. $n - 1$ and $n$.

The integrals in $\boldsymbol{R}$ are given in Appendix C and summarized in Tables 4.2 and 4.3, immediately below, where the row and column indices of $\boldsymbol{R}$ run from *0* through *n*, whereas the first indices of the design points run from *1* through *n.*

SMS: Throughout this paper, the initials "SMS" stand for symbolic-manipulation software [10].

Without loss of generality, the factor $\sigma_z^2$ has been set to unity in subsequent equations of this paper. Reconstitution of any of $\boldsymbol{L}, \boldsymbol{V}, \boldsymbol{R}$ or the $IMSPE$ is straightforward, using the definitions, above, although readers should be aware that matrices $\boldsymbol{L}$ and $\boldsymbol{R}$ are defined as dimensionally inhomogeneous.

| Class param. | Eq. | $R_{0.i} \quad 1 \leq i \leq n$ |
|---|---|---|
| $p = 1$ | C.1 | $\prod_{k=1}^{d} \dfrac{1 - e^{-\theta_k} \cosh(\theta_k x_{i,k})}{\theta_k}$ |
| $\nu = 3/2$ | C.7 | $\prod_{k=1}^{d} \dfrac{1}{2\sqrt{3\theta_k}} \left\{ 2\begin{bmatrix} 1 - e^{-\sqrt{3\theta_k}\,(1+x_{i,k})} \\ +1 - e^{-\sqrt{3\theta_k}\,(1-x_{i,k})} \end{bmatrix} - \sqrt{3\theta_k} \begin{bmatrix} (1+x_{i,k})e^{-\sqrt{3\theta_k}\,(1+x_{i,k})} \\ +(1-x_{i,k})e^{-\sqrt{3\theta_k}\,(1-x_{i,k})} \end{bmatrix} \right\}$ |
| $\nu = 5/2$ | C.9 | $\prod_{k=1}^{d} \dfrac{1}{6\sqrt{5\theta_k}} \left\{ 8\begin{bmatrix} 1 - e^{-\sqrt{5\theta_k}\,(1+x_{i,k})} \\ +1 - e^{-\sqrt{5\theta_k}\,(1-x_{i,k})} \end{bmatrix} - 5\sqrt{5\theta_k}\begin{bmatrix} (1+x_{i,k})e^{-\sqrt{5\theta_k}\,(1+x_{i,k})} \\ +(1-x_{i,k})e^{-\sqrt{5\theta_k}\,(1-x_{i,k})} \end{bmatrix} - 5\theta_k \begin{bmatrix} (1+x_{i,k})^2 e^{-\sqrt{5\theta_k}\,(1+x_{i,k})} \\ +(1-x_{i,k})^2 e^{-\sqrt{5\theta_k}\,(1-x_{i,k})} \end{bmatrix} \right\}$ |
| $p = 2$ | C.5 | $\prod_{k=1}^{d} \sqrt{\dfrac{\pi}{16\theta_k}} \left\{ \begin{array}{l} erf\!\left[\sqrt{\theta_k}(1+x_{i,k})\right] \\ +erf\!\left[\sqrt{\theta_k}(1-x_{i,k})\right] \end{array} \right\}$ |

Table 4.2. For the class parameters $[p = 1, \nu = 3/2, \nu = 5/2, p = 2]$, this table gives the Appendix-C equation number for the corresponding matrix elements $R_{0.i} \quad 1 \leq i \leq n$, as well as the expressions for these elements.

| Class param. | Eq. | $R_{i,j} \quad 1 \leq i,j \leq n$ |
|---|---|---|
| $p = 1$ | C.3 | $\prod_{k=1}^{d} \left\{ \dfrac{e^{-\theta_k |x_{i,k}-x_{j,k}|} - e^{-2\theta_k} \cosh[\theta_k(x_{i,k}+x_{j,k})]}{2\theta_k} + \dfrac{|x_{i,k}-x_{j,k}|\, e^{-\theta_k|x_{i,k}-x_{j,k}|}}{2} \right\}$ |
| $\nu = 3/2$ | C.8 | MRSE |
| $\nu = 5/2$ | C.10 | MRSE |
| $p = 2$ | C.6 | $\prod_{k=1}^{d} \sqrt{\dfrac{\pi}{32\theta_k}} \left\{ erf\left[\sqrt{2\theta_k}\left(1 + \dfrac{x_{i,k}+x_{j,k}}{2}\right)\right] + erf\left[\sqrt{2\theta_k}\left(1 - \dfrac{x_{i,k}+x_{j,k}}{2}\right)\right] \right\} e^{-\dfrac{\theta_k(x_{i,k}-x_{j,k})^2}{2}}$ |

Table 4.3. For the set of class parameters $[p = 1, \nu = 3/2, \nu = 5/2, p = 2]$, this table gives the Appendix-C equation number for the corresponding matrix elements $R_{i,j} \quad 1 \leq i,j \leq n$, as well as the expressions for these elements, unless such expression is too long to be reasonable written here, in which case "MRSE" appears. MRSE stands for "machine-readable symbolic expression."

# 5.  $N = 1$ design examples, including optimal designs

The *IMSPE*'s for designs with $d = 1$ and $n = 1$ are given in Table 5.1, below, which also lists the corresponding appendix where each formula is developed, as well as noting that each optimal design is a central point.

| Correlation-function parameter | Eq. | *IMSPE* formula | Optimal design: $[x_1]$ |
|---|---|---|---|
| $p = 1$ | L.1 | $2\left[1 - \dfrac{1 - e^{-\theta}\cosh(\theta x_1)}{\theta}\right]$ | [0] |
| $\nu = 3/2$ | O.1 | $2\left(1 - \dfrac{1}{2\sqrt{3\theta}}\left\{ \begin{array}{l} 2\left[\begin{array}{l}1 - e^{-\sqrt{3\theta}\,(1+x_1)} \\ +1 - e^{-\sqrt{3\theta}\,(1-x_1)}\end{array}\right] \\ -\sqrt{3\theta}\left[\begin{array}{l}(1+x_1)e^{-\sqrt{3\theta}\,(1+x_1)} \\ +(1-x_1)e^{-\sqrt{3\theta}\,(1-x_1)}\end{array}\right] \end{array}\right\}\right)$ | [0] |
| $\nu = 5/2$ | P.1 | $2\left(1 - \dfrac{1}{6\sqrt{5\theta}}\left\{ \begin{array}{l} 8\left[\begin{array}{l}1 - e^{-\sqrt{5\theta}\,(1+x_1)} \\ +1 - e^{-\sqrt{5\theta}\,(1-x_1)}\end{array}\right] \\ -5\sqrt{5\theta}\left[\begin{array}{l}(1+x_1)e^{-\sqrt{5\theta}\,(1+x_1)} \\ +(1-x_1)e^{-\sqrt{5\theta}\,(1-x_1)}\end{array}\right] \\ -5\theta\left[\begin{array}{l}(1+x_1)^2 e^{-\sqrt{5\theta}\,(1+x_1)} \\ +(1-x_1)^2 e^{-\sqrt{5\theta}\,(1-x_1)}\end{array}\right] \end{array}\right\}\right)$ | [0] |
| $p = 2$ | N.1 | $2\left(1 - \sqrt{\dfrac{\pi}{16\theta}}\left\{\begin{array}{l}erf\left[\sqrt{\theta}\,(1+x_1)\right] \\ +erf\left[\sqrt{\theta}\,(1-x_1)\right]\end{array}\right\}\right)$ | [0] |

Table 5.1. For each of four correlation-function class parameters, this table gives the equation number for the corresponding *IMSPE*, as well as the optimal design, which is a perfectly centered point, in each case.

Plots of $IMSPE$ vs. $x_1$, parameterized by $\theta$, are given in Fig. 5.1, below. That $x_1 = 0$ is the optimal design, for the exponential covariance function and any constant $\theta > 0$, is demonstrated by setting the derivative of $IMSPE$ with respect to $x_1$ to zero, solving for the value of $x_1$, and confirming that the second derivative of $IMSPE$ at this point is positive, as follows: $\frac{d\,IMSPE}{d\,x_1} = 2e^{-\theta}\sinh(\theta x_1) = 0$ has unique solution $x_1 = 0$, and $\left.\frac{d^2 IMSPE}{dx_1^2}\right|_{x_1=0} = 2\theta e^{-\theta}\cosh(\theta x_1)\big|_{x_1=0} = 2\theta e^{-\theta} > 0$, for $\theta > 0$. Similar demonstrations of the optimal designs for the Matérn- and Gaussian-covariance cases are left to the reader.

A just slightly more sophisticated demonstration that the optimal designs are always $x_1 = 0$ follows by noting $IMSPE = 2\left(1 - \frac{1}{2}\int_{-1}^{1} v_1 dx_1\right)$, as in Eq. L.1. This demonstration is again left as an exercise.

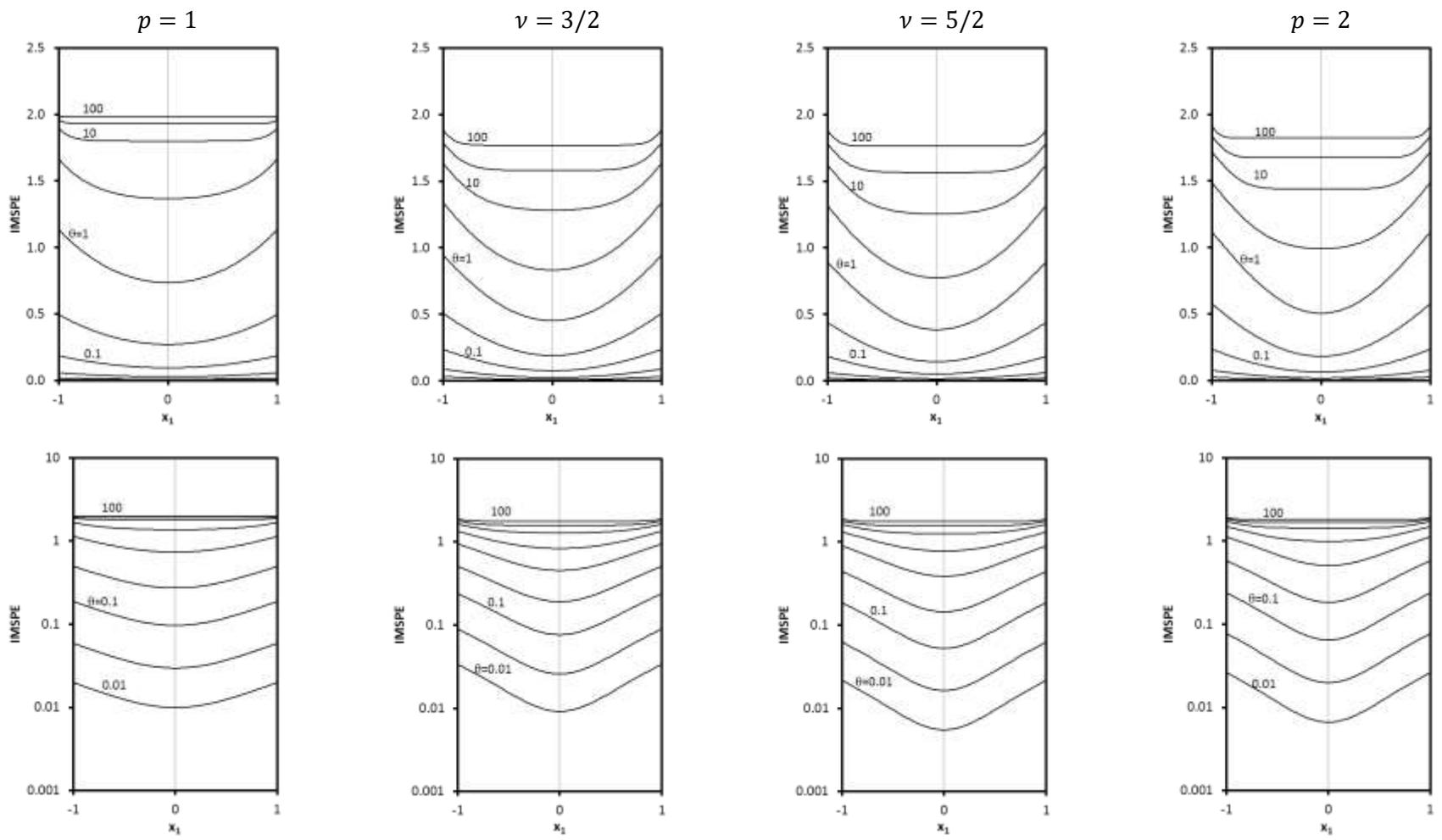

Fig. 5.1. For each of four correlation-function class parameters, given as column headings, this figure shows plots of $IMSPE$ vs. the location of the singleton design point $x_1$, parameterized by $\theta$. The vertical scale is linear (resp., logarithmic) in the upper (lower) row of plots.

# 6. $N = 2$ design examples, including optimal designs

The *IMSPE* formulas for designs with $[d, n] = [1,2]$ are given in Table 6.1, below, which also lists the equation number in the appendices for each formula. As demonstrated graphically in Fig. 6.1, below, each optimal design is a pair of points symmetrically disposed about the center of the design domain, with the coordinates of the points all falling roughly in the central three-fifths of the design domain. The numerical ranges of the design points are noted in the right-most column of Table 6.1. Plots of the *IMSPE*, vs. $x_1$ and parameterized by $\theta$, are given in Fig. 6.2, below, which also demonstrates the lowest-degree discontinuous derivative of the *IMSPE* for each correlation-function-class parameter, as discussed by Stein [9].

| Cov. function param. | Eq. | IMSPE formula | Optimal design: $[x_1, x_2 = -x_1]$ |
| --- | --- | --- | --- |
| | | | Range of $x_1's$ for $0.01 \leq \theta \leq 100$ |
| $p = 1$ | Q.1.1 | $\dfrac{3 + e^{-\theta\|x_1 - x_2\|}}{2}$ $+ \dfrac{e^{-\theta\|x_1-x_2\|} - e^{-2\theta}\cosh[\theta(x_1 + x_2)] + \theta\|x_1 - x_2\|e^{-\theta\|x_1-x_2\|}}{2\theta(1 - e^{-\theta\|x_1-x_2\|})}$ $- \dfrac{1 - e^{-\theta}\cosh(\theta x_1)}{\theta} - \dfrac{1 - e^{-2\theta}\cosh(2\theta x_1)}{4\theta(1 - e^{-\theta\|x_1-x_2\|})}$ $- \dfrac{1 - e^{-\theta}\cosh(\theta x_2)}{\theta} - \dfrac{1 - e^{-2\theta}\cosh(2\theta x_2)}{4\theta(1 - e^{-\theta\|x_1-x_2\|})}$ | $0.35 \leq \|x_1\| \leq 0.60$ |
| $\nu = 3/2$ | S.1 | *LSAE* | $0.42 \leq \|x_1\| \leq 0.59$ |
| $\nu = 5/2$ | T.1 | *LSAE* | $0.41 \leq \|x_1\| \leq 0.58$ |
| $p = 2$ | R.1.1 | *LSAE* | $0.42 \leq \|x_1\| \leq 0.58$ |

Table 6.1. For each of four correlation-function class parameters, this table gives the equation number for, and expression of, the corresponding *IMSPE* formula, as well as the range of coordinate values, $x_1$, of the two-point optimal designs. When the *IMSPE* formula is too long to fit the table, "LSAE" appears. LSAE stands for "long symbolic-algebra expression."

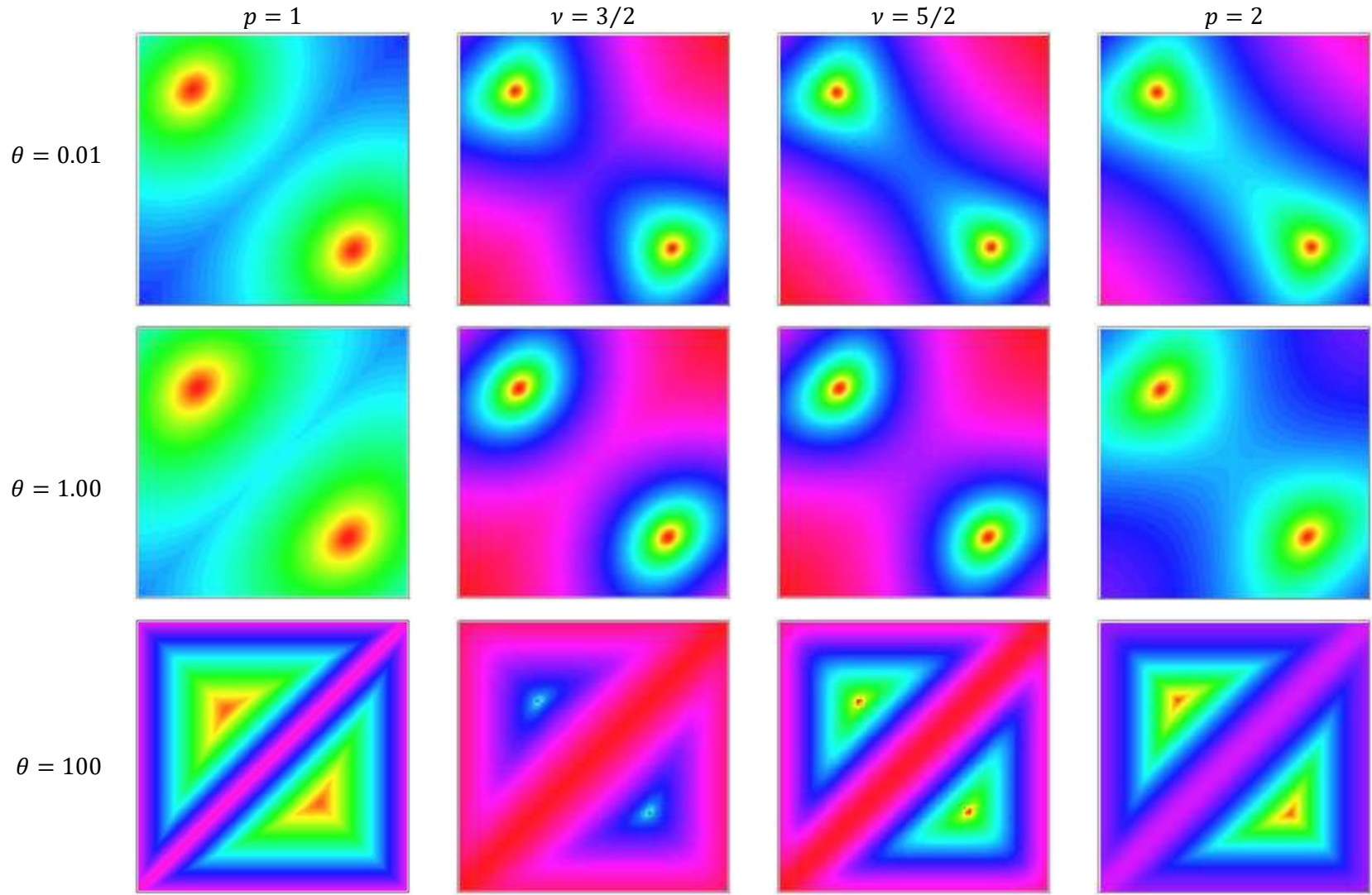

Fig. 6.1. For each of four correlation-function class parameters, given as column headings, this figure shows rainbow-coded hue plots of *IMSPE* vs. design-point locations, with $x_1$ as abscissa and $x_2$ as ordinate, for various $\theta$. The global minima, denoted by paired, small red regions, always occur on upper-left-to-lower-right diagonals. The figure demonstrates putatively that all the $[d, n] = [1,2]$ optimal designs are symmetric.

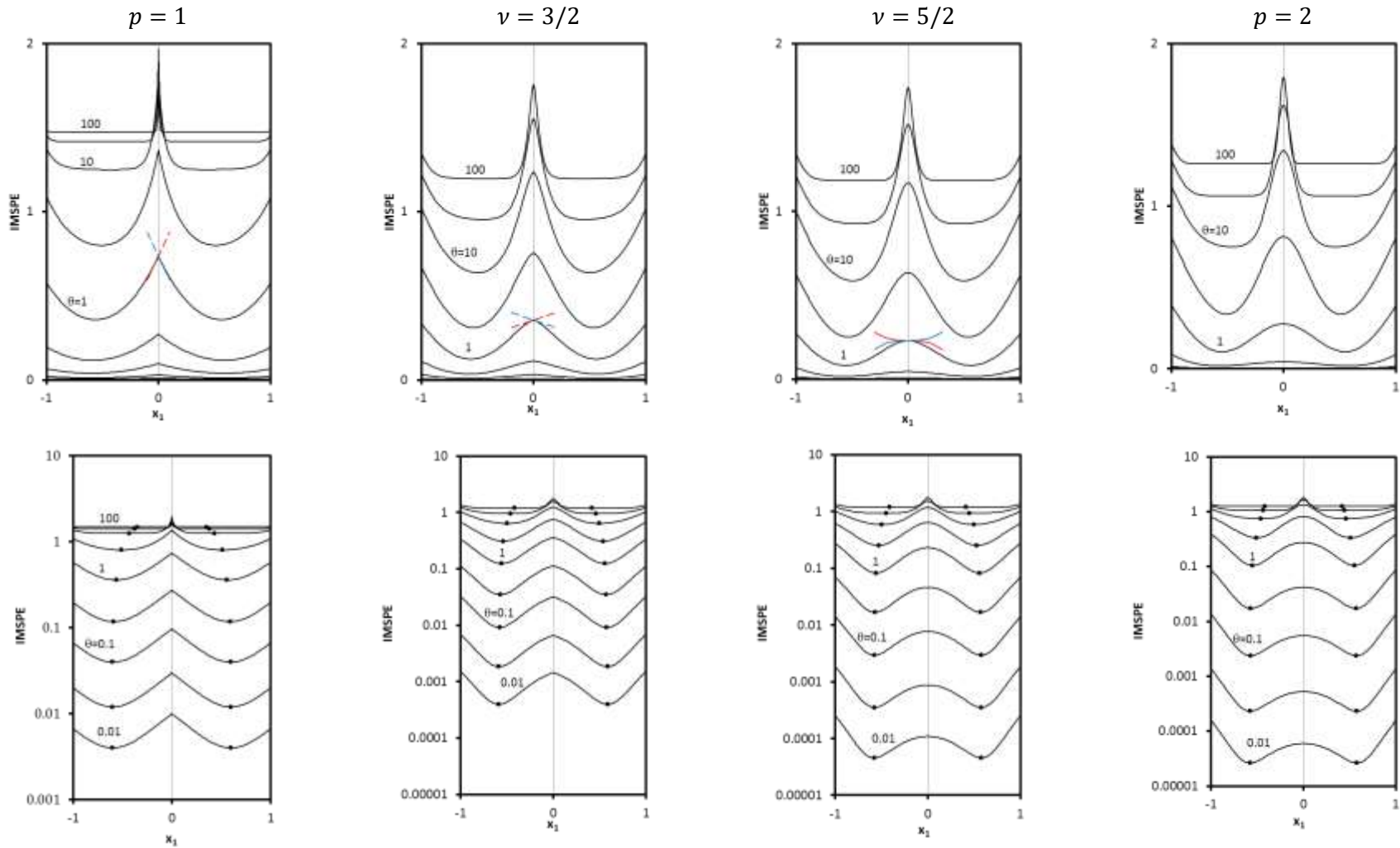

Fig. 6.2. For each of four correlation-function class parameters, given as column headings, this figure shows plots of $IMSPE$, parameterized by $\theta$, vs. location of one of the symmetrically disposed design points, $x_1$. The vertical scale is linear (resp., log) in the upper (resp., lower) row of plots. In the upper-row sub-plot with $\theta = 1$ and $p = 1$ (resp., $\nu = 3/2$, $\nu = 5/2$), colored lines are extrapolations at fixed first (resp., 2'nd, 3'rd) derivative of the $IMSPE$ from the origin, demonstrating discontinuous first; (resp., 2'nd, 3'rd) derivatives of the $IMSPE$. Dots in the lower plots show the loci of the minima of the curves, demonstrating that no optimal-design point is near the design-domain boundary.

We now demonstrate a simple technique for exploring whether any of the $p = 2$ optimal designs are twin-point designs. From Eq. R.3.2, we have the following Taylor-series expansion for $IMSPE$:

$$IMSPE = \begin{pmatrix} +\frac{1}{4}\begin{bmatrix}(1+x_t)e^{-2\theta(1+x_t)^2} \\ +(1-x_t)e^{-2\theta(1-x_t)^2}\end{bmatrix}^2 \\ -\sqrt{\frac{\pi}{4\theta}}\begin{Bmatrix} erf[\sqrt{\theta}(1+x_t)] \\ +erf[\sqrt{\theta}(1-x_t)]\end{Bmatrix} \\ -\sqrt{\frac{\pi}{128\theta}}\begin{Bmatrix} erf[\sqrt{2\theta}(1+x_t)] \\ +erf[\sqrt{2\theta}(1-x_t)]\end{Bmatrix}\end{pmatrix} + \begin{pmatrix} +\left[\frac{1}{4}+\frac{\theta}{3}(1+x_t)^2\right](1+x_t)e^{-2\theta(1+x_t)^2} \\ +\left[\frac{1}{4}+\frac{\theta}{3}(1-x_t)^2\right](1-x_t)e^{-2\theta(1-x_t)^2} \\ +\begin{bmatrix}(1+x_t)e^{-\theta(1+x_t)^2} \\ +(1-x_t)e^{-\theta(1-x_t)^2}\end{bmatrix} \\ -\sqrt{\frac{\pi}{128\theta}}\begin{Bmatrix} erf[\sqrt{2\theta}(1+x_t)] \\ +erf[\sqrt{2\theta}(1-x_t)]\end{Bmatrix}\end{pmatrix}^{-2}\theta\delta^2 + O(\theta^2\delta^4)$$

$$= IMSPE|_{\delta=0} + \left.\frac{\partial^2 IMSPE}{\partial(\theta\delta^2)}\right|_{\theta\delta^2=0} \cdot \theta\delta^2 + O(\theta^2\delta^4).$$

Optimal-design searches for this problem always yield designs symmetric about the origin, as shown graphically in Fig. 6.1, above, so $x_t = 0$, and thus the value of the optimal design $IMSPE_0$ is given by the following quadratic expression, for small $\sqrt{\theta}\delta$:

$$IMSPE_0 = 2 + \frac{e^{-2\theta}}{2} - \sqrt{\frac{\pi}{\theta}}erf(\sqrt{\theta}) - \sqrt{\frac{\pi}{32\theta}}erf(\sqrt{2\theta})$$

$$+ \left[-2 + \left(\frac{1}{2} + \frac{2\theta}{3}\right)e^{-2\theta} + 2e^{-\theta} - \sqrt{\frac{\pi}{32\theta}}erf(\sqrt{2\theta})\right]\theta\delta^2 + O(\theta^2\delta^4). \quad (6.1)$$

The following plot shows the term in square brackets of Eq. 6.1 is always negative, leading to the conclusion there can be no global minimum of $IMSPE$ for two proximal design points, thus ruling out twin-point optimal designs for the case $[d, n, p] = [1,2,2]$.

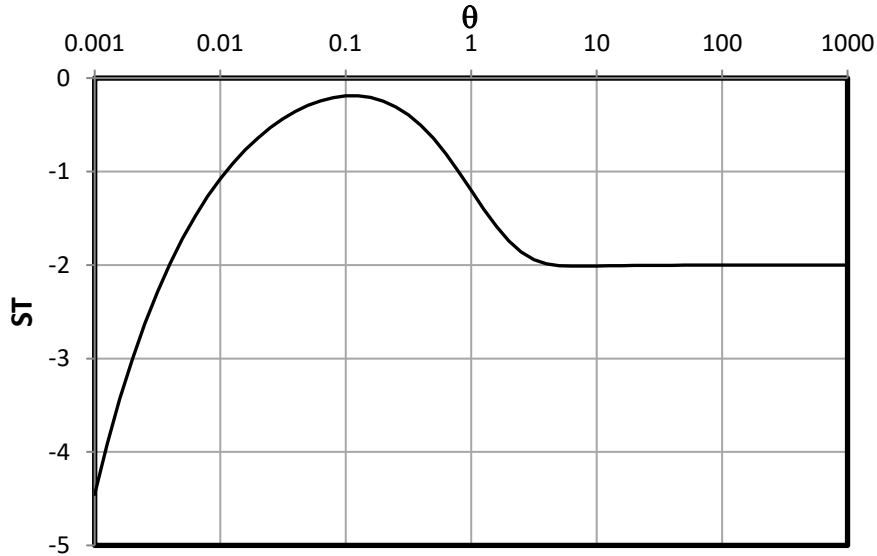

Fig. 6.3. The second term (ST) of the Taylor-series expansion is plotted, versus $\theta$, in this figure. ST is always negative, which supports the conclusion that there are no twin-point designs for $[d, n, p] = [1,2,2]$.

# 7. Attributes of the optimal designs

## 7.1 Optimal designs with $[d, n, p] = [any, 1, any]$

For fixed vector $\boldsymbol{\theta}$ comprised of all positive elements, simple extensions of Eq. L.1 show $IMSPE = 2(1 - R_{0,1})$. An optimal design will be centered if $\frac{d(IMSPE)}{dx_1} = -2\frac{dR_{0,1}}{dx_1} = 0$ has a unique solution at the center of the design domain and $\frac{d^2(IMSPE)}{dx_1^2}\Big|_{x_1\ center} = -2\frac{dR_{0,1}}{dx_1}\Big|_{x_1\ center} > 0$. For exponential covariance, i.e. $p = 1$, $R_{0,1} = \frac{1 - e^{-\theta}\cosh(\theta x_1)}{\theta}$, for which we saw, in Section 5, the optimal design was a centered point. For Gaussian covariance, i.e. $p = 2$, $R_{0,1} = \left(\frac{\pi}{16\theta}\right)^{1/2}\left\{\begin{array}{l}erf[\sqrt{\theta}(1 + x_1)] \\ +erf[\sqrt{\theta}(1 - x_1)]\end{array}\right\}$, for which we saw, in Section 5, the optimal design was, again, a centered point. Extensions to other covariance functions, such as those of the Matérn class, are left to the reader.

## 7.2 A possible centering effect

A notable attribute of the $N = 1$ optimal designs of Section 5 is the central location, within their common design domain, of their singleton points. This may indicate a centering effect for optimal-design points, even for problems with $N > 1$.

## 7.3 Designs are not required to include boundary points

$IMSPE$-optimal designs for $[d, n, p] = [1,2, any]$ do not include boundary points, in contrast with the mistaken notion that designs for $[d, n, p] = [1,2,2]$ must be points on the boundary, as expressed in Thm. 4.2 of [11]. We identified the problem as a conflation of an informal rule-of-thumb in Sub-section 3.1 of [11] with a rigorous mathematical statement.

# 8. Summary and concluding comment

This paper provides a detailed algebraic basis for future research on clustered designs, under a variety of correlation functions, as well as the beginning of a basis for the development of the concept of the Nu-class of low-degree-truncated rational functions.

# 9. Research reproducibility

We support the recommendations of ICERM's Workshop on Reproducibility in Computational and Experimental Mathematics Workshop [13]. In addition to the extensive appendices in this paper, all other data, figure-generation files, and SMS codes used in this research are available to responsible parties from the author at selden_crary (at) yahoo (dot) com.

# 10. Revision history

V2: On Page 6, the top row of the definition of matrix $\boldsymbol{L}$ was corrected. Sec. 9 on research reproducibility was added. Appendix H was rewritten to discuss the removable singularity in the SMS result.

V3: In each of Table 4.2, Table 5.1, and Eq. B.7, a "6" was corrected to a "2" in expressions for the case $\nu = 3/2$. In Appendix C, the second line was corrected to take into account all cases of $p$ and $\nu$, and two missing summation symbols were added for each of the two Matérn cases.

V4: Major changes were made to Sec. R.5. Ref. 6 was corrected. "Generalized" was added to the title.

## Acknowledgments

We thank Prof. Max Morris of Iowa State Univ. for encouraging inclusion of Matérn covariance functions; Prof. Rafe Mazzeo of Stanford Univ. and Bruce Pourciau of Lawrence Univ. for helpful and encouraging conversations; Prof. David H. Bailey of the Univ. of California, Davis for pointing us to Ref. 13; and Prof. Milan Stehlík of Johannes Kepler Univ., Linz, Austria for correcting an algebra error.

## Appendix A. Matrix identities

**A1:** The trace of the product of $n \times n$ ($n \geq 1$) matrices $\mathbf{A}$ and $\mathbf{B}$, with $\mathbf{B}$ symmetric, equals the sum of their element-by-element products, i.e. $tr(\mathbf{AB}) = \sum_{i,j=1}^{n} A_{i,j} B_{i,j}$. (A.1)

Demonstration: $tr(\mathbf{AB}) = tr(\sum_{j=1}^{n} A_{i,j} B_{j,k}) = \sum_{i=1}^{n} \sum_{j=1}^{n} A_{i,j} B_{j,i} = \sum_{i,j=1}^{n} A_{i,j} B_{i,j}$.

**A2:** The inverse of the symmetric 3x3 matrix $A \equiv \begin{bmatrix} a & b & c \\ b & d & e \\ c & e & f \end{bmatrix}$ is given by the following, which was generated via SMS:

$$A^{-1} = \frac{1}{det(A)} \begin{bmatrix} df - e^2 & -bf + ce & be - cd \\ -bf + ce & af - c^2 & -ae + bc \\ be - cd & -ae + bc & ad - b^2 \end{bmatrix}, \text{ where } (A) = adf - ae^2 - b^2 f + 2bce - c^2 d.$$

(A.2)

## Appendix B. Basic exponential-, Matérn-, and Gaussian-correlation integrals

For $-1 \leq a \leq b \leq 1$, we have the following ten basic integrals:

*Exponential-covariance-function class, exponential-covariance sub-class, as detailed in Appendices D and E:*

$$I_1 \equiv \frac{1}{2} \int_{-1}^{1} e^{-\theta|a-x|} dx = \frac{1 - e^{-\theta} \cosh(\theta a)}{\theta}.$$ (B.1)

$$J_1 \equiv \int_{0}^{1} e^{-\theta|a-x|} dx = \frac{2 - e^{-\theta a} - e^{-\theta(1-a)}}{\theta}.$$ (B.2)

$$I_2 \equiv \frac{1}{2} \int_{-1}^{1} e^{-\theta(|a-x|+|b-x|)} dx = \frac{e^{-\theta(b-a)} - e^{-2\theta} \cosh[\theta(a+b)]}{2\theta} + \frac{(b-a) e^{-\theta(b-a)}}{2}.$$ (B.3)

$$J_2 \equiv \int_{0}^{1} e^{-(|a-x|+|b-x|)} dx = \frac{2 e^{-\theta(b-a)} - e^{-\theta(a+b)} - e^{-\theta(2-a-b)}}{2\theta} + (b-a) e^{-\theta(b-a)}.$$ (B.4)

*Exponential-covariance-function class, Gaussian-covariance sub-class, as detailed in Appendices F and G:*

$$I_3 \equiv \frac{1}{2} \int_{-1}^{1} e^{-\theta(a-x)^2} dx = \sqrt{\frac{\pi}{16\theta}} \left\{ \begin{array}{l} erf[\sqrt{\theta}(1+a)] \\ +erf[\sqrt{\theta}(1-a)] \end{array} \right\}.$$ (B.5)

$$I_4 \equiv \frac{1}{2} \int_{-1}^{1} e^{-\theta[(a-x)^2 + (b-x)^2]} dx = \sqrt{\frac{\pi}{32\theta}} \left\{ \begin{array}{l} erf\left[\sqrt{2\theta}\left(1 + \frac{a+b}{2}\right)\right] \\ +erf\left[\sqrt{2\theta}\left(1 - \frac{a+b}{2}\right)\right] \end{array} \right\} e^{-\frac{\theta(a-b)^2}{2}}.$$ (B.6)

*Matérn-covariance-function class, parameter 3/2 sub-class, as detailed in Appendices H and I:*

$$I_5 \equiv \tfrac{1}{2}\int_{-1}^{1}\left[1+\sqrt{3\theta(a-x)^2}\right]e^{-\sqrt{3\theta(a-x)^2}}\,dx$$

$$= \frac{1}{2\sqrt{3\theta}}\left(2\left\{\begin{matrix}\left[1-e^{-\sqrt{3\theta}\,(1+a)}\right]\\ +\left[1-e^{-\sqrt{3\theta}\,(1-a)}\right]\end{matrix}\right\} - \sqrt{3\theta}\left\{\begin{matrix}(1+a)e^{-\sqrt{3\theta}\,(1+a)}\\ +(1-a)e^{-\sqrt{3\theta}\,(1-a)}\end{matrix}\right\}\right). \quad (B.7)$$

$$I_6 \equiv \tfrac{1}{2}\int_{-1}^{1}\left[1+\sqrt{3\theta(a-x)^2}\right]\left[1+\sqrt{3\theta(b-x)^2}\right]e^{-\sqrt{3\theta[(a-x)^2+(b-x)^2]}}\,dx$$

$$= \text{long expression given as Eq. I.1.} \quad (B.8)$$

*Matérn-covariance-function class, parameter 5/2 subclass, as detailed in Appendices J and K:*

$$I_7 \equiv \tfrac{1}{2}\int_{-1}^{1}\left[1+\sqrt{5\theta(a-x)^2}+\tfrac{5\theta(a-x)^2}{3}\right]e^{-\sqrt{5\theta(a-x)^2}}\,dx$$

$$= \frac{1}{6\sqrt{5\theta}}\left\{8\left[\begin{matrix}1-e^{-\sqrt{5\theta}\,(1+a)}\\ +1-e^{-\sqrt{5\theta}\,(1-a)}\end{matrix}\right] - 5\sqrt{5\theta}\left[\begin{matrix}(1+a)e^{-\sqrt{5\theta}\,(1+a)}\\ +(1-a)e^{-\sqrt{5\theta}\,(1-a)}\end{matrix}\right] - 5\theta\left[\begin{matrix}(1+a)^2 e^{-\sqrt{5\theta}\,(1+a)}\\ +(1-a)^2 e^{-\sqrt{5\theta}\,(1-a)}\end{matrix}\right]\right\}. \quad (B.9)$$

$$I_8 \equiv \tfrac{1}{2}\int_{-1}^{1}\left[1+\sqrt{5\theta(a-x)^2}+\tfrac{5\theta(a-x)^2}{3}\right]\left[1+\sqrt{5\theta(b-x)^2}+\tfrac{5\theta(b-x)^2}{3}\right]e^{-\sqrt{5\theta[(a-x)^2+(b-x)^2]}}\,dx$$

$$= \text{long expression given as Eq. K.1.} \quad (B.10)$$

## Appendix C.  Integrals that appear in matrix $R$

From integrals $I_1$ through $I_8$, as well as integrals $J_1$ and $J_2$, all of which were defined in Appendix B; and using the definitions for $v_{i,k}/\sigma_z^2$ given in Table 4.1, but, going forward, with the hyper-parameter $\sigma_z^2$ suppressed; we have the following ten integrals that appear in matrix $R$, for the examples in this paper. Absolute value symbols appear in some cases, to overcome Appendix B's overly restrictive assumption, $-1 \le a \le b \le 1$. N.B.: Careful attention must be given to the use of absolute values in the generation of machine-readable symbolic expressions for $I_{14}(x_i, x_j; \theta)$ and $I_{16}(x_i, x_j; \theta)$. Maple input and output code is available from the author for these moderately complex integrals. Throughout this section, the indices $i$ and $j$ each run from 1 through $n$, and the index $k$ runs from 1 through $d$.

*Exponential-covariance-function class, exponential-covariance sub-class, and making use of Eqs. B.1-B.4:*

$$I_9(x_i;\theta) \equiv \tfrac{1}{2^d}\int_{-1}^{1}\int_{-1}^{1}\cdots\int_{-1}^{1} v_{i,k}\,dx_1 dx_2\cdots dx_d$$

$$= \tfrac{1}{2^d}\int_{-1}^{1}\int_{-1}^{1}\cdots\int_{-1}^{1} e^{-\sum_{k=1}^{d}\theta_k|x_{i,k}-x_k|}\,dx_1 dx_2\cdots dx_d$$

$$= \prod_{k=1}^{d} \frac{1-e^{-\theta_k}\cosh(\theta_k x_{i,k})}{\theta_k}. \quad (C.1)$$

$$J_9(x_i;\theta) \equiv \int_0^1\int_0^1\cdots\int_0^1 v_{i,k}\,dx_1 dx_2\cdots dx_d = \prod_{k=1}^{d}\frac{2-e^{-\theta_k x_{i,k}}-e^{-\theta_k(1-x_{i,k})}}{\theta_k}. \quad (C.2)$$

$$I_{10}(x_i,x_j;\theta) \equiv \tfrac{1}{2^d}\int_{-1}^{1}\int_{-1}^{1}\cdots\int_{-1}^{1} v_{i,k}\,v_{j,k}\,dx_1 dx_2\cdots dx_d$$

$$= \prod_{k=1}^{d} \left\{ \frac{e^{-\theta_k |x_{i,k}-x_{j,k}|} - e^{-2\theta_k} \cosh[\theta_k(x_{i,k}+x_{j,k})]}{2\theta_k} + \frac{|x_{i,k}-x_{j,k}| \, e^{-\theta_k|x_{i,k}-x_{j,k}|}}{2} \right\}. \tag{C.3}$$

$$J_{10}(\pmb{x_i}, \pmb{x_j}; \pmb{\theta}) \equiv \int_0^1 \int_0^1 \cdots \int_0^1 v_{i,k} \, v_{j,k} dx_1 dx_2 \cdots dx_d$$

$$= \prod_{k=1}^{d} \left[ \frac{2e^{-\theta_k|x_{i,k}-x_{j,k}|} - e^{-\theta_k(x_{i,k}+x_{j,k})} - e^{-\theta_k(2-x_{i,k}-x_{j,k})}}{2\theta_k} + |x_{i,k} - x_{j,k}| \, e^{-\theta|x_{i,k}-x_{j,k}|} \right]. \tag{C.4}$$

*Exponential-covariance-function class, Gaussian-covariance sub-class, and making use of Eqs. B.5 and B.6:*

$$I_{11}(\pmb{x_i}; \pmb{\theta}) \equiv \frac{1}{2^d} \int_{-1}^{1} \int_{-1}^{1} \cdots \int_{-1}^{1} v_{i,k} \, dx_1 dx_2 \cdots dx_d$$

$$= \frac{1}{2^d} \int_{-1}^{1} \int_{-1}^{1} \cdots \int_{-1}^{1} e^{-\sum_{k=1}^{d} \theta_k (x_{i,k}-x_k)^2} dx_1 dx_2 \cdots dx_d$$

$$= \prod_{k=1}^{d} \sqrt{\frac{\pi}{16\theta_k}} \left\{ \begin{array}{l} erf[\sqrt{\theta_k}(1+x_{i,k})] \\ +erf[\sqrt{\theta_k}(1-x_{i,k})] \end{array} \right\}. \tag{C.5}$$

$$I_{12}(\pmb{x_i}, \pmb{x_j}; \pmb{\theta}) \equiv \frac{1}{2^d} \int_{-1}^{1} \int_{-1}^{1} \cdots \int_{-1}^{1} v_{i,k} \, v_{j,k} dx_1 dx_2 \cdots dx_d$$

$$= \frac{1}{2^d} \int_{-1}^{1} \int_{-1}^{1} \cdots \int_{-1}^{1} e^{-\sum_{k=1}^{d} \theta_k \left[(x_{i,k}-x_k)^2 + (x_{j,k}-x_k)^2\right]} dx_1 dx_2 \cdots dx_d$$

$$= \prod_{k=1}^{d} \sqrt{\frac{\pi}{32\theta_k}} \left\{ \begin{array}{l} erf\left[\sqrt{2\theta_k}\left(1 + \frac{x_{i,k}+x_{j,k}}{2}\right)\right] \\ +erf\left[\sqrt{2\theta_k}\left(1 - \frac{x_{i,k}+x_{j,k}}{2}\right)\right] \end{array} \right\} e^{\frac{-\theta_k(x_{i,k}-x_{j,k})^2}{2}}. \tag{C.6}$$

*Matérn-covariance-function class, parameter 3/2 sub-class, and making use of Eqs. B.7 and B.8:*

$$I_{13}(\pmb{x_i}; \pmb{\theta}) \equiv \frac{1}{2^d} \int_{-1}^{1} \int_{-1}^{1} \cdots \int_{-1}^{1} v_{i,k} \, dx_1 dx_2 \cdots dx_d$$

$$= \frac{1}{2^d} \int_{-1}^{1} \int_{-1}^{1} \cdots \int_{-1}^{1} \left\{ \prod_{k=1}^{d} \left[ 1 + \sqrt{3\theta_k (x_{i,k}-x_k)^2} \right] e^{-\sqrt{3\theta_k(x_{i,k}-x_k)^2}} \right\} dx_1 dx_2 \cdots dx_d$$

$$= \prod_{k=1}^{d} \frac{1}{6\sqrt{3\theta_k}} \left\{ \begin{array}{l} 2\left[ \begin{array}{l} 1 - e^{-\sqrt{3\theta_k}(1+x_{i,k})} \\ +1 - e^{-\sqrt{3\theta_k}(1-x_{i,k})} \end{array} \right] \\ -\sqrt{3\theta_k} \left[ \begin{array}{l} (1+x_{i,k})e^{-\sqrt{3\theta_k}(1+x_{i,k})} \\ +(1-x_{i,k})e^{-\sqrt{3\theta_k}(1-x_{i,k})} \end{array} \right] \end{array} \right\}. \tag{C.7}$$

$$I_{14}(\pmb{x_i}, \pmb{x_j}; \pmb{\theta}) \equiv \frac{1}{2^d} \int_{-1}^{1} \int_{-1}^{1} \cdots \int_{-1}^{1} v_{i,k} \, v_{j,k} dx_1 dx_2 \cdots dx_d$$

$$= \frac{1}{2^d} \int_{-1}^{1} \int_{-1}^{1} \cdots \int_{-1}^{1} \left( \prod_{k=1}^{d} \left\{ \begin{array}{l} \left[1 + \sqrt{3\theta_k (x_{i,k} - x_k)^2}\right] \\ \cdot \left[1 + \sqrt{3\theta_k (x_{j,k} - x_k)^2}\right] \\ \cdot e^{-\sqrt{3\theta_k \left[(x_{i,k}-x_k)^2 + (x_{j,k}-x_k)^2\right]}} \end{array} \right\} \right) dx_1 dx_2 \cdots dx_d$$

$$= a\ machine-readable\ symbolic\ expression. \tag{C.8}$$

*Matérn-covariance-function class, parameter 5/2 subclass, and making use of Eqs. B.9 and B.10:*

$$I_{15}(\boldsymbol{x_i}; \boldsymbol{\theta}) \equiv \frac{1}{2^d} \int_{-1}^{1} \int_{-1}^{1} \cdots \int_{-1}^{1} v_{i,k}\, dx_1 dx_2 \cdots dx_d$$

$$= \frac{1}{2^d} \int_{-1}^{1} \int_{-1}^{1} \cdots \int_{-1}^{1} \left\{ \prod_{k=1}^{d} \left[1 + \sqrt{5\theta_k (x_{i,k} - x_k)^2} + \frac{5\theta_k(x_{i,k}-x_k)^2}{3}\right] e^{-\sqrt{5\theta_k(x_{i,k}-x_k)^2}} \right\} dx_1 dx_2 \cdots dx_d$$

$$= \prod_{k=1}^{d} \frac{1}{6\sqrt{5\theta_k}} \left( \begin{array}{l} 8 \left\{ \begin{array}{l} \left[1 - e^{-\sqrt{5\theta_k}\,(1+x_{i,k})}\right] \\ + \left[1 - e^{-\sqrt{5\theta_k}\,(1-x_{i,k})}\right] \end{array} \right\} \\ -5\sqrt{5\theta_k} \left\{ \begin{array}{l} (1 + x_{i,k}) e^{-\sqrt{5\theta_k}\,(1+x_{i,k})} \\ +(1 - x_{i,k}) e^{-\sqrt{5\theta_k}\,(1-x_{i,k})} \end{array} \right\} \\ -5\theta_k \left\{ \begin{array}{l} (1 + x_{i,k})^2 e^{-\sqrt{5\theta_k}\,(1+x_{i,k})} \\ +(1 - x_{i,k})^2 e^{-\sqrt{5\theta_k}\,(1-x_{i,k})} \end{array} \right\} \end{array} \right). \tag{C.9}$$

$$I_{16}(\boldsymbol{x_i}, \boldsymbol{x_j}; \boldsymbol{\theta}) \equiv \frac{1}{2^d} \int_{-1}^{1} \int_{-1}^{1} \cdots \int_{-1}^{1} v_{i,k}\, v_{j,k}\, dx_1 dx_2 \cdots dx_d$$

$$= \frac{1}{2^d} \int_{-1}^{1} \int_{-1}^{1} \cdots \int_{-1}^{1} \prod_{k=1}^{d} \left( \left\{ \begin{array}{l} \left[1 + \sqrt{5\theta_k (x_{i,k} - x_k)^2} + \frac{5\theta_k(x_{i,k}-x_k)^2}{3}\right] \\ \cdot \left[1 + \sqrt{5\theta_k (x_{j,k} - x_k)^2} + \frac{5\theta_k(x_{j,k}-x_k)^2}{3}\right] \\ \cdot e^{-\sqrt{5\theta_k\left[(x_{i,k}-x_k)^2 + (x_{j,k}-x_k)^2\right]}} \end{array} \right\} \right) dx_1 dx_2 \cdots dx_d.$$

$$= a\ machine-readable\ symbolic\ expression. \tag{C.10}$$

## Appendix D.  Hand and SMS algebra evaluating integrals $I_1$ and $J_1$

Hand algebra:

$$I_1(a) \equiv \frac{1}{2} \int_{-1}^{1} e^{-\theta|a-x|}\, dx = \frac{1}{2} \left[ \begin{array}{l} \int_{-1}^{a} e^{-\theta(a-x)}\, dx \\ + \int_{a}^{1} e^{-\theta(x-a)}\, dx \end{array} \right] = \frac{1}{2\theta} \left[ \begin{array}{l} 1 - e^{-\theta(a+1)} \\ -e^{-\theta(1-a)} + 1 \end{array} \right] = \frac{1 - e^{-\theta}\cosh(\theta a)}{\theta}. \tag{D.1}$$

$$J_1(a) \equiv \int_{0}^{1} e^{-\theta|a-x|}\, dx = \left[ \begin{array}{l} \int_{0}^{a} e^{-\theta(a-x)}\, dx \\ + \int_{a}^{1} e^{-\theta(x-a)}\, dx \end{array} \right] = \frac{1}{\theta} \left[ \begin{array}{l} 1 - e^{-\theta a} \\ -e^{-\theta(1-a)} + 1 \end{array} \right] = \frac{2 - e^{-\theta a} - e^{-\theta(1-a)}}{\theta}. \tag{D.2}$$

SMS algebra: Maple inputs and outputs, finished with hand algebra, follow:

```
> I1:=(1/2)*int(exp(-theta*abs(a-x)),x=-1..1) assuming -1<a
assuming a<1;
```

$$I1 := \frac{1}{2} \frac{\left(-1 - e^{2\theta a} + 2 e^{a\theta + \theta}\right) e^{-a\theta - \theta}}{\theta}$$

$$= \frac{[-1 - e^{2\theta a} + 2e^{\theta(1+a)}]e^{-\theta(1+a)}}{2\theta} = \frac{-e^{-\theta(1+a)} - e^{2\theta a}e^{-\theta(1+a)} + 2}{2\theta} = \frac{-e^{-\theta(1+a)} - e^{-\theta(1-a)} + 2}{2\theta} = \frac{1 - e^{-\theta}\cosh(\theta a)}{\theta},$$

which matches Eq. D.1 exactly.

```
> J1:=int(exp(-theta*abs(a-x)),x=0..1)  assuming  0<a  assuming
a<1;
```

$$J1 := \frac{\left(-1 - e^{2a\theta - \theta} + 2 e^{\theta a}\right) e^{-\theta a}}{\theta}$$

$$= \frac{(-1 - e^{2\theta a - \theta} + 2e^{\theta a})e^{-\theta a}}{\theta} = \frac{-e^{-\theta a} - e^{2\theta a - \theta}e^{-\theta a} + 2e^{\theta a}e^{-\theta a}}{\theta} = \frac{-e^{-\theta a} - e^{\theta a - \theta} + 2}{\theta} = \frac{2 - e^{-\theta a} - e^{-\theta(1-a)}}{\theta},$$

which matches Eq. D.2 exactly.

## Appendix E. Hand and SMS algebra evaluating integrals $I_2$ and $J_2$

Hand algebra:

$$\begin{aligned}
I_2(a,b) &\equiv \frac{1}{2}\int_{-1}^{1} e^{-\theta(|a-x|+|b-x|)}\, dx \\
&= \frac{1}{2}\left\{\int_{-1}^{a} e^{-\theta[(a-x)+(b-x)]}\, dx + \int_{a}^{b} e^{-\theta[(x-a)+(b-x)]}\, dx + \int_{b}^{1} e^{-\theta[(x-a)+(x-b)]}\, dx\right\} \\
&= \frac{1}{2}\left[\int_{-1}^{a} e^{-\theta(a+b-2x)}\, dx + \int_{a}^{b} e^{-\theta(b-a)}\, dx + \int_{b}^{1} e^{-\theta(2x-a-b)}\, dx\right] \\
&= \frac{1}{2}\left[\frac{e^{-\theta(b-a)}}{2\theta} - \frac{e^{-\theta(2+a+b)}}{2\theta} + (b-a)e^{-\theta(b-a)} - \frac{e^{-\theta(2-a-b)}}{2\theta} + \frac{e^{-\theta(b-a)}}{2\theta}\right] \\
&= \frac{1}{4\theta}\left[e^{-\theta(b-a)} - e^{-\theta(2+a+b)} + 2\theta(b-a)e^{-\theta(b-a)} - e^{-\theta(2-a-b)} + e^{-\theta(b-a)}\right] \\
&= \frac{1}{4\theta}\left\{2[1+\theta(b-a)]e^{-\theta(b-a)} - e^{-2\theta}\left[e^{-\theta(a+b)} + e^{\theta(a+b)}\right]\right\} \\
&= \frac{1}{2\theta}\left\{[1+\theta(b-a)]e^{-\theta(b-a)} - e^{-2\theta}\cosh[\theta(a+b)]\right\} \\
&= \frac{e^{-\theta(b-a)} - e^{-2\theta}\cosh[\theta(a+b)]}{2\theta} + \frac{(b-a)e^{-\theta(b-a)}}{2}. \tag{E.1}
\end{aligned}$$

$$J_2(a,b) \equiv \int_0^1 e^{-\theta(|a-x|+|b-x|)}\,dx$$

$$= \int_0^a e^{-\theta[(a-x)+(b-x)]}\,dx + \int_a^b e^{-\theta[(x-a)+(b-x)]}\,dx + \int_b^1 e^{-\theta[(x-a)+(x-b)]}\,dx$$

$$= \int_0^a e^{-\theta(a+b-2x)}\,dx + \int_a^b e^{-\theta(b-a)}\,dx + \int_b^1 e^{-\theta(2x-a-b)}\,dx$$

$$= \frac{e^{-\theta(b-a)}}{2\theta} - \frac{e^{-\theta(a+b)}}{2\theta} + (b-a)e^{-\theta(b-a)} - \frac{e^{-\theta(2-a-b)}}{2\theta} + \frac{e^{-\theta(b-a)}}{2\theta}$$

$$= \frac{2e^{-\theta(b-a)} - e^{-\theta(a+b)} - e^{-\theta(2-a-b)}}{2\theta} + (b-a)e^{-\theta(b-a)}. \tag{E.2}$$

SMS algebra: Maple inputs and outputs, finished with hand algebra, follow:

```
> I2:=(1/2)*int(exp(-theta*(abs(a-x)+abs(b-x))),x=-1..1)
assuming -1<a assuming a<b assuming a<1 assuming -1<b assuming
b<1;
```

$$I2 := -\frac{1}{4}\,\frac{2e^{\theta(-b+a)}a\theta - 2e^{\theta(-b+a)}b\theta + e^{-(a+2+b)\theta} + e^{(a-2+b)\theta} - 2e^{\theta(-b+a)}}{\theta}$$

$$= \frac{-[2a\theta e^{-\theta(b-a)} - 2b\theta e^{-\theta(b-a)} + e^{-\theta(2+a+b)} + e^{-\theta(2-a-b)} - 2e^{-\theta(b-a)}]}{4\theta}$$

$$= \frac{2\theta(b-a)e^{-\theta(b-a)} - e^{-2\theta}[e^{-\theta(a+b)} + e^{\theta(a+b)}] + 2e^{-\theta(b-a)}}{4\theta}$$

$$= \frac{e^{-\theta(b-a)} - e^{-2\theta}\cosh[\theta(a+b)]}{2\theta} + \frac{(b-a)e^{-\theta(b-a)}}{2}, \text{ which matches Eq. E.1 exactly.}$$

```
>   J2:=int(exp(-theta*(abs(a-x)+abs(b-x))),x=0..1)   assuming   0<a
assuming a<b assuming a<1 assuming 0<b assuming b<1;
```

$$J2 := -\frac{1}{2}\,\frac{2e^{\theta(-b+a)}a\theta - 2e^{\theta(-b+a)}b\theta + e^{-(a+b)\theta} + e^{(a-2+b)\theta} - 2e^{\theta(-b+a)}}{\theta}$$

$$= \frac{-2a\theta e^{-\theta(b-a)} + 2b\theta e^{-\theta(b-a)} - e^{-\theta(a+b)} - e^{-\theta(2-a-b)} + 2e^{-\theta(b-a)}}{2\theta}$$

$$= \frac{2e^{-\theta(b-a)} - e^{-\theta(a+b)} - e^{-\theta(2-a-b)}}{2\theta} + (b-a)e^{-\theta(b-a)}, \text{ which matches Eq. E.2 exactly.}$$

## Appendix F.  Hand and SMS algebra evaluating integral $I_3$

Hand algebra:

From Dwight 590 [12], $erf(x) \equiv \frac{2}{\sqrt{\pi}}\int_{t=0}^{x} e^{-t^2}\,dt$, the integrand of which is an even function of $t$, we have

$erf(x) = \frac{1}{\sqrt{\pi}}\int_{t=-x}^{x} e^{-t^2}\,dt$, which is an odd function of $x$. Changing variables via $t = \sqrt{\theta}(a - \tilde{x})$ gives $erf(x) = \sqrt{\frac{\theta}{\pi}} \int_{\tilde{x}=a-x/\sqrt{\theta}}^{-a+x/\sqrt{\theta}} e^{-\theta(a-\tilde{x})^2} d\tilde{x}$. Thus, $erf[\sqrt{\theta}(1+a)] = \sqrt{\frac{\theta}{\pi}} \int_{\tilde{x}=-1}^{1} e^{-\theta(a-\tilde{x})^2} d\tilde{x}$, and

$erf[\sqrt{\theta}(1-a)] = \sqrt{\frac{\theta}{\pi}} \int_{\tilde{x}=-1+2a}^{1-2a} e^{-\theta(a-\tilde{x})^2} d\tilde{x}$. Summing these last two equations and dropping the dummy tildes gives the following:

$$\sqrt{\frac{\pi}{16\theta}} \left\{ \begin{array}{l} erf[\sqrt{\theta}(1+a)] \\ +erf[\sqrt{\theta}(1-a)] \end{array} \right\}$$

$$= \frac{1}{4}\left[ \int_{x=-1}^{1} e^{-\theta(a-x)^2} dx + \int_{x=-1+2a}^{1-2a} e^{-\theta(a-x)^2} dx \right]$$

$$= \frac{1}{4} \left[ \begin{array}{l} \int_{x=-1}^{1} e^{-\theta(a-x)^2} dx + \int_{x=-1+2a}^{-1} e^{-\theta(a-x)^2} dx \\ + \int_{x=-1}^{1} e^{-\theta(a-x)^2} dx + \int_{x=1}^{1-2a} e^{-\theta(a-x)^2} dx \end{array} \right], \text{ where the 2'nd and 4'th terms cancel,}$$

$$= \frac{1}{2} \int_{x=-1}^{1} e^{-\theta(a-x)^2} dx. \text{ Thus,}$$

$I_3(a) \quad = \frac{1}{2} \int_{x=-1}^{1} e^{-\theta(a-x)^2} dx = \sqrt{\frac{\pi}{16\theta}} \left\{ \begin{array}{l} erf[\sqrt{\theta}(1+a)] \\ +erf[\sqrt{\theta}(1-a)] \end{array} \right\}.$ \hfill (F.1)

SMS algebra: Maple input and output, finished with hand algebra, that agree with Eq. F.1, follow:

```
> I3:=(1/2)*int(exp(-theta*(a-x)^2),x=-1..1);
```

$I3 := -\frac{1}{4} \frac{\sqrt{\pi}\left(-\text{erf}\left(\sqrt{\theta}+\sqrt{\theta}\,a\right) + \text{erf}\left(-\sqrt{\theta}+\sqrt{\theta}\,a\right)\right)}{\sqrt{\theta}}$

$= \sqrt{\frac{\pi}{16\theta}} \left\{ \begin{array}{l} erf[\sqrt{\theta}(1+a)] \\ +erf[\sqrt{\theta}(1-a)] \end{array} \right\}$, which uses the fact that $erf(x)$ is an odd function of its argument.

## Appendix G. Hand and SMS algebra evaluating integral $I_4$

Hand algebra is left as an exercise for the reader. N.B.: $erf(x)$ is an odd function of its argument. The result, Eq. G, agrees with the SMS algebra, below:

$I_4(a,b) \quad \equiv \frac{1}{2}\int_{-1}^{1} e^{-\theta[(a-x)^2+(b-x)^2]} dx = \sqrt{\frac{\pi}{32\theta}} \left\{ \begin{array}{l} erf\left[\sqrt{2\theta}\left(1+\frac{a+b}{2}\right)\right] \\ +erf\left[\sqrt{2\theta}\left(1-\frac{a+b}{2}\right)\right] \end{array} \right\} e^{\frac{-\theta(a-b)^2}{2}}.$ \hfill (G.1)

SMS algebra: Maple input and output that agree, by inspection, with Eq. G.1, follow:

```
> I4:=(1/2)*int(exp(-theta*((a-x)^2+(b-x)^2)),x=-1..1);
```

$$I4 := -\frac{1}{8}\frac{1}{\sqrt{\theta}}\left(\sqrt{\pi}\,e^{-\frac{1}{2}a^2\theta-\frac{1}{2}b^2\theta+\theta ab}\sqrt{2}\left(-\mathrm{erf}\left(\sqrt{2}\sqrt{\theta}+\frac{1}{2}\sqrt{2}\sqrt{\theta}\,a\right.\right.\right.$$
$$\left.\left.\left.+\frac{1}{2}\sqrt{2}\sqrt{\theta}\,b\right)+\mathrm{erf}\left(-\sqrt{2}\sqrt{\theta}+\frac{1}{2}\sqrt{2}\sqrt{\theta}\,a+\frac{1}{2}\sqrt{2}\sqrt{\theta}\,b\right)\right)\right)$$

## Appendix H.   Hand and SMS algebra evaluating integral $I_5$

*Hand algebra:* For $-1 \le a \le 1$, and using the definition of $Q_{3/2}$ from Sec. 4:

$$I_5 \equiv \tfrac{1}{2}\int_{-1}^{1}\left[1+\sqrt{3\theta(a-x)^2}\right]e^{-\sqrt{3\theta(a-x)^2}}\,dx$$

$$= \tfrac{1}{2}\int_{-1}^{a}\left[1+\sqrt{3\theta(a-x)^2}\right]e^{-\sqrt{3\theta(a-x)^2}}\,dx + \tfrac{1}{2}\int_{a}^{1}\left[1+\sqrt{3\theta(a-x)^2}\right]e^{-\sqrt{3\theta(a-x)^2}}\,dx\,.$$

Change variables:   for $x < a$, $\tilde{x} \equiv \sqrt{3\theta}\,(a-x)$, which gives $d\tilde{x} = -\sqrt{3\theta}\,dx$, and for $x \ge a$, $\tilde{\tilde{x}} \equiv \sqrt{3\theta}\,(x-a)$, which gives $d\tilde{\tilde{x}} = \sqrt{3\theta}\,dx$. Then,

$$I_5 = -\tfrac{1}{2\sqrt{3\theta}}\int_{\tilde{x}=\sqrt{3\theta}\,(1+a)}^{0}(1+\tilde{x})e^{-\tilde{x}}\,d\tilde{x} + \tfrac{1}{2\sqrt{3\theta}}\int_{\tilde{\tilde{x}}=0}^{\sqrt{3\theta}\,(1-a)}(1+\tilde{\tilde{x}})e^{-\tilde{\tilde{x}}}\,d\tilde{\tilde{x}}\,.$$

Then, by Dwight 565.2 and 567.1 [12],

$$I_5 = -\tfrac{1}{2\sqrt{3\theta}}\left(\{[-1+(-\tilde{x}-1)]e^{-\tilde{x}}\}\Big|_{\tilde{x}=\sqrt{3\theta}\,(1+a)}^{\tilde{x}=0} - \{[-1+(-\tilde{\tilde{x}}-1)]e^{-\tilde{\tilde{x}}}\}\Big|_{\tilde{\tilde{x}}=0}^{\tilde{\tilde{x}}=\sqrt{3\theta}\,(1-a)}\right)$$

$$= -\tfrac{1}{2\sqrt{3\theta}}\left(\{[-1+(-\tilde{x}-1)]e^{-\tilde{x}}\}\Big|_{\sqrt{3\theta}\,(1+a)}^{0} + \{[-1+(-\tilde{\tilde{x}}-1)]e^{-\tilde{\tilde{x}}}\}\Big|_{\sqrt{3\theta}\,(1-a)}^{0}\right)$$

$$= \tfrac{1}{2\sqrt{3\theta}}\left\{[(2+\tilde{x})e^{-\tilde{x}}]\Big|_{\sqrt{3\theta}\,(1+a)}^{0} + [(2+\tilde{\tilde{x}})e^{-\tilde{\tilde{x}}}]\Big|_{\sqrt{3\theta}\,(1-a)}^{0}\right\}$$

$$= \tfrac{1}{2\sqrt{3\theta}}\left[4 - 2e^{-\sqrt{3\theta}\,(1+a)} - 2e^{-\sqrt{3\theta}\,(1-a)} - \sqrt{3\theta}\,(1+a)e^{-\sqrt{3\theta}\,(1+a)} - \sqrt{3\theta}\,(1-a)e^{-\sqrt{3\theta}\,(1-a)}\right]$$

$$= \tfrac{1}{2\sqrt{3\theta}}\left(\begin{matrix}2\left\{\begin{matrix}[1-e^{-\sqrt{3\theta}\,(1+a)}]\\+[1-e^{-\sqrt{3\theta}\,(1-a)}]\end{matrix}\right\}\\-\sqrt{3\theta}\left\{\begin{matrix}(1+a)e^{-\sqrt{3\theta}\,(1+a)}\\+(1-a)e^{-\sqrt{3\theta}\,(1-a)}\end{matrix}\right\}\end{matrix}\right).\quad\text{(Hand-algebra result)}\qquad\text{(H.1)}$$

*SMS algebra:*

```
> Q:=3*theta*((a-x)^2);
> I5Maple:=(1/2)*int((1+sqrt(Q))*exp(-sqrt(Q)),x=-1..1) assuming
-1<a assuming a<1;
```

$$Q := 3\,\theta\,(a-x)^2$$

$$I5Maple := -\frac{3}{2}\left(\left(\sqrt{3}\,\theta^2 e^{2\sqrt{3}\sqrt{\theta}\,a}a + 15\sqrt{3}\,\theta^3 e^{2\sqrt{3}\sqrt{\theta}\,a}a\right.\right.$$

$$-3\sqrt{3}\,\theta^3 e^{2\sqrt{3}\sqrt{\theta}\,a}a^2 - 3\sqrt{3}\,\theta^3 e^{2\sqrt{3}\sqrt{\theta}\,a}a^3$$

$$-12\sqrt{3}\,e^{\sqrt{3}\sqrt{\theta}\,a+\sqrt{\theta}\sqrt{3}}a\theta^3 - 4\sqrt{3}\,e^{\sqrt{3}\sqrt{\theta}\,a+\sqrt{\theta}\sqrt{3}}a\theta^2$$

$$+12\sqrt{3}\,e^{\sqrt{3}\sqrt{\theta}\,a+\sqrt{\theta}\sqrt{3}}a^3\theta^3 - 12\sqrt{3}\,e^{\sqrt{3}\sqrt{\theta}\,a+\sqrt{\theta}\sqrt{3}}a^2\theta^3 + 3\sqrt{3}\,\theta^2$$

$$-4\sqrt{3}\,e^{\sqrt{3}\sqrt{\theta}\,a+\sqrt{\theta}\sqrt{3}}\theta^2 + 12\sqrt{3}\,e^{\sqrt{3}\sqrt{\theta}\,a+\sqrt{\theta}\sqrt{3}}\theta^3$$

$$-24\,\theta^{5/2}e^{\sqrt{3}\sqrt{\theta}\,a+\sqrt{\theta}\sqrt{3}}a + 12\,\theta^{5/2}e^{\sqrt{3}\sqrt{\theta}\,a+\sqrt{\theta}\sqrt{3}}a^2$$

$$+18\,\theta^{7/2}e^{2\sqrt{3}\sqrt{\theta}\,a}a - 18\,\theta^{7/2}e^{2\sqrt{3}\sqrt{\theta}\,a}a^3 + 9\,\theta^{7/2}e^{2\sqrt{3}\sqrt{\theta}\,a}a^4$$

$$+3\sqrt{3}\,\theta^2 e^{2\sqrt{3}\sqrt{\theta}\,a} + 12\,\theta^{5/2}e^{2\sqrt{3}\sqrt{\theta}\,a}a - 9\sqrt{3}\,\theta^3 e^{2\sqrt{3}\sqrt{\theta}\,a}$$

$$-9\,\theta^{5/2}e^{2\sqrt{3}\sqrt{\theta}\,a}a^2 + 3\sqrt{3}\,a\theta^2 - 9\sqrt{3}\,\theta^3 a^3 + 9\sqrt{3}\,\theta^3 a^2 + 9\sqrt{3}\,\theta^3 a$$

$$-9\,\theta^{7/2} - 3\,\theta^{5/2} + 2\,\theta^{3/2} - 9\,\theta^{7/2}e^{2\sqrt{3}\sqrt{\theta}\,a} + 18\,\theta^{5/2}a - 3\,\theta^{5/2}a^2$$

$$+18\,\theta^{7/2}a^2 - 9\,\theta^{7/2}a^4 + 2\,\theta^{3/2}e^{2\sqrt{3}\sqrt{\theta}\,a} - 3\,\theta^{5/2}e^{2\sqrt{3}\sqrt{\theta}\,a}$$

$$\left.-4\,\theta^{3/2}e^{\sqrt{3}\sqrt{\theta}\,a+\sqrt{\theta}\sqrt{3}} + 12\,\theta^{5/2}e^{\sqrt{3}\sqrt{\theta}\,a+\sqrt{\theta}\sqrt{3}} - 9\sqrt{3}\,\theta^3\right)$$

$$\left.e^{-\sqrt{3}\sqrt{\theta}\,a-\sqrt{\theta}\sqrt{3}}\right)\bigg/\left(\left(\sqrt{\theta}\sqrt{3}+3\theta a+3\theta\right)\left(\sqrt{\theta}\sqrt{3}+3\theta a\right.\right.$$

$$\left.\left.-3\theta\right)\left(\sqrt{\theta}\sqrt{3}-3\theta a+3\theta\right)\sqrt{\theta}\right)$$

A slight clipping of $\theta^2$, on the right-hand side of the fourth row of the I5Maple, above, is of no concern.

The denominator of the SMS result, above, contains the following three factors: $D_1 \equiv \sqrt{3\theta} + 3\theta(1+a)$, $D_2 \equiv \sqrt{3\theta} + 3\theta(1-a)$, and $D_3 \equiv \sqrt{3\theta} - 3\theta(1-a)$, the first two of which are non-zero over the domain of interest, viz. $-1 \leq a \leq 1$. However, $D_3$ can be zero in this domain. We will comment on this possibly zero denominator, at the conclusion of this section. For now, we expand symbolically the product $D_1 D_2 D_3$, as follows:

$$D_1 D_2 D_3 = \left[(3\theta)^{1/2} + 3\theta + 3\theta a\right]\left[(3\theta)^{1/2} + 3\theta - 3\theta a\right]\left[(3\theta)^{1/2} - 3\theta + 3\theta a\right]$$

$$= \left[3\theta + (3\theta)^2 - (3\theta)^2 a^2 + 2(3\theta)^{3/2}\right]\left[(3\theta)^{1/2} - 3\theta + 3\theta a\right]$$

$$= \begin{bmatrix} (3\theta)^{3/2} + (3\theta)^{5/2} - (3\theta)^{5/2}a^2 + 2(3\theta)^2 \\ -(3\theta)^2 - (3\theta)^3 + (3\theta)^3 a^2 - 2(3\theta)^{5/2} \\ +(3\theta)^2 a + (3\theta)^3 a - (3\theta)^3 a^3 + 2(3\theta)^{5/2}a \end{bmatrix}, \text{ where colors denote collectible terms,}$$

$$= \begin{bmatrix} (3\theta)^{3/2} \\ +(3\theta)^2(1+a) \\ +(3\theta)^{5/2}(-1+2a-a^2) \\ +(3\theta)^3(-1+a+a^2-a^3) \end{bmatrix}.$$

The above SMS result, typeset while maintaining the order of the terms, noting well the correct order in the denominator, i.e. $D_1 D_3 D_2$, is the following:

$$I_5 = \frac{-3e^{-\sqrt{3\theta}(1+a)}}{2D_1 D_3 D_2 \sqrt{\theta}} \begin{bmatrix} \sqrt{3}\theta^2 a e^{2\sqrt{3\theta}a} + 15\sqrt{3}\theta^3 a e^{2\sqrt{3\theta}a} - 3\sqrt{3}\theta^3 a^2 e^{2\sqrt{3\theta}a} - 3\sqrt{3}\theta^3 a^3 e^{2\sqrt{3\theta}a} \\ -12\sqrt{3}\theta^3 a e^{\sqrt{3\theta}(1+a)} - 4\sqrt{3}\theta^2 a e^{\sqrt{3\theta}(1+a)} + 12\sqrt{3}\theta^3 a^3 e^{\sqrt{3\theta}(1+a)} \\ -12\sqrt{3}\theta^3 a^2 e^{\sqrt{3\theta}(1+a)} + 3\sqrt{3}\theta^2 - 4\sqrt{3}\theta^2 e^{\sqrt{3\theta}(1+a)} + 12\sqrt{3}\theta^3 e^{\sqrt{3\theta}(1+a)} \\ -24\theta^{5/2} a e^{\sqrt{3\theta}(1+a)} + 12\theta^{5/2} a^2 e^{\sqrt{3\theta}(1+a)} + 18\theta^{7/2} a e^{2\sqrt{3\theta}} - 18\theta^{7/2} a^3 e^{2\sqrt{3\theta}a} \\ +9\theta^{7/2} a^4 e^{2\sqrt{3\theta}a} + 3\sqrt{3}\theta^2 e^{2\sqrt{3\theta}a} + 12\theta^{5/2} a e^{2\sqrt{3\theta}a} - 9\sqrt{3}\theta^3 e^{2\sqrt{3\theta}a} \\ -9\theta^{5/2} a^2 e^{2\sqrt{3\theta}a} + 3\sqrt{3}\theta^2 a - 9\sqrt{3}\theta^3 a^3 + 9\sqrt{3}\theta^3 a^2 + 9\sqrt{3}\theta^3 a - 9\theta^{7/2} - 3\theta^{5/2} \\ +2\theta^{3/2} - 9\theta^{7/2} e^{2\sqrt{3\theta}a} + 18\theta^{5/2} a - 3\theta^{5/2} a^2 + 18\theta^{7/2} a^2 - 9\theta^{7/2} a^4 \\ +2\theta^{3/2} e^{2\sqrt{3\theta}a} - 3\theta^{5/2} e^{2\sqrt{3\theta}a} - 4\theta^{3/2} e^{\sqrt{3\theta}(1+a)} + 12\theta^{5/2} e^{\sqrt{3\theta}(1+a)} - 9\sqrt{3}\theta^3 \end{bmatrix}.$$

We now organize vertically the terms of the ultimate equation by their exponential factor, viz. one of $e^{-\sqrt{3\theta}(1+a)}$, $e^{-\sqrt{3\theta}(1-a)}$, or unity, as follows:

$$I_5 = \frac{3}{2D_1 D_2 D_3 \sqrt{\theta}} \begin{bmatrix} -e^{-\sqrt{3\theta}(1+a)} \begin{pmatrix} 3\sqrt{3}\theta^2 + 3\sqrt{3}\theta^2 a - 9\sqrt{3}\theta^3 a^3 + 9\sqrt{3}\theta^3 a^2 \\ +9\sqrt{3}\theta^3 a - 9\theta^{7/2} - 3\theta^{5/2} + 2\theta^{3/2} \\ +18\theta^{5/2} a - 3\theta^{5/2} a^2 + 18\theta^{7/2} a^2 - 9\theta^{7/2} a^4 \\ -9\sqrt{3}\theta^3 \end{pmatrix} \\ -e^{-\sqrt{3\theta}(1-a)} \begin{pmatrix} \sqrt{3}\theta^2 a + 15\sqrt{3}\theta^3 a - 3\sqrt{3}\theta^3 a^2 - 3\sqrt{3}\theta^3 a^3 \\ +18\theta^{7/2} a - 18\theta^{7/2} a^3 + 9\theta^{7/2} a^4 + 3\sqrt{3}\theta^2 \\ +12\theta^{5/2} a - 9\sqrt{3}\theta^3 - 9\theta^{5/2} a^2 - 9\theta^{7/2} \\ +2\theta^{3/2} - 3\theta^{5/2} \end{pmatrix} \\ + \begin{pmatrix} 12\sqrt{3}\theta^3 a + 4\sqrt{3}\theta^2 a - 12\sqrt{3}\theta^3 a^3 + 12\sqrt{3}\theta^3 a^2 \\ +4\sqrt{3}\theta^2 - 12\sqrt{3}\theta^3 + 24\theta^{5/2} a - 12\theta^{5/2} a^2 \\ +4\theta^{3/2} - 12\theta^{5/2} \end{pmatrix} \end{bmatrix}.$$

Multiplying terms in the square bracket by $3\sqrt{3}$ and dividing overall by the same quantity gives the following:

$$I_5 = \frac{3}{2D_1 D_2 D_3 \sqrt{\theta}} \left[ \begin{array}{l} -e^{-\sqrt{3\theta}(1+a)} \begin{pmatrix} 27\theta^2 + 27\theta^2 a - 81\theta^3 a^3 + 81\theta^3 a^2 \\ +81\theta^3 a - 27\sqrt{3}\theta^{7/2} - 9\sqrt{3}\theta^{5/2} + 6\sqrt{3}\theta^{3/2} \\ +54\sqrt{3}\theta^{5/2} a - 9\sqrt{3}\theta^{5/2} a^2 + 54\sqrt{3}\theta^{7/2} a^2 - 27\sqrt{3}\theta^{7/2} a^4 \\ -81\theta^3 \end{pmatrix} \\ -e^{-\sqrt{3\theta}(1-a)} \begin{pmatrix} 9\theta^2 a + 135\theta^3 a - 27\theta^3 a^2 - 27\theta^3 a^3 \\ +54\sqrt{3}\theta^{7/2} a - 54\sqrt{3}\theta^{7/2} a^3 + 27\sqrt{3}\theta^{7/2} a^4 + 27\theta^2 \\ +36\sqrt{3}\theta^{5/2} a - 81\theta^3 - 27\sqrt{3}\theta^{5/2} a^2 - 27\sqrt{3}\theta^{7/2} \\ +6\sqrt{3}\theta^{3/2} - 9\sqrt{3}\theta^{5/2} \end{pmatrix} \\ + \begin{pmatrix} 108\theta^3 a + 36\theta^2 a - 108\theta^3 a^3 + 108\theta^3 a^2 + 36\theta^2 \\ -108\theta^3 + 72\sqrt{3}\theta^{5/2} a - 36\sqrt{3}\theta^{5/2} a^2 + 12\sqrt{3}\theta^{3/2} - 36\sqrt{3}\theta^{5/2} \end{pmatrix} \end{array} \right].$$

Then, organizing further by powers of $(3\theta)^{1/2}$ gives the final SMS result:

$$I_5 = \frac{1}{2D_1 D_2 D_3 \sqrt{3\theta}} \left[ \begin{array}{l} -e^{-(3\theta)^{1/2}(1+a)} \begin{bmatrix} (3\theta)^{3/2} 2 \\ +(3\theta)^2 (3 + 3a) \\ -(3\theta)^{5/2}(1 - 6a + a^2) \\ -(3\theta)^3 (3 - 3a - 3a^2 + 3a^3) \\ -(3\theta)^{7/2}(1 - 2a^2 + a^4) \end{bmatrix} \\ -e^{-(3\theta)^{1/2}(1-a)} \begin{bmatrix} (3\theta)^{3/2} 2 \\ +(3\theta)^2 (3 + a) \\ -(3\theta)^{5/2}(1 - 4a + 3a^2) \\ -(3\theta)^3 (3 - 5a + a^2 + a^3) \\ -(3\theta)^{7/2}(1 - 2a + 2a^3 - a^4) \end{bmatrix} \\ + \begin{bmatrix} (3\theta)^{3/2} 4 \\ +(3\theta)^2 (4 + 4a) \\ -(3\theta)^{5/2}(4 - 8a + a^2) \\ -(3\theta)^3 (4 - 4a - 4a^2 + 4a^3) \end{bmatrix} \end{array} \right]. \quad \text{(SMS-algebra result)} \quad \text{(H.2)}$$

*Reconciliation of hand-algebra result, Eq. H.1, with SMS-algebra result, Eq. H2:* We now show the hand-algebra result, Eq. H.1, matches the SMS-algebra result, Eq. H.2. The former equation may be rearranged as follows:

$$I_5 = \frac{1}{2\sqrt{3\theta}} \left\{ \begin{array}{l} -e^{-(3\theta)^{1/2}(1+a)} \begin{bmatrix} 2 \\ +(3\theta)^{1/2}(1+a) \end{bmatrix} \\ -e^{-(3\theta)^{1/2}(1-a)} \begin{bmatrix} 2 \\ +(3\theta)^{1/2}(1-a) \end{bmatrix} \\ +4 \end{array} \right\}.$$

We symbolically multiply and divide the ultimate equation by the product $D_1 D_2 D_3$, cognizant that $D_3$ may be zero, to obtain the following:

$$I_5 = \frac{[(3\theta)^{1/2} + 3\theta + 3\theta a][(3\theta)^{1/2} + 3\theta - 3\theta a][(3\theta)^{1/2} - 3\theta + 3\theta a]}{2(3\theta)^{1/2} D_1 D_2 D_3} \left\{ \begin{array}{l} -e^{-(3\theta)^{3/2}(1+a)} \begin{bmatrix} 2 \\ +(3\theta)^{1/2}(1+a) \end{bmatrix} \\ -e^{-(3\theta)^{3/2}(1-a)} \begin{bmatrix} 2 \\ +(3\theta)^{1/2}(1-a) \end{bmatrix} \\ +4 \end{array} \right\}$$

$$= \frac{1}{2(3\theta)^{1/2} D_1 D_2 D_3} \begin{bmatrix} (3\theta)^{3/2} \\ +(3\theta)^2(1+a) \\ -(3\theta)^{5/2}(1-2a+a^2) \\ -(3\theta)^3(1-a-a^2+a^3) \end{bmatrix} \left\{ \begin{matrix} -e^{-\sqrt{3\theta}\,(1+a)} \begin{bmatrix} 2 \\ (3\theta)^{1/2}(1+a) \end{bmatrix} \\ -e^{-\sqrt{3\theta}\,(1-a)} \begin{bmatrix} 2 \\ (3\theta)^{1/2}(1-a) \end{bmatrix} \\ +4 \end{matrix} \right\}$$

$$= \frac{1}{2(3\theta)^{1/2} D_1 D_2 D_3} \left( \begin{matrix} -e^{-(3\theta)^{1/2}(1+a)} \left\{ \begin{matrix} 2 \begin{bmatrix} (3\theta)^{3/2} \\ +(3\theta)^2(1+a) \\ -(3\theta)^{5/2}(1-2a+a^2) \\ -(3\theta)^3(1-a-a^2+a^3) \end{bmatrix} \\ +(3\theta)^{1/2}(1+a) \begin{bmatrix} (3\theta)^{3/2} \\ +(3\theta)^2(1+a) \\ -(3\theta)^{5/2}(1-2a+a^2) \\ -(3\theta)^3(1-a-a^2+a^3) \end{bmatrix} \end{matrix} \right\} \\ -e^{-(3\theta)^{1/2}(1-a)} \left\{ \begin{matrix} 2 \begin{bmatrix} (3\theta)^{3/2} \\ +(3\theta)^2(1+a) \\ -(3\theta)^{5/2}(1-2a+a^2) \\ -(3\theta)^3(1-a-a^2+a^3) \end{bmatrix} \\ +(3\theta)^{1/2}(1-a) \begin{bmatrix} (3\theta)^{3/2} \\ +(3\theta)^2(1+a) \\ -(3\theta)^{5/2}(1-2a+a^2) \\ -(3\theta)^3(1-a-a^2+a^3) \end{bmatrix} \end{matrix} \right\} \\ +4 \begin{bmatrix} (3\theta)^{3/2} \\ +(3\theta)^2(1+a) \\ -(3\theta)^{5/2}(1-2a+a^2) \\ -(3\theta)^3(1-a-a^2+a^3) \end{bmatrix} \end{matrix} \right)$$

$$= \frac{1}{2(3\theta)^{1/2} D_1 D_2 D_3} \left( \begin{matrix} -e^{-(3\theta)^{1/2}(1+a)} \left\{ \begin{matrix} (3\theta)^{3/2} 2 \\ +(3\theta)^2[2(1+a)+(1+a)] \\ -(3\theta)^{5/2}[2(1-2a+a^2)-(1+a)(1+a)] \\ -(3\theta)^3[2(1-a-a^2+a^3)+(1+a)(1-2a+a^2)] \\ -(3\theta)^{7/2}(1+a)(1-a-a^2+a^3) \end{matrix} \right\} \\ -e^{-(3\theta)^{1/2}(1-a)} \left\{ \begin{matrix} (3\theta)^{3/2} 2 \\ +(3\theta)^2[2(1+a)+(1-a)] \\ -(3\theta)^{5/2}[2(1-2a+a^2)-(1+a)(1-a)] \\ -(3\theta)^3[2(1-a-a^2+a^3)+(1-a)(1-2a+a^2)] \\ -(3\theta)^{7/2}(1-a)(1-a-a^2+a^3) \end{matrix} \right\} \\ +4 \begin{bmatrix} (3\theta)^{3/2} \\ +(3\theta)^2(1+a) \\ -(3\theta)^{5/2}(1-2a+a^2) \\ -(3\theta)^3(1-a-a^2+a^3) \end{bmatrix} \end{matrix} \right)$$

$$= \frac{1}{2(3\theta)^{1/2} D_1 D_2 D_3} \left( \begin{array}{l} -e^{-(3\theta)^{1/2}(1+a)} \begin{bmatrix} (3\theta)^{3/2} 2 \\ +(3\theta)^2 (3+3a) \\ -(3\theta)^{5/2}(1-6a+a^2) \\ -(3\theta)^3 (3-3a-3a^2+3a^3) \\ -(3\theta)^{7/2}(1-2a^2+a^4) \end{bmatrix} \\ -e^{-(3\theta)^{1/2}(1-a)} \begin{bmatrix} (3\theta)^{3/2} 2 \\ +(3\theta)^2 (3+a) \\ -(3\theta)^{5/2}(1-4a+3a^2) \\ -(3\theta)^3 (3-5a+a^2+a^3) \\ -(3\theta)^{7/2}(1-2a+2a^3-a^4) \end{bmatrix} \\ +4 \begin{bmatrix} (3\theta)^{3/2} \\ +(3\theta)^2 (1+a) \\ -(3\theta)^{5/2}(1-2a+a^2) \\ -(3\theta)^3 (1-a-a^2+a^3) \end{bmatrix} \end{array} \right)$$, which matches Eq. H.2.

We now comment on the denominator factor $D_3$, which can be zero in the domain $-1 \leq a \leq 1$, in the ultimate equation. Because Eq. H.1 does not contain $D_3$ and has a non-zero denominator, we may consider that the algebra in this section is equivalent to factoring out $D_3$. Thus, the locus of all zeros of the denominator of Eq. H.2 is shared by a locus of zeros of this equation's numerator. This is conceptually not a problem, as any singularity is removable. However, if Eq. H.2 is used to evaluate $I_5$, the evaluation may fail or suffer ill-conditioning. Thus, the preferred equation for evaluation of $I_5$ is Eq. H.1.

## Appendix I.  SMS algebra evaluating integral $I_6$

$I_6 \equiv \frac{1}{2}\int_{-1}^{1} \left[1+\sqrt{3\theta(a-x)^2}\right]\left[1+\sqrt{3\theta(b-x)^2}\right] e^{-\sqrt{3\theta[(a-x)^2+(b-x)^2]}}\, dx$.

Hand algebra is skipped. The result from SMS algebra, using the definition of $Q_{3/2}$ from Sec. 4, is given as Eq. I.1, below.

SMS algebra: Maple input and output, follow:

```
> Q:=3*theta*((a-x)^2);
> R:=3*theta*((b-x)^2);
> I611:=(1/2)*int(exp(-sqrt(Q)-sqrt(R)),x=-1..1) assuming -1<a
assuming a<b assuming a<1 assuming -1<b assuming b<1;
> I612:=(1/2)*int(sqrt(R)*exp(-sqrt(Q)-sqrt(R)),x=-1..1)
assuming -1<a assuming a<b assuming a<1 assuming -1<b assuming
b<1;
> I621:=(1/2)*int(sqrt(Q)*exp(-sqrt(Q)-sqrt(R)),x=-1..1)
assuming -1<a assuming a<b assuming a<1 assuming -1<b assuming
b<1;
> I622:=(1/2)*int(sqrt(Q*R)*exp(-sqrt(Q)-sqrt(R)),x=-1..1)
assuming -1<a assuming a<b assuming a<1 assuming -1<b assuming
b<1 assuming 0<theta;
> NormalExpandI611:=normal(expand(I611));
> NormalExpandI612:=normal(expand(I612));
```

```
> NormalExpandI621:=normal(expand(I621));
> NormalExpandI622:=normal(expand(I622));
>
Sum4NormalExpandTerms:=NormalExpandI611+NormalExpandI612+NormalE
xpandI621+NormalExpandI622;
> I6:=normal(expand(Sum4NormalExpandTerms));
```

$$I6 := -\frac{1}{24} \frac{1}{e^{\sqrt{3}\sqrt{\theta}\,a} \, e^{\sqrt{3}\sqrt{\theta}\,b} \left(e^{\sqrt{3}\sqrt{\theta}}\right)^2 \sqrt{\theta}} \Big( 5\sqrt{3} + 18\sqrt{\theta} + 6\sqrt{3}\,b\theta$$

$$+ 24\sqrt{3}\left(e^{\sqrt{3}\sqrt{\theta}\,a}\right)^2 \theta a b \left(e^{\sqrt{3}\sqrt{\theta}}\right)^2$$

$$+ 6\sqrt{3}\left(e^{\sqrt{3}\sqrt{\theta}\,a}\right)^2 \left(e^{\sqrt{3}\sqrt{\theta}\,b}\right)^2 a b\theta + 6\sqrt{3}\,a\theta$$

$$- 9\left(e^{\sqrt{3}\sqrt{\theta}\,a}\right)^2 \left(e^{\sqrt{3}\sqrt{\theta}\,b}\right)^2 a\sqrt{\theta} + 6\left(e^{\sqrt{3}\sqrt{\theta}\,a}\right)^2 \left(e^{\sqrt{3}\sqrt{\theta}}\right)^2 a^3 \theta^{3/2}$$

$$- 6\left(e^{\sqrt{3}\sqrt{\theta}\,a}\right)^2 \left(e^{\sqrt{3}\sqrt{\theta}}\right)^2 b^3 \theta^{3/2} + 30\,a\left(e^{\sqrt{3}\sqrt{\theta}\,a}\right)^2 \left(e^{\sqrt{3}\sqrt{\theta}}\right)^2 \sqrt{\theta}$$

$$- 30\left(e^{\sqrt{3}\sqrt{\theta}\,a}\right)^2 \left(e^{\sqrt{3}\sqrt{\theta}}\right)^2 b\sqrt{\theta} + 6\sqrt{3}\left(e^{\sqrt{3}\sqrt{\theta}\,a}\right)^2 \left(e^{\sqrt{3}\sqrt{\theta}\,b}\right)^2 \theta$$

$$- 9\left(e^{\sqrt{3}\sqrt{\theta}\,a}\right)^2 \left(e^{\sqrt{3}\sqrt{\theta}\,b}\right)^2 b\sqrt{\theta} + 6\sqrt{3}\,ab\theta + 6\sqrt{3}\,\theta$$

$$- 18\left(e^{\sqrt{3}\sqrt{\theta}\,a}\right)^2 \left(e^{\sqrt{3}\sqrt{\theta}}\right)^2 a^2 b\theta^{3/2}$$

$$+ 18\left(e^{\sqrt{3}\sqrt{\theta}\,a}\right)^2 \left(e^{\sqrt{3}\sqrt{\theta}}\right)^2 a b^2 \theta^{3/2}$$

$$- 12\sqrt{3}\left(e^{\sqrt{3}\sqrt{\theta}\,a}\right)^2 \theta a^2 \left(e^{\sqrt{3}\sqrt{\theta}}\right)^2$$

$$- 12\sqrt{3}\left(e^{\sqrt{3}\sqrt{\theta}\,a}\right)^2 \theta b^2 \left(e^{\sqrt{3}\sqrt{\theta}}\right)^2$$

$$- 6\sqrt{3}\left(e^{\sqrt{3}\sqrt{\theta}\,a}\right)^2 \left(e^{\sqrt{3}\sqrt{\theta}\,b}\right)^2 a\theta - 6\sqrt{3}\left(e^{\sqrt{3}\sqrt{\theta}\,a}\right)^2 \left(e^{\sqrt{3}\sqrt{\theta}\,b}\right)^2 b\theta$$

$$+ 18\left(e^{\sqrt{3}\sqrt{\theta}\,a}\right)^2 \left(e^{\sqrt{3}\sqrt{\theta}\,b}\right)^2 \sqrt{\theta} + 5\sqrt{3}\left(e^{\sqrt{3}\sqrt{\theta}\,a}\right)^2 \left(e^{\sqrt{3}\sqrt{\theta}\,b}\right)^2$$

$$- 10\sqrt{3}\left(e^{\sqrt{3}\sqrt{\theta}\,a}\right)^2 \left(e^{\sqrt{3}\sqrt{\theta}}\right)^2 + 9\,a\sqrt{\theta} + 9\,b\sqrt{\theta} \Big)$$

(I.1)

## Appendix J.  Hand and SMS Algebra Demonstrating Integral $I_7$

Hand algebra, using the definition of $Q_{5/2}$ from Sec. 4:

$$I_7 \equiv \frac{1}{2}\int_{-1}^{1}\left[1 + \sqrt{5\theta(a-x)^2} + \frac{5\theta(a-x)^2}{3}\right]e^{-\sqrt{5\theta(a-x)^2}}\,dx$$

$$= \frac{1}{2}\int_{-1}^{a}\left[1 + \sqrt{5\theta(a-x)^2} + \frac{5\theta(a-x)^2}{3}\right]e^{-\sqrt{5\theta(a-x)^2}}\,dx$$

$$+ \frac{1}{2}\int_{a}^{1}\left[1 + \sqrt{5\theta(a-x)^2} + \frac{5\theta(a-x)^2}{3}\right]e^{-\sqrt{5\theta(a-x)^2}}\,dx.$$

Change variables: For $x < a$, define $\tilde{x} \equiv \sqrt{5\theta}\,(a-x)$, which gives $d\tilde{x} = -\sqrt{5\theta}\,dx$, and for $x \geq a$, define $\tilde{\tilde{x}} \equiv \sqrt{5\theta}\,(x-a)$, which gives $d\tilde{\tilde{x}} = \sqrt{5\theta}\,dx$. Then,

$$I_7 = -\frac{1}{2\sqrt{5\theta}}\int_{\tilde{x}=\sqrt{5\theta}\,(1+a)}^{0}\left(1 + \tilde{x} + \frac{1}{3}\tilde{x}^2\right)e^{-\tilde{x}}\,d\tilde{x} + \frac{1}{2\sqrt{5\theta}}\int_{\tilde{\tilde{x}}=0}^{\sqrt{5\theta}\,(1-a)}\left(1 + \tilde{\tilde{x}} + \frac{1}{3}\tilde{\tilde{x}}^2\right)e^{-\tilde{\tilde{x}}}\,d\tilde{\tilde{x}}.$$

Then, by Dwight 565.2, 567.1, and 567.2 [12],

$$I_7 = -\frac{1}{2\sqrt{5\theta}}\left\{\begin{array}{l}\left[-1 + (-\tilde{x} - 1) + \frac{1}{3}(-\tilde{x}^2 - 2\tilde{x} - 2)\right]e^{-\tilde{x}}\bigg|_{\tilde{x}=\sqrt{5\theta}\,(1+a)}^{\tilde{x}=0} \\ -\left[-1 + (-\tilde{\tilde{x}} - 1) + \frac{1}{3}(-\tilde{\tilde{x}}^2 - 2\tilde{\tilde{x}} - 2)\right]e^{-\tilde{\tilde{x}}}\bigg|_{\tilde{\tilde{x}}=0}^{\tilde{\tilde{x}}=\sqrt{5\theta}\,(1-a)}\end{array}\right\}$$

$$= -\frac{1}{2\sqrt{5\theta}}\left\{\begin{array}{l}\left[-1 + (-\tilde{x} - 1) + \frac{1}{3}(-\tilde{x}^2 - 2\tilde{x} - 2)\right]e^{-\tilde{x}}\bigg|_{\sqrt{5\theta}\,(1+a)}^{0} \\ +\left[-1 + (-\tilde{\tilde{x}} - 1) + \frac{1}{3}(-\tilde{\tilde{x}}^2 - 2\tilde{\tilde{x}} - 2)\right]e^{-\tilde{\tilde{x}}}\bigg|_{\sqrt{5\theta}\,(1-a)}^{0}\end{array}\right\}$$

$$= \frac{1}{2\sqrt{5\theta}}\left\{\begin{array}{l}\left[\left(\frac{8}{3} + \frac{5}{3}\tilde{x} + \frac{1}{3}\tilde{x}^2\right)e^{-\tilde{x}}\right]\bigg|_{\sqrt{5\theta}\,(1+a)}^{0} \\ +\left[\left(\frac{8}{3} + \frac{5}{3}\tilde{\tilde{x}} + \frac{1}{3}\tilde{\tilde{x}}^2\right)e^{-\tilde{\tilde{x}}}\right]\bigg|_{\sqrt{5\theta}\,(1-a)}^{0}\end{array}\right\}$$

$$= \frac{1}{6\sqrt{5\theta}}\left[\begin{array}{l}16 - 8e^{-\sqrt{5\theta}\,(1+a)} - 8e^{-\sqrt{5\theta}\,(1-a)} \\ -5\sqrt{5\theta}\,(1+a)e^{-\sqrt{5\theta}\,(1+a)} - 5\sqrt{5\theta}\,(1-a)e^{-\sqrt{5\theta}\,(1-a)} \\ -5\theta\,(1+a)^2 e^{-\sqrt{5\theta}\,(1+a)} - 5\theta\,(1-a)^2 e^{-\sqrt{5\theta}\,(1-a)}\end{array}\right] \quad (\text{J.1})$$

$$= \frac{1}{6\sqrt{5\theta}}\left[\begin{array}{l}16 - 8e^{-\sqrt{5\theta}\,(1+a)} - 8e^{-\sqrt{5\theta}\,(1-a)} \\ +\left[-5\sqrt{5}\,(1+a)e^{-\sqrt{5\theta}\,(1+a)} - 5\sqrt{5}\,(1-a)e^{-\sqrt{5\theta}\,(1-a)}\right]\sqrt{\theta} \\ +\left[-5(1+a)^2 e^{-\sqrt{5\theta}\,(1+a)} - 5(1-a)^2 e^{-\sqrt{5\theta}\,(1-a)}\right]\theta\end{array}\right]$$

$$= \frac{1}{6\sqrt{5\theta}}\left\{\begin{array}{l}8\left[1 - e^{-\sqrt{5\theta}\,(1+a)}\right] + 8\left[1 - e^{-\sqrt{5\theta}\,(1-a)}\right] \\ +\left[-5\sqrt{5}\,(1+a)e^{-\sqrt{5\theta}\,(1+a)} - 5\sqrt{5}\,(1-a)e^{-\sqrt{5\theta}\,(1-a)}\right]\sqrt{\theta} \\ +\left[-5(1+a)^2 e^{-\sqrt{5\theta}\,(1+a)} - 5(1-a)^2 e^{-\sqrt{5\theta}\,(1-a)}\right]\theta\end{array}\right\}$$

$$= \frac{1}{6\sqrt{5\theta}}\left(8\left\{\begin{array}{l}\left[1 - e^{-\sqrt{5\theta}\,(1+a)}\right] \\ +\left[1 - e^{-\sqrt{5\theta}\,(1-a)}\right]\end{array}\right\} - 5\sqrt{5\theta}\left\{\begin{array}{l}(1+a)e^{-\sqrt{5\theta}\,(1+a)} \\ +(1-a)e^{-\sqrt{5\theta}\,(1-a)}\end{array}\right\} - 5\theta\left\{\begin{array}{l}(1+a)^2 e^{-\sqrt{5\theta}\,(1+a)} \\ +(1-a)^2 e^{-\sqrt{5\theta}\,(1-a)}\end{array}\right\}\right)$$

$$= \frac{1}{6\sqrt{5\theta}} \begin{pmatrix} 8 \left\{ \begin{matrix} \left[1 - e^{-\sqrt{5\theta}\,(1+a)}\right] \\ + \left[1 - e^{-\sqrt{5\theta}\,(1-a)}\right] \end{matrix} \right\} \\ -5\sqrt{5\theta} \left\{ \begin{matrix} (1+a)e^{-\sqrt{5\theta}\,(1+a)} \\ +(1-a)e^{-\sqrt{5\theta}\,(1-a)} \end{matrix} \right\} \\ -5\theta \left\{ \begin{matrix} (1+a)^2 e^{-\sqrt{5\theta}\,(1+a)} \\ +(1-a)^2 e^{-\sqrt{5\theta}\,(1-a)} \end{matrix} \right\} \end{pmatrix}. \tag{J.2}$$

SMS algebra: Maple input and output, finished with hand algebra, follow:

```
> Q:=5*theta*((a-x)^2);
> I7:=(1/2)*int((1+sqrt(Q)+Q/3)*exp(-sqrt(Q)),x=-1..1) assuming
-1<a assuming a<1;
```

$$I7 := \frac{1}{30} \frac{1}{\sqrt{\theta}} \Biggl( \Bigl( -8\sqrt{5} - 25\sqrt{\theta}\,a - 25\sqrt{\theta} - 8\,e^{2\sqrt{\theta}\sqrt{5}\,a}\sqrt{5}$$

$$- 25\sqrt{\theta}\,e^{2\sqrt{\theta}\sqrt{5}\,a} + 16\,e^{\sqrt{\theta}\sqrt{5}\,a+\sqrt{\theta}\sqrt{5}}\sqrt{5} - 5\sqrt{5}\,e^{2\sqrt{\theta}\sqrt{5}\,a}\theta$$

$$+ 25\sqrt{\theta}\,e^{2\sqrt{\theta}\sqrt{5}\,a}\,a - 5\sqrt{5}\,e^{2\sqrt{\theta}\sqrt{5}\,a}a^2\theta + 10\,e^{2\sqrt{\theta}\sqrt{5}\,a}\sqrt{5}\,a\theta$$

$$- 5\sqrt{5}\,\theta - 5\sqrt{5}\,\theta a^2 - 10\sqrt{5}\,\theta a \Bigr) e^{-\sqrt{\theta}\sqrt{5}\,a-\sqrt{\theta}\sqrt{5}} \Biggr)$$

$$= \frac{1}{30\sqrt{\theta}} \left\{ \begin{matrix} -8\sqrt{5}\left[1 - 2e^{\sqrt{5\theta}(1+a)} + e^{2\sqrt{5\theta}a}\right] \\ -25\left[(1+a) + (1-a)e^{2\sqrt{5\theta}a}\right]\sqrt{\theta} \\ -5\sqrt{5}\left[1 + 2a + a^2 + e^{2\sqrt{5\theta}a}(1 - 2a + a^2)\right]\theta \end{matrix} \right\} e^{-\sqrt{5\theta}(1+a)}$$

$$= \frac{1}{6\sqrt{5\theta}} \left\{ \begin{matrix} -8\left[1 - 2e^{\sqrt{5\theta}(1+a)} + e^{2\sqrt{5\theta}a}\right] \\ -5\sqrt{5}\left[(1+a) + (1-a)e^{2\sqrt{5\theta}a}\right]\sqrt{\theta} \\ -5\left[(1+a)^2 + (1-a)^2 e^{2\sqrt{5\theta}a}\right]\theta \end{matrix} \right\} e^{-\sqrt{5\theta}(1+a)}$$

$$= \frac{1}{6\sqrt{5\theta}} \left\{ \begin{matrix} \left[-8\left(1 - 2e^{\sqrt{5\theta}(1+a)} + e^{2\sqrt{5\theta}a}\right)e^{-\sqrt{5\theta}(1+a)}\right] \\ +\left[-5\sqrt{5}(1+a)e^{-\sqrt{5\theta}(1+a)} - 5\sqrt{5}(1-a)e^{2\sqrt{5\theta}a}e^{-\sqrt{5\theta}(1+a)}\right]\sqrt{\theta} \\ +\left[-5(1+a)^2 e^{-\sqrt{5\theta}(1+a)} - 5(1-a)^2 e^{2\sqrt{5\theta}a}e^{-\sqrt{5\theta}(1+a)}\right]\theta \end{matrix} \right\}$$

$$= \frac{1}{6\sqrt{5\theta}} \left\{ \begin{matrix} \left[-8\left(e^{-\sqrt{5\theta}(1+a)} - 2 + e^{2\sqrt{5\theta}a}e^{-\sqrt{5\theta}(1+a)}\right)\right] \\ +\left[-5\sqrt{5}(1+a)e^{-\sqrt{5\theta}(1+a)} - 5\sqrt{5}(1-a)e^{-\sqrt{5\theta}(1-a)}\right]\sqrt{\theta} \\ +\left[-5(1+a)^2 e^{-\sqrt{5\theta}(1+a)} - 5(1-a)^2 e^{-\sqrt{5\theta}(1-a)}\right]\theta \end{matrix} \right\}$$

$$= \frac{1}{6\sqrt{5\theta}} \left\{ \begin{array}{l} \left[16 - 8e^{-\sqrt{5\theta}\,(1+a)} - 8e^{-\sqrt{5\theta}\,(1-a)}\right] \\ + \left[-5\sqrt{5}\,(1+a)e^{-\sqrt{5\theta}\,(1+a)} - 5\sqrt{5}\,(1-a)e^{-\sqrt{5\theta}\,(1-a)}\right]\sqrt{\theta} \\ + \left[-5(1+a)^2 e^{-\sqrt{5\theta}\,(1+a)} - 5(1-a)^2 e^{-\sqrt{5\theta}\,(1-a)}\right]\theta \end{array} \right\}, \text{ which matches Eq. J.1.}$$

## Appendix K.   SMS algebra demonstrating integral $I_8$

$$I_8 \equiv \frac{1}{2}\int_{-1}^{1}\left[1 + \sqrt{5\theta(a-x)^2} + \frac{5\theta(a-x)^2}{3}\right]\left[1 + \sqrt{5\theta(b-x)^2} + \frac{5\theta(b-x)^2}{3}\right]e^{-\sqrt{5\theta[(a-x)^2+(b-x)^2]}}\,dx\,.$$

Hand algebra is skipped. The result from SMS algebra, using the definition of $Q_{5/2}$ from Sec. 4, is given as Eq. K.1, below.

SMS algebra: Maple input and output follow:

```
> Q:=5*theta*((a-x)^2);
> R:=5*theta*((b-x)^2);
> I811:=(1/2)*int((1)*(1)*exp(-sqrt(Q))*exp(-sqrt(R)),x=-1..1)
assuming -1<a assuming a<b assuming a<1 assuming -1<b assuming
b<1;
> I812:=(1/2)*int((1)*(sqrt(R))*exp(-sqrt(Q))*exp(-sqrt(R)),x=-
1..1) assuming -1<a assuming a<b assuming a<1 assuming -1<b
assuming b<1;
> I813:=(1/2)*int((1)*(R/3)*exp(-sqrt(Q))*exp(-sqrt(R)),x=-1..1)
assuming -1<a assuming a<b assuming a<1 assuming -1<b assuming
b<1;
> I821:=(1/2)*int((sqrt(Q))*(1)*exp(-sqrt(Q))*exp(-sqrt(R)),x=-
1..1) assuming -1<a assuming a<b assuming a<1 assuming -1<b
assuming b<1;
> I822:=(1/2)*int((sqrt(Q))*(sqrt(R))*exp(-sqrt(Q))*exp(-
sqrt(R)),x=-1..1) assuming -1<a assuming a<b assuming a<1
assuming -1<b assuming b<1;
> I823:=(1/2)*int((sqrt(Q))*(R/3)*exp(-sqrt(Q))*exp(-
sqrt(R)),x=-1..1) assuming -1<a assuming a<b assuming a<1
assuming -1<b assuming b<1;
> I831:=(1/2)*int((Q/3)*(1)*exp(-sqrt(Q))*exp(-sqrt(R)),x=-1..1)
assuming -1<a assuming a<b assuming a<1 assuming -1<b assuming
b<1;
> I832:=(1/2)*int((Q/3)*(sqrt(R))*exp(-sqrt(Q))*exp(-
sqrt(R)),x=-1..1) assuming -1<a assuming a<b assuming a<1
assuming -1<b assuming b<1;
> I833:=(1/2)*int((Q/3)*(R/3)*exp(-sqrt(Q))*exp(-sqrt(R)),x=-
1..1) assuming -1<a assuming a<b assuming a<1 assuming -1<b
assuming b<1;
> NormalExpandI811:=normal(expand(I811));
> NormalExpandI812:=normal(expand(I812));
```

```
> NormalExpandI813:=normal(expand(I813));
> NormalExpandI821:=normal(expand(I821));
> NormalExpandI822:=normal(expand(I822));
> NormalExpandI823:=normal(expand(I823));
> NormalExpandI831:=normal(expand(I831));
> NormalExpandI832:=normal(expand(I832));
> NormalExpandI833:=normal(expand(I833));
>
Sum9NormalExpandTerms:=NormalExpandI811+NormalExpandI812+NormalE
xpandI813+NormalExpandI821+NormalExpandI822+NormalExpandI823+Nor
malExpandI831+NormalExpandI832+NormalExpandI833;
> I8:=normal(expand(Sum9NormalExpandTerms));
```

$$I8 := -\frac{1}{1080} \frac{1}{e^{\sqrt{\theta}\sqrt{5}\,b} e^{\sqrt{\theta}\sqrt{5}\,a} \sqrt{\theta} \left(e^{\sqrt{\theta}\sqrt{5}}\right)^2} \Big(189\sqrt{5} + 600\,b^2\,\theta^{3/2}$$

$$+ 1800\,b\,\theta^{3/2} + 150\left(e^{\sqrt{\theta}\sqrt{5}\,a}\right)^2 \left(e^{\sqrt{\theta}\sqrt{5}\,b}\right)^2 \sqrt{5}\,\theta^2$$

$$- 675\left(e^{\sqrt{\theta}\sqrt{5}\,a}\right)^2 \left(e^{\sqrt{\theta}\sqrt{5}\,b}\right)^2 a\sqrt{\theta} + 150\sqrt{5}\,a^2\,\theta + 810\sqrt{5}\,\theta$$

$$+ 150\sqrt{5}\,b^2\,\theta + 600\,a^2\,b\,\theta^{3/2} + 600\,a\,b^2\,\theta^{3/2} + 2400\,a\,b\,\theta^{3/2}$$

$$+ 150\sqrt{5}\,b^2\,\theta^2 + 1200\,\theta^{3/2} + 1350\sqrt{\theta} + 810\sqrt{5}\,b\theta + 675\,b\sqrt{\theta}$$

$$+ 150\sqrt{5}\,a^2\,\theta^2 + 810\sqrt{5}\,a\theta + 300\,b\sqrt{5}\,\theta^2 + 300\,a\sqrt{5}\,\theta^2$$

$$+ 510\sqrt{5}\,ab\theta + 150\sqrt{5}\,\theta^2 + 1890\left(e^{\sqrt{\theta}\sqrt{5}\,a}\right)^2 a\sqrt{\theta}\left(e^{\sqrt{\theta}\sqrt{5}}\right)^2$$

$$- 1890\left(e^{\sqrt{\theta}\sqrt{5}\,a}\right)^2 b\sqrt{\theta}\left(e^{\sqrt{\theta}\sqrt{5}}\right)^2$$

$$+ 810\sqrt{5}\left(e^{\sqrt{\theta}\sqrt{5}\,a}\right)^2 \left(e^{\sqrt{\theta}\sqrt{5}\,b}\right)^2 \theta$$

$$- 675\left(e^{\sqrt{\theta}\sqrt{5}\,a}\right)^2 \left(e^{\sqrt{\theta}\sqrt{5}\,b}\right)^2 b\sqrt{\theta}$$

$$+ 1050\left(e^{\sqrt{\theta}\sqrt{5}\,a}\right)^2 \left(e^{\sqrt{\theta}\sqrt{5}}\right)^2 a^3\,\theta^{3/2} + 600\sqrt{5}\,ab\theta^2 + 150\sqrt{5}\,a^2\,b^2\,\theta^2$$

$$+ 300\sqrt{5}\,a^2\,b\theta^2 + 300\sqrt{5}\,ab^2\,\theta^2 + 675\,a\sqrt{\theta}$$

$$- 1050\left(e^{\sqrt{\theta}\sqrt{5}\,a}\right)^2 \left(e^{\sqrt{\theta}\sqrt{5}}\right)^2 b^3\,\theta^{3/2} - 50\left(e^{\sqrt{\theta}\sqrt{5}\,a}\right)^2 \left(e^{\sqrt{\theta}\sqrt{5}}\right)^2 b^5\,\theta^{5/2}$$

$$+ 50\,a^5\left(e^{\sqrt{\theta}\sqrt{5}\,a}\right)^2 \left(e^{\sqrt{\theta}\sqrt{5}}\right)^2 \theta^{5/2} + 600\,a^2\,\theta^{3/2}$$

$$+ 1200\left(e^{\sqrt{\theta}\sqrt{5}\,a}\right)^2 \left(e^{\sqrt{\theta}\sqrt{5}\,b}\right)^2 \theta^{3/2}$$

$$- 250\,a^4\left(e^{\sqrt{\theta}\sqrt{5}\,a}\right)^2 \left(e^{\sqrt{\theta}\sqrt{5}}\right)^2 b\theta^{5/2}$$

$$+ 500\,a^3\left(e^{\sqrt{\theta}\sqrt{5}\,a}\right)^2 \left(e^{\sqrt{\theta}\sqrt{5}}\right)^2 b^2\,\theta^{5/2}$$

$$+ 150\sqrt{5}\left(e^{\sqrt{\theta}\sqrt{5}\,a}\right)^2 \left(e^{\sqrt{\theta}\sqrt{5}\,b}\right)^2 b^2\,\theta^2$$

$$+ 150\sqrt{5}\left(e^{\sqrt{\theta}\sqrt{5}\,a}\right)^2 \left(e^{\sqrt{\theta}\sqrt{5}\,b}\right)^2 a^2\,\theta^2$$

$$+ 250\left(e^{\sqrt{\theta}\sqrt{5}\,a}\right)^2 \left(e^{\sqrt{\theta}\sqrt{5}}\right)^2 ab^4\,\theta^{5/2}$$

$$- 500\left(e^{\sqrt{\theta}\sqrt{5}\,a}\right)^2 \left(e^{\sqrt{\theta}\sqrt{5}}\right)^2 a^2\,b^3\,\theta^{5/2}$$

$$+ 600\,\theta^{3/2}\left(e^{\sqrt{\theta}\sqrt{5}\,a}\right)^2 \left(e^{\sqrt{\theta}\sqrt{5}\,b}\right)^2 a^2$$

A slight clipping of $\theta^{5/2}$, on the right-hand side of the thirteenth row of the I8, above, is of no concern.

$$-150\sqrt{5}\,\theta^2\left(e^{\sqrt{\theta}\sqrt{5}\,a}\right)^2 a^4 \left(e^{\sqrt{\theta}\sqrt{5}}\right)^2$$

$$-150\sqrt{5}\,\theta^2\left(e^{\sqrt{\theta}\sqrt{5}\,a}\right)^2 b^4 \left(e^{\sqrt{\theta}\sqrt{5}}\right)^2$$

$$+600\,\theta^{3/2}\left(e^{\sqrt{\theta}\sqrt{5}\,a}\right)^2 \left(e^{\sqrt{\theta}\sqrt{5}\,b}\right)^2 b^2$$

$$-1800\,\theta^{3/2}\left(e^{\sqrt{\theta}\sqrt{5}\,a}\right)^2 \left(e^{\sqrt{\theta}\sqrt{5}\,b}\right)^2 a$$

$$-1800\,\theta^{3/2}\left(e^{\sqrt{\theta}\sqrt{5}\,a}\right)^2 \left(e^{\sqrt{\theta}\sqrt{5}\,b}\right)^2 b$$

$$+150\sqrt{5}\left(e^{\sqrt{\theta}\sqrt{5}\,a}\right)^2 \left(e^{\sqrt{\theta}\sqrt{5}\,b}\right)^2 a^2 \theta$$

$$-300\left(e^{\sqrt{\theta}\sqrt{5}\,a}\right)^2 \left(e^{\sqrt{\theta}\sqrt{5}\,b}\right)^2 a\sqrt{5}\,\theta^2$$

$$-300\left(e^{\sqrt{\theta}\sqrt{5}\,a}\right)^2 \left(e^{\sqrt{\theta}\sqrt{5}\,b}\right)^2 b\sqrt{5}\,\theta^2$$

$$+1350\left(e^{\sqrt{\theta}\sqrt{5}\,a}\right)^2 \left(e^{\sqrt{\theta}\sqrt{5}\,b}\right)^2 \sqrt{\theta} - 378\left(e^{\sqrt{\theta}\sqrt{5}\,a}\right)^2 \sqrt{5}\left(e^{\sqrt{\theta}\sqrt{5}}\right)^2$$

$$+189\left(e^{\sqrt{\theta}\sqrt{5}\,b}\right)^2 \left(e^{\sqrt{\theta}\sqrt{5}\,a}\right)^2 \sqrt{5}$$

$$-810\sqrt{5}\left(e^{\sqrt{\theta}\sqrt{5}\,a}\right)^2 \left(e^{\sqrt{\theta}\sqrt{5}\,b}\right)^2 a\theta$$

$$-3150\left(e^{\sqrt{\theta}\sqrt{5}\,a}\right)^2 \left(e^{\sqrt{\theta}\sqrt{5}}\right)^2 a^2 b\,\theta^{3/2}$$

$$+3150\left(e^{\sqrt{\theta}\sqrt{5}\,a}\right)^2 \left(e^{\sqrt{\theta}\sqrt{5}}\right)^2 a b^2 \theta^{3/2}$$

$$+150\sqrt{5}\left(e^{\sqrt{\theta}\sqrt{5}\,a}\right)^2 \left(e^{\sqrt{\theta}\sqrt{5}\,b}\right)^2 b^2 \theta$$

$$-810\sqrt{5}\left(e^{\sqrt{\theta}\sqrt{5}\,a}\right)^2 \left(e^{\sqrt{\theta}\sqrt{5}\,b}\right)^2 b\theta$$

$$-840\sqrt{5}\left(e^{\sqrt{\theta}\sqrt{5}\,a}\right)^2 \theta a^2 \left(e^{\sqrt{\theta}\sqrt{5}}\right)^2$$

$$-840\sqrt{5}\left(e^{\sqrt{\theta}\sqrt{5}\,a}\right)^2 \theta b^2 \left(e^{\sqrt{\theta}\sqrt{5}}\right)^2 + 1800\,a\,\theta^{3/2}$$

$$+150\sqrt{5}\left(e^{\sqrt{\theta}\sqrt{5}\,a}\right)^2 \left(e^{\sqrt{\theta}\sqrt{5}\,b}\right)^2 a^2 b^2 \theta^2$$

$$+600\sqrt{5}\left(e^{\sqrt{\theta}\sqrt{5}\,a}\right)^2 \left(e^{\sqrt{\theta}\sqrt{5}\,b}\right)^2 a b\,\theta^2$$

$$-300\sqrt{5}\left(e^{\sqrt{\theta}\sqrt{5}\,a}\right)^2 \left(e^{\sqrt{\theta}\sqrt{5}\,b}\right)^2 a b^2 \theta^2$$

$$-300\sqrt{5}\left(e^{\sqrt{\theta}\sqrt{5}\,a}\right)^2 \left(e^{\sqrt{\theta}\sqrt{5}\,b}\right)^2 a^2 b\,\theta^2$$

$$-600\,\theta^{3/2}\left(e^{\sqrt{\theta}\sqrt{5}\,a}\right)^2 \left(e^{\sqrt{\theta}\sqrt{5}\,b}\right)^2 a^2 b$$

$$+ 600\sqrt{5}\,\theta^2 \left(e^{\sqrt{\theta}\sqrt{5}\,a}\right)^2 a^3 b \left(e^{\sqrt{\theta}\sqrt{5}}\right)^2$$

$$- 900\sqrt{5}\,\theta^2 \left(e^{\sqrt{\theta}\sqrt{5}\,a}\right)^2 a^2 b^2 \left(e^{\sqrt{\theta}\sqrt{5}}\right)^2$$

$$+ 600\sqrt{5}\,\theta^2 \left(e^{\sqrt{\theta}\sqrt{5}\,a}\right)^2 a b^3 \left(e^{\sqrt{\theta}\sqrt{5}}\right)^2$$

$$- 600\,\theta^{3/2} \left(e^{\sqrt{\theta}\sqrt{5}\,a}\right)^2 \left(e^{\sqrt{\theta}\sqrt{5}\,b}\right)^2 a b^2$$

$$+ 2400\,\theta^{3/2} \left(e^{\sqrt{\theta}\sqrt{5}\,a}\right)^2 \left(e^{\sqrt{\theta}\sqrt{5}\,b}\right)^2 a b$$

$$+ 510\sqrt{5} \left(e^{\sqrt{\theta}\sqrt{5}\,a}\right)^2 \left(e^{\sqrt{\theta}\sqrt{5}\,b}\right)^2 a b \theta$$

$$+ 1680\sqrt{5} \left(e^{\sqrt{\theta}\sqrt{5}\,a}\right)^2 \theta a b \left(e^{\sqrt{\theta}\sqrt{5}}\right)^2 \Big)$$

(K.1)

## Appendix L.   Support for Section 5, when $[d, n, p] = [1, 1, 1]$

For the case of a single factor, a single design point, and exponential covariance, matrices $\boldsymbol{L}$, $\boldsymbol{L^{-1}}$, and $\boldsymbol{R}$ can be determined, using Eqs. C.1, with $d = 1$ and $x_i = x_1$, as well as with Eq. C.3, with $d = 1$ and $x_i = x_j = x_1$, as follows, where "·" is used for some elements in symmetric matrices, and "∗" denotes ordinary matrix multiplication:

$$\boldsymbol{L} = \begin{bmatrix} 0 & 1 \\ \cdot & 1 \end{bmatrix},\ \boldsymbol{L^{-1}} = \begin{bmatrix} -1 & 1 \\ \cdot & 0 \end{bmatrix},\ \boldsymbol{R} = \begin{bmatrix} 1 & | & \frac{1-e^{-\theta}\cosh(\theta x_1)}{\theta} \\ -- & | & --------- \\ \cdot & | & \frac{1-e^{-2\theta}\cosh(2\theta x_1)}{2\theta} \end{bmatrix}.$$ Then, using Matrix Identity A1,

$$IMSPE = 1 - tr(\boldsymbol{L^{-1}R}) = 1 - tr\left\{\begin{bmatrix} -1 & 1 \\ \cdot & 0 \end{bmatrix} * \begin{bmatrix} 1 & \frac{1-e^{-\theta}\cosh(\theta x_1)}{\theta} \\ \cdot & \frac{1-e^{-2\theta}\cosh(2\theta x_1)}{2\theta} \end{bmatrix}\right\} = 2\left[1 - \frac{1-e^{-\theta}\cosh(\theta x_1)}{\theta}\right]. \quad \text{(L.1)}$$

## Appendix M.   Transformation required, when using different design domains

We begin with the following example: For the example $[d, n, p] = [1,1,1]$ and design domain $[-1,1]$, and using Eq. L.1, we obtain

$$IMSPE(x_1, \theta) = 2\left[1 - \frac{1 - e^{-\theta}\cosh(\theta x_1)}{\theta}\right],$$

whereas for design domain $[0,1]$, using tilded variables and Eqs. C.2 and C.4, we have

$$\tilde{\boldsymbol{L}} = \begin{bmatrix} 0 & 1 \\ \cdot & 1 \end{bmatrix}, \tilde{\boldsymbol{L}}^{-1} = \begin{bmatrix} -1 & 1 \\ \cdot & 0 \end{bmatrix}, \text{and } \tilde{\boldsymbol{R}} = \begin{bmatrix} 1 & | & \frac{2-e^{-\tilde{\theta}\tilde{x}_i}-e^{-\tilde{\theta}(1-\tilde{x}_i)}}{\tilde{\theta}} \\ -- & | & --------- \\ \cdot & | & \frac{2 - e^{-2\tilde{\theta}\tilde{x}_i} - e^{-2\tilde{\theta}(1-\tilde{x}_i)}}{2\tilde{\theta}} \end{bmatrix}, \text{giving}$$

$$IMSPE(\widetilde{x_1}, \tilde{\theta}) = 1 - tr(\widetilde{LR}) = 1 - tr\left\{\begin{bmatrix} -1 & 1 \\ . & 0 \end{bmatrix} * \begin{bmatrix} 1 & \frac{2-e^{-\tilde{\theta}\widetilde{x_i}} - e^{-\tilde{\theta}(1-\widetilde{x_i})}}{\tilde{\theta}} \\ . & \frac{2-e^{-2\tilde{\theta}\widetilde{x_i}} - e^{-2\tilde{\theta}(1-\widetilde{x_i})}}{2\tilde{\theta}} \end{bmatrix}\right\}$$

$$= 2\left[1 - \frac{2-e^{-\tilde{\theta}\widetilde{x_i}} - e^{-\tilde{\theta}(1-\widetilde{x_i})}}{\tilde{\theta}}\right].$$

The following transformation between the tilded variables and parameters and their untilded counterparts,

$\tilde{\theta} = 2\theta$ and $\widetilde{x_i} = \frac{1}{2}(x+1)$, leads to

$$IMSPE(\widetilde{x_1}, \tilde{\theta}) = 2\left[1 - \frac{2-e^{-\tilde{\theta}\widetilde{x_i}} - e^{-\tilde{\theta}(1-\widetilde{x_i})}}{\tilde{\theta}}\right] = 2\left[1 - \frac{2-e^{-2\theta\frac{1+x_1}{2}} - e^{-2\theta\left(1-\frac{x_1+1}{2}\right)}}{2\theta}\right]$$

$$= 2\left[1 - \frac{1-e^{-\theta}\cosh(\theta x_1)}{\theta}\right] = IMSPE(x_1, \theta)\text{, so this transformation is correct, in this}$$

case, as it leads to invariance of $IMSPE$. It is evident that the general transformation between design domains $[a,b]$ and $[\tilde{a}, \tilde{b}]$ is the following:

$\tilde{\theta} = \frac{b-a}{\tilde{b}-\tilde{a}}\theta$ and $\widetilde{x_i} = \frac{\tilde{b}-\tilde{a}}{b-a}(x + \tilde{b} - \tilde{a} - b + a)$, the simple proof of which is left to the reader.

## Appendix N.   Support for Section 5, when $[d, n, p] = [1, 1, 2]$

For the case of a single factor, a single design point, and Gaussian covariance, $L$ and $L^{-1}$ are the same as in Appendix M, above. Then $R$ can be determined using Eqs. C.5 and C.6, as follows:

$$R = \begin{bmatrix} 1 & | & \sqrt{\frac{\pi}{16\theta}}\left\{\begin{array}{l}erf[\sqrt{\theta}\,(1+x_1)] \\ +erf[\sqrt{\theta}\,(1-x_1)]\end{array}\right\} \\ -- & | & ---------- \\ . & | & \sqrt{\frac{\pi}{32\theta}}\left\{\begin{array}{l}erf[\sqrt{2\theta}\,(1+x_1)] \\ +erf[\sqrt{2\theta}\,(1-x_1)]\end{array}\right\} \end{bmatrix}.$$ Then, using Matrix Identity A1,

$$IMSPE = 1 - tr(L^{-1}R) = 1 - tr\left\{\begin{bmatrix}-1 & 1 \\ . & 0\end{bmatrix} * \begin{bmatrix} 1 & | & \sqrt{\frac{\pi}{16\theta}}\left\{\begin{array}{l}erf[\sqrt{\theta}\,(1+x_1)] \\ +erf[\sqrt{\theta}\,(1-x_1)]\end{array}\right\} \\ -- & | & ---------- \\ . & | & \sqrt{\frac{\pi}{32\theta}}\left\{\begin{array}{l}erf[\sqrt{2\theta}\,(1+x_1)] \\ +erf[\sqrt{2\theta}\,(1-x_1)]\end{array}\right\} \end{bmatrix}\right\}.$$

$$= 2\left(1 - \sqrt{\frac{\pi}{16\theta}}\left\{\begin{array}{l}erf[\sqrt{\theta}\,(1+x_1)] \\ +erf[\sqrt{\theta}\,(1-x_1)]\end{array}\right\}\right). \tag{N.1}$$

## Appendix O.  Support for Section 5, when $[d, n, \nu] = [1, 1, 3/2]$

For the case of a single factor, a single design point, and Matérn covariance with parameter $\nu = 3/2$, $\boldsymbol{L}$ and $\boldsymbol{L}^{-1}$ are the same as in Appendix M, above. Then $\boldsymbol{R}$ can be determined using Eqs. C.7 and C.8, as follows:

$$\boldsymbol{R} = \begin{bmatrix} 1 & | & \frac{1}{2\sqrt{3\theta}}\left\{ 2\begin{bmatrix} 1 - e^{-\sqrt{3\theta}\,(1+x_1)} \\ +1 - e^{-\sqrt{3\theta}\,(1-x_1)} \end{bmatrix} - \sqrt{3\theta}\begin{bmatrix} (1+x_1)e^{-\sqrt{3\theta}\,(1+x_1)} \\ +(1-x_1)e^{-\sqrt{3\theta}\,(1-x_1)} \end{bmatrix}\right\} \\ -- & | & ---------------- \\ \cdot & | & MRSE \end{bmatrix},$$

where $MRSE$ stands for a machine-readable symbolic expression. Then, using Matrix Identity A1,

$$IMSPE = 1 - tr(\boldsymbol{L}^{-1}\boldsymbol{R}) = 1 - tr\left\{ \begin{bmatrix} -1 & 1 \\ \cdot & 0 \end{bmatrix} * \begin{bmatrix} 1 & | & \frac{1}{2\sqrt{3\theta}}\left\{ 2\begin{bmatrix} 1 - e^{-\sqrt{3\theta}\,(1+x_1)} \\ +1 - e^{-\sqrt{3\theta}\,(1-x_1)} \end{bmatrix} - \sqrt{3\theta}\begin{bmatrix} (1+x_1)e^{-\sqrt{3\theta}\,(1+x_1)} \\ +(1-x_1)e^{-\sqrt{3\theta}\,(1-x_1)} \end{bmatrix}\right\} \\ -- & | & ---------------- \\ \cdot & | & MRSE \end{bmatrix} \right\}$$

$$= 2\left(1 - \frac{1}{2\sqrt{3\theta}}\left\{ 2\begin{bmatrix} 1 - e^{-\sqrt{3\theta}\,(1+x_1)} \\ +1 - e^{-\sqrt{3\theta}\,(1-x_1)} \end{bmatrix} - \sqrt{3\theta}\begin{bmatrix} (1+x_1)e^{-\sqrt{3\theta}\,(1+x_1)} \\ +(1-x_1)e^{-\sqrt{3\theta}\,(1-x_1)} \end{bmatrix}\right\}\right). \tag{O.1}$$

## Appendix P.  Support for Section 5, when $[d, n, \nu] = [1, 1, 5/2]$

For the case of a single factor, a single design point, and Matérn covariance with parameter $5/2$, $\boldsymbol{L}$ and $\boldsymbol{L}^{-1}$ are the same as in Appendix L, above. Then $\boldsymbol{R}$ can be determined using Eqs. C.9 and C.10, as follows:

$$\boldsymbol{R} = \begin{bmatrix} 1 & | & \frac{1}{6\sqrt{5\theta}}\left( 8\left\{ \begin{bmatrix} 1 - e^{-\sqrt{5\theta}\,(1+x_1)} \end{bmatrix} + \begin{bmatrix} 1 - e^{-\sqrt{5\theta}\,(1-x_1)} \end{bmatrix}\right\} -5\sqrt{5\theta}\left\{ \begin{matrix} (1+x_1)e^{-\sqrt{5\theta}\,(1+x_1)} \\ +(1-x_1)e^{-\sqrt{5\theta}\,(1-x_1)} \end{matrix}\right\} -5\theta\left\{ \begin{matrix} (1+x_1)^2 e^{-\sqrt{5\theta}\,(1+x_1)} \\ +(1-x_1)^2 e^{-\sqrt{5\theta}\,(1-x_1)} \end{matrix}\right\} \right) \\ -- & | & ---------------- \\ \cdot & | & MRSE \end{bmatrix},$$

where $MRSE$ stands for a machine-readable symbolic expression. Then, using Matrix Identity A1,

$$IMSPE = 1 - tr(\boldsymbol{L}^{-1}\boldsymbol{R}) = 1 - tr\left\{\begin{bmatrix}-1 & 1\\ \cdot & 0\end{bmatrix} * \begin{bmatrix}1 & | & \frac{1}{6\sqrt{5\theta}}\begin{pmatrix}8\left\{\begin{matrix}\left[1-e^{-\sqrt{5\theta}\,(1+x_1)}\right]\\+\left[1-e^{-\sqrt{5\theta}\,(1-x_1)}\right]\end{matrix}\right\}\\-5\sqrt{5\theta}\left\{\begin{matrix}(1+x_1)e^{-\sqrt{5\theta}\,(1+x_1)}\\+(1-x_1)e^{-\sqrt{5\theta}\,(1-x_1)}\end{matrix}\right\}\\-5\theta\left\{\begin{matrix}(1+x_1)^2 e^{-\sqrt{5\theta}\,(1+x_1)}\\+(1-x_1)^2 e^{-\sqrt{5\theta}\,(1-x_1)}\end{matrix}\right\}\end{pmatrix}\\ -- & | & -----------------\\ \cdot & | & MRSE\end{bmatrix}\right\}$$

$$= 2\left[1 - \frac{1}{6\sqrt{5\theta}}\begin{pmatrix}8\left\{\begin{matrix}\left[1-e^{-\sqrt{5\theta}\,(1+x_1)}\right]\\+\left[1-e^{-\sqrt{5\theta}\,(1-x_1)}\right]\end{matrix}\right\}\\-5\sqrt{5\theta}\left\{\begin{matrix}(1+x_1)e^{-\sqrt{5\theta}\,(1+x_1)}\\+(1-x_1)e^{-\sqrt{5\theta}\,(1-x_1)}\end{matrix}\right\}\\-5\theta\left\{\begin{matrix}(1+x_1)^2 e^{-\sqrt{5\theta}\,(1+x_1)}\\+(1-x_1)^2 e^{-\sqrt{5\theta}\,(1-x_1)}\end{matrix}\right\}\end{pmatrix}\right].$$
(P.1)

## Appendix Q. Support for Section 6, when $[d, n, p] = [1, 2, 1]$

### Q.1 Customary design variables

For the case of two design points in one factor, the points can be denoted $x_1$ and $x_2$. Matrices $\boldsymbol{L}$, $\boldsymbol{L}^{-1}$, and $\boldsymbol{R}$, as well as objective $IMSPE$, are the following:

$$\boldsymbol{L} = \begin{bmatrix}0 & | & 1 & 1\\ -- & | & -- & -----\\ \cdot & | & 1 & e^{-\theta|x_1-x_2|}\\ \cdot & | & \cdot & 1\end{bmatrix},$$

$$\boldsymbol{L}^{-1} = \frac{1}{2} * \begin{bmatrix}-1-e^{-\theta|x_1-x_2|} & | & 1 & 1\\ --------- & | & ----- & -----\\ \cdot & | & \frac{1}{1-e^{-\theta|x_1-x_2|}} & \frac{-1}{1-e^{-\theta|x_1-x_2|}}\\ \cdot & | & \cdot & \frac{1}{1-e^{-\theta|x_1-x_2|}}\end{bmatrix}, \text{ as confirmed by SMS, and}$$

$$\boldsymbol{R} = \begin{bmatrix}1 & \frac{1-e^{-\theta}\cosh(\theta x_1)}{\theta} & \frac{1-e^{-\theta}\cosh(\theta x_2)}{\theta}\\ \cdot & \frac{1-e^{-2\theta}\cosh(2\theta x_1)}{2\theta} & \frac{e^{-\theta|x_1-x_2|}-e^{-2\theta}\cosh[\theta(x_1+x_2)]}{2\theta} + \frac{|x_1-x_2|}{2}e^{-\theta|x_1-x_2|}\\ \cdot & \cdot & \frac{1-e^{-2\theta}\cosh(2\theta x_2)}{2\theta}\end{bmatrix}, \text{ from Eqs. C.1 and C.3.}$$

Then, using Matrix Identity A1,

$$IMSPE = 1 - tr\left\{\begin{array}{l}\frac{1}{2}*\begin{bmatrix}-1-e^{-\theta|x_1-x_2|} & | & 1 & 1 \\ ------- & | & ----- & ----- \\ . & | & \frac{1}{1-e^{-\theta|x_1-x_2|}} & \frac{-1}{1-e^{-\theta|x_1-x_2|}} \\ . & | & . & \frac{1}{1-e^{-\theta|x_1-x_2|}}\end{bmatrix} \\ *\begin{bmatrix}1 & \frac{1-e^{-\theta}\cosh(\theta x_1)}{\theta} & & \frac{1-e^{-\theta}\cosh(\theta x_2)}{\theta} \\ . & \frac{1-e^{-2\theta}\cosh(2\theta x_1)}{2\theta} & \frac{e^{-\theta|x_1-x_2|}-e^{-2\theta}\cosh[\theta(x_1+x_2)]}{2\theta}+\frac{|x_1-x_2|e^{-\theta|x_1-x_2|}}{2} \\ . & . & \frac{1-e^{-2\theta}\cosh(2\theta x_2)}{2\theta}\end{bmatrix}\end{array}\right\}$$

$$= 1-\frac{1}{2}\left\{\begin{array}{l}-1-e^{-\theta|x_1-x_2|} \\ +\frac{1}{1-e^{-\theta|x_1-x_2|}}\left[\frac{1-e^{-2\theta}\cosh(2\theta x_1)}{2\theta}\right] \\ +\frac{1}{1-e^{-\theta|x_1-x_2|}}\left[\frac{1-e^{-2\theta}\cosh(2\theta x_2)}{2\theta}\right] \\ +2\frac{1-e^{-\theta}\cosh(\theta x_1)}{\theta}+2\frac{1-e^{-\theta}\cosh(\theta x_2)}{\theta} \\ +2\left(\frac{-1}{1-e^{-\theta|x_1-x_2|}}\right)\left[\frac{e^{-\theta|x_1-x_2|}-e^{-2\theta}\cosh[\theta(x_1+x_2)]}{2\theta}+\frac{|x_1-x_2|e^{-\theta|x_1-x_2|}}{2}\right]\end{array}\right\}$$

$$= \frac{3}{2}+\frac{e^{-\theta|x_1-x_2|}}{2}-\frac{1-e^{-2\theta}\cosh(2\theta x_1)}{4\theta(1-e^{-\theta|x_1-x_2|})}-\frac{1-e^{-2\theta}\cosh(2\theta x_2)}{4\theta(1-e^{-\theta|x_1-x_2|})}-\frac{1-e^{-\theta}\cosh(\theta x_1)}{\theta}-\frac{1-e^{-\theta}\cosh(\theta x_2)}{\theta}$$

$$+\frac{e^{-\theta|x_1-x_2|}-e^{-2\theta}\cosh[\theta(x_1+x_2)]+\theta|x_1-x_2|e^{-\theta|x_1-x_2|}}{2\theta(1-e^{-\theta|x_1-x_2|})}$$

$$= \frac{3+e^{-\theta|x_1-x_2|}}{2}+\frac{e^{-\theta|x_1-x_2|}-e^{-2\theta}\cosh[\theta(x_1+x_2)]+\theta|x_1-x_2|e^{-\theta|x_1-x_2|}}{2\theta(1-e^{-\theta|x_1-x_2|})}$$

$$-\frac{1-e^{-\theta}\cosh(\theta x_1)}{\theta}-\frac{1-e^{-2\theta}\cosh(2\theta x_1)}{4\theta(1-e^{-\theta|x_1-x_2|})}$$

$$-\frac{1-e^{-\theta}\cosh(\theta x_2)}{\theta}-\frac{1-e^{-2\theta}\cosh(2\theta x_2)}{4\theta(1-e^{-\theta|x_1-x_2|})}.\qquad\text{(Q.1.1)}$$

The following SMS result is a perfect match to Eq. Q.1.1

$$IMSPE for debugging := \frac{3}{2}+\frac{1}{2}e^{-\theta|x1-x2|}-\frac{1-e^{-\theta}\cosh(\theta x1)}{\theta}-\frac{1-e^{-\theta}\cosh(\theta x2)}{\theta}$$

$$+\frac{1}{4}\frac{1-e^{-2\theta}\cosh(2\theta x1)}{(-1+e^{-\theta|x1-x2|})\theta}$$

$$-\frac{\frac{1}{2}\frac{e^{-\theta|x1-x2|}-e^{-2\theta}\cosh(\theta(x1+x2))}{\theta}+\frac{1}{2}|x1-x2|e^{-\theta|x1-x2|}}{-1+e^{-\theta|x1-x2|}}$$

$$+\frac{1}{4}\frac{1-e^{-2\theta}\cosh(2\theta x2)}{(-1+e^{-\theta|x1-x2|})\theta}$$

## Q.2 Cluster variables

When two design points are proximal, it is helpful to change the variables from $x_1$ and $x_2$ to the following: $x_t \equiv \frac{x_1+x_2}{2}$, i.e. the points' center, and $\delta \equiv x_1 - x_t$, i.e. the signed distance from the points' center to the point with the lower-subscript value. In these variables, which we name "cluster variables,"

$$L = \begin{bmatrix} 0 & | & 1 & 1 \\ -- & | & -- & ---- \\ \cdot & | & 1 & e^{-2\theta|\delta|} \\ \cdot & | & \cdot & 1 \end{bmatrix},$$

$$L^{-1} = \frac{1}{2} * \begin{bmatrix} -1 - e^{-2\theta|\delta|} & | & 1 & 1 \\ ------ & | & ---- & ------ \\ \cdot & | & \frac{1}{1-e^{-2\theta|\delta|}} & \frac{-1}{1-e^{-2\theta|\delta|}} \\ \cdot & | & \cdot & \frac{1}{1-e^{-2\theta|\delta|}} \end{bmatrix}, \text{ as confirmed by SMS,}$$

$$R = \begin{bmatrix} 1 & \frac{1-e^{-\theta}\cosh[\theta(x_t+\delta)]}{\theta} & \frac{1-e^{-\theta}\cosh[\theta(x_t-\delta)]}{\theta} \\ \cdot & \frac{1-e^{-2\theta}\cosh[2\theta(x_t+\delta)]}{2\theta} & \frac{e^{-2\theta|\delta|}-e^{-2\theta}\cosh(2\theta x_t)}{2\theta} + |\delta|e^{-2\theta|\delta|} \\ \cdot & \cdot & \frac{1-e^{-2\theta}\cosh[2\theta(x_t-\delta)]}{2\theta} \end{bmatrix}, \text{ and}$$

$$IMSPE = 1 - tr\left( \frac{1}{2} * \begin{bmatrix} -1 - e^{-2\theta|\delta|} & | & 1 & 1 \\ ------ & | & ---- & ------ \\ \cdot & | & \frac{1}{1-e^{-2\theta|\delta|}} & \frac{-1}{1-e^{-2\theta|\delta|}} \\ \cdot & | & \cdot & \frac{1}{1-e^{-2\theta|\delta|}} \end{bmatrix} \right.$$
$$\left. * \begin{Bmatrix} 1 & \frac{1-e^{-\theta}\cosh[\theta(x_t+\delta)]}{\theta} & \frac{1-e^{-\theta}\cosh[\theta(x_t-\delta)]}{\theta} \\ \cdot & \frac{1-e^{-2\theta}\cosh[2\theta(x_t+\delta)]}{2\theta} & \frac{e^{-2\theta|\delta|}-e^{-2\theta}\cosh(2\theta x_t)}{2\theta} + |\delta|e^{-2\theta|\delta|} \\ \cdot & \cdot & \frac{1-e^{-2\theta}\cosh[2\theta(x_t-\delta)]}{2\theta} \end{Bmatrix} \right).$$

# Appendix R. Support for Section 6 $[d, n, p] = [1, 2, 2]$

### R.1 Customary design variables

For the case of one factor, two design points, and Gaussian covariance, with design points denoted $x_1$ and $x_2$, we have the following, via Matrix Identity A2:

$$L = \begin{bmatrix} 0 & | & 1 & 1 \\ -- & | & -- & ------ \\ \cdot & | & 1 & e^{-\theta(x_1-x_2)^2} \\ \cdot & | & \cdot & 1 \end{bmatrix} \text{ and}$$

$$L^{-1} = \frac{1}{2} * \begin{bmatrix} -1-e^{-\theta(x_1-x_2)^2} & | & 1 & 1 \\ \text{---} & | & \text{---} & \text{---} \\ \cdot & | & \frac{1}{1-e^{-\theta(x_1-x_2)^2}} & \frac{-1}{1-e^{-\theta(x_1-x_2)^2}} \\ \cdot & | & \cdot & \frac{1}{1-e^{-\theta(x_1-x_2)^2}} \end{bmatrix}, \text{ as confirmed by SMS. Then, via}$$

Eqs. C.5 and C.6,

$$R = \begin{pmatrix} 1 & \sqrt{\frac{\pi}{16\theta}} \begin{Bmatrix} erf[\sqrt{\theta}(1+x_1)] \\ +erf[\sqrt{\theta}(1-x_1)] \end{Bmatrix} & \sqrt{\frac{\pi}{16\theta}} \begin{Bmatrix} erf[\sqrt{\theta}(1+x_2)] \\ +erf[\sqrt{\theta}(1-x_2)] \end{Bmatrix} \\ \cdot & \sqrt{\frac{\pi}{32\theta}} \begin{Bmatrix} erf[\sqrt{2\theta}(1+x_1)] \\ +erf[\sqrt{2\theta}(1-x_1)] \end{Bmatrix} & \sqrt{\frac{\pi}{32\theta}} \begin{Bmatrix} erf\left[\sqrt{2\theta}\left(1+\frac{x_1+x_2}{2}\right)\right] \\ +erf\left[\sqrt{2\theta}\left(1-\frac{x_1+x_2}{2}\right)\right] \end{Bmatrix} e^{\frac{-\theta(x_1-x_2)^2}{2}} \\ \cdot & \cdot & \sqrt{\frac{\pi}{32\theta}} \begin{Bmatrix} erf[\sqrt{2\theta}(1+x_2)] \\ +erf[\sqrt{2\theta}(1-x_2)] \end{Bmatrix} \end{pmatrix}, \text{ and}$$

$$IMSPE = 1 - \frac{1}{2} tr \left[ \begin{bmatrix} -1-e^{-\theta(x_1-x_2)^2} & 1 & 1 \\ \cdot & \frac{1}{1-e^{-\theta(x_1-x_2)^2}} & \frac{-1}{1-e^{-\theta(x_1-x_2)^2}} \\ \cdot & \cdot & \frac{1}{1-e^{-\theta(x_1-x_2)^2}} \end{bmatrix} \\ * \begin{pmatrix} 1 & \sqrt{\frac{\pi}{16\theta}} \begin{Bmatrix} erf[\sqrt{\theta}(1+x_1)] \\ +erf[\sqrt{\theta}(1-x_1)] \end{Bmatrix} & \sqrt{\frac{\pi}{16\theta}} \begin{Bmatrix} erf[\sqrt{\theta}(1+x_2)] \\ +erf[\sqrt{\theta}(1-x_2)] \end{Bmatrix} \\ \cdot & \sqrt{\frac{\pi}{32\theta}} \begin{Bmatrix} erf[\sqrt{2\theta}(1+x_1)] \\ +erf[\sqrt{2\theta}(1-x_1)] \end{Bmatrix} & \sqrt{\frac{\pi}{32\theta}} \begin{Bmatrix} erf\left[\sqrt{2\theta}\left(1+\frac{x_1+x_2}{2}\right)\right] \\ +erf\left[\sqrt{2\theta}\left(1-\frac{x_1+x_2}{2}\right)\right] \end{Bmatrix} e^{\frac{-\theta(x_1-x_2)^2}{2}} \\ \cdot & \cdot & \sqrt{\frac{\pi}{32\theta}} \begin{Bmatrix} erf[\sqrt{2\theta}(1+x_2)] \\ +erf[\sqrt{2\theta}(1-x_2)] \end{Bmatrix} \end{pmatrix} \right].$$

(R.1.1)

**R.2 Cluster variables**

In cluster variables,

$$L = \begin{bmatrix} 0 & | & 1 & 1 \\ \text{---} & | & \text{---} & \text{---} \\ \cdot & | & 1 & e^{-4\theta\delta^2} \\ \cdot & | & \cdot & 1 \end{bmatrix},$$

$$L^{-1} = \frac{1}{2} * \begin{bmatrix} -1-e^{-4\theta\delta^2} & | & 1 & 1 \\ \text{---} & | & \text{---} & \text{---} \\ \cdot & | & \frac{1}{1-e^{-4\theta\delta^2}} & \frac{-1}{1-e^{-4\theta\delta^2}} \\ \cdot & | & \cdot & \frac{1}{1-e^{-4\theta\delta^2}} \end{bmatrix}, \text{ as confirmed by SMS,}$$

$$R = \begin{bmatrix} 1 & | & \sqrt{\frac{\pi}{16\theta}}\left\{\begin{array}{l} erf[\sqrt{\theta}\,(1+x_t+\delta)] \\ +erf[\sqrt{\theta}\,(1-x_t-\delta)]\end{array}\right\} & \sqrt{\frac{\pi}{16\theta}}\left\{\begin{array}{l} erf[\sqrt{\theta}\,(1+x_t-\delta)] \\ +erf[\sqrt{\theta}\,(1-x_t+\delta)]\end{array}\right\} \\ -- & | & ---------- & ---------- \\ \cdot & | & \sqrt{\frac{\pi}{32\theta}}\left\{\begin{array}{l} erf[\sqrt{2\theta}(1+x_t+\delta)] \\ +erf[\sqrt{2\theta}(1-x_t-\delta)]\end{array}\right\} & \sqrt{\frac{\pi}{32\theta}}\left\{\begin{array}{l} erf[\sqrt{2\theta}(1+x_t)] \\ +erf[\sqrt{2\theta}(1-x_t)]\end{array}\right\} e^{-2\theta\delta^2} \\ \cdot & | & \cdot & \sqrt{\frac{\pi}{32\theta}}\left\{\begin{array}{l} erf[\sqrt{2\theta}(1+x_t-\delta)] \\ +erf[\sqrt{2\theta}(1-x_t+\delta)]\end{array}\right\} \end{bmatrix}, \text{ and}$$

$$IMSPE = 1 - \tfrac{1}{2} tr\left(\begin{bmatrix} -1-e^{-4\theta\delta^2} & | & 1 & 1 \\ ----- & | & ----- & ----- \\ \cdot & | & \frac{-1}{-1+e^{-4\theta\delta^2}} & \frac{1}{-1+e^{-4\theta\delta^2}} \\ \cdot & | & \cdot & \frac{-1}{-1+e^{-4\theta\delta^2}} \end{bmatrix} * \begin{bmatrix} 1 & \sqrt{\frac{\pi}{16\theta}}\left\{\begin{array}{l} erf[\sqrt{\theta}\,(1+x_t+\delta)] \\ +erf[\sqrt{\theta}\,(1-x_t-\delta)]\end{array}\right\} & \sqrt{\frac{\pi}{16\theta}}\left\{\begin{array}{l} erf[\sqrt{\theta}\,(1+x_t-\delta)] \\ +erf[\sqrt{\theta}\,(1-x_t+\delta)]\end{array}\right\} \\ \cdot & \sqrt{\frac{\pi}{32\theta}}\left\{\begin{array}{l} erf[\sqrt{2\theta}(1+x_t+\delta)] \\ +erf[\sqrt{2\theta}(1-x_t-\delta)]\end{array}\right\} & \sqrt{\frac{\pi}{32\theta}}\left\{\begin{array}{l} erf[\sqrt{2\theta}(1+x_t)] \\ +erf[\sqrt{2\theta}(1-x_t)]\end{array}\right\} e^{-2\theta\delta^2} \\ \cdot & \cdot & \sqrt{\frac{\pi}{32\theta}}\left\{\begin{array}{l} erf[\sqrt{2\theta}(1+x_t-\delta)] \\ +erf[\sqrt{2\theta}(1-x_t+\delta)]\end{array}\right\} \end{bmatrix}\right). \quad \text{(R.2.1)}$$

### R.3 Expansions in powers of $\sqrt{\theta}\delta$, using hand algebra

We expand $L^{-1}$ and $R$ of the last section, via Taylor series, in powers of $\sqrt{\theta}\delta$. Through power $\theta\delta^2$, we have

$$L^{-1} = \begin{bmatrix} -1+2\theta\delta^2+O(\delta^4) & | & \tfrac{1}{2} & \tfrac{1}{2} \\ ---------- & | & ---------- & ---------- \\ \cdot & | & \frac{1}{8\theta\delta^2}+\tfrac{1}{4}+\tfrac{\theta\delta^2}{6}+O(\delta^4) & -\left(\frac{1}{8\theta\delta^2}+\tfrac{1}{4}+\tfrac{\theta\delta^2}{6}\right)+O(\delta^4) \\ \cdot & | & \cdot & \frac{1}{8\theta\delta^2}+\tfrac{1}{4}+\tfrac{\theta\delta^2}{6}+O(\delta^4) \end{bmatrix},$$

as confirmed by SMS or by hand matrix multiplication of the above $L^{-1}$ and $L$.

Taylor-series expansion of $R$ requires derivatives of the error function. The definition of the error function [12] and its first four derivatives are useful:

$$erf(x) \equiv \tfrac{2}{\sqrt{\pi}} \int_{t=0}^{x} e^{-t^2} dt,$$

$$\frac{derf(x)}{dx} = \tfrac{2}{\sqrt{\pi}}\, e^{-x^2},$$

$$\frac{d^2 erf(x)}{dx^2} = \tfrac{-4}{\sqrt{\pi}} x e^{-x^2},$$

$$\frac{d^3 erf(x)}{dx^3} = \tfrac{-4}{\sqrt{\pi}}\, e^{-x^2} + \tfrac{8}{\sqrt{\pi}} x^2 e^{-x^2} = \tfrac{-4}{\sqrt{\pi}}\,(1-2x^2) e^{-x^2}, \text{ and}$$

$$\frac{d^4 erf(x)}{dx^4} = \frac{24}{\sqrt{\pi}} x e^{-x^2} - \frac{16}{\sqrt{\pi}} x^3 e^{-x^2} = \frac{24}{\sqrt{\pi}} x \left(1 - \frac{2}{3} x^2\right) e^{-x^2}.$$

The error function enters Eq. R.2.1 in the form $\left\{\begin{array}{c} erf[\sqrt{c\theta}(1 + x_t + \delta)] \\ +erf[\sqrt{c\theta}(1 - x_t - \delta)] \end{array}\right\}$, with variants of the signs, the value $1$ or $2$ assigned to $c$, and sometimes $\delta = 0$. Expanding in powers of $\sqrt{c\theta}\delta$, via Taylor series about $\sqrt{c\theta}(1 \pm x_t)$, as appropriate, gives the following generic expansion:

$$erf[\sqrt{c\theta}(1 + x_t + \delta)] + erf[\sqrt{c\theta}(1 - x_t - \delta)]$$
$$= erf[\sqrt{c\theta}(1 + x_t)] + erf[\sqrt{c\theta}(1 - x_t)]$$
$$+ \sqrt{\frac{4}{\pi}} \left[\begin{array}{c} e^{-c\theta(1+x_t)^2} \\ -e^{-c\theta(1-x_t)^2} \end{array}\right] \sqrt{c\theta}\delta$$
$$- \sqrt{\frac{4c\theta}{\pi}} \left[\begin{array}{c} (1 + x_t)e^{-c\theta(1+x_t)^2} \\ +(1 - x_t)e^{-c\theta(1-x_t)^2} \end{array}\right] c\theta\delta^2$$
$$- \sqrt{\frac{4}{9\pi}} \left\{\begin{array}{c} [1 - 2c\theta(1 + x_t)^2]e^{-c\theta(1+x_t)^2} \\ -[1 - 2c\theta(1 - x_t)^2]e^{-c\theta(1-x_t)^2} \end{array}\right\} (c\theta)^{3/2}\delta^3$$
$$+ \sqrt{\frac{c\theta}{\pi}} \left\{\begin{array}{c} \left[(1 + x_t) - \frac{2c\theta}{3}(1 + x_t)^3\right] e^{-c\theta(1+x_t)^2} \\ + \left[(1 - x_t) - \frac{2c\theta}{3}(1 - x_t)^3\right] e^{-c\theta(1-x_t)^2} \end{array}\right\} c^2\theta^2\delta^4$$
$$+ O[(c\theta)^{5/2}\delta^5], \text{ from which we have the following elements of } \mathbf{R}:$$

$R_{0,0} = 1$

$$R_{0,1} = \sqrt{\frac{\pi}{16\theta}} \left(\begin{array}{c} erf[\sqrt{\theta}(1 + x_t)] \\ +erf[\sqrt{\theta}(1 - x_t)] \\ + \sqrt{\frac{4}{\pi}} \left[\begin{array}{c} e^{-\theta(1+x_t)^2} \\ -e^{-\theta(1-x_t)^2} \end{array}\right] \sqrt{\theta}\delta \\ - \sqrt{\frac{4\theta}{\pi}} \left[\begin{array}{c} (1 + x_t)e^{-\theta(1+x_t)^2} \\ +(1 - x_t)e^{-\theta(1-x_t)^2} \end{array}\right] \theta\delta^2 \\ - \sqrt{\frac{4}{9\pi}} \left\{\begin{array}{c} [1 - 2\theta(1 + x_t)^2]e^{-\theta(1+x_t)^2} \\ -[1 - 2\theta(1 - x_t)^2]e^{-\theta(1-x_t)^2} \end{array}\right\} \theta^{3/2}\delta^3 \\ + \sqrt{\frac{\theta}{\pi}} \left\{\begin{array}{c} \left[(1 + x_t) - \frac{2\theta}{3}(1 + x_t)^3\right] e^{-\theta(1+x_t)^2} \\ + \left[(1 - x_t) - \frac{2\theta}{3}(1 - x_t)^3\right] e^{-\theta(1-x_t)^2} \end{array}\right\} \theta^2\delta^4 \\ + O[\theta^{5/2}\delta^5] \end{array}\right) \quad (R.3.1)$$

$$R_{0,2} = \sqrt{\frac{\pi}{16\theta}} \begin{pmatrix} erf[\sqrt{\theta}(1+x_t)] \\ +erf[\sqrt{\theta}(1-x_t)] \\ -\sqrt{\frac{4}{\pi}}\begin{bmatrix} e^{-\theta(1+x_t)^2} \\ -e^{-\theta(1-x_t)^2} \end{bmatrix}\sqrt{\theta}\delta \\ -\sqrt{\frac{4\theta}{\pi}}\begin{bmatrix} (1+x_t)e^{-\theta(1+x_t)^2} \\ +(1-x_t)e^{-\theta(1-x_t)^2} \end{bmatrix}\theta\delta^2 \\ +\sqrt{\frac{4}{9\pi}}\begin{Bmatrix} [1-2\theta(1+x_t)^2]e^{-\theta(1+x_t)^2} \\ -[1-2\theta(1-x_t)^2]e^{-\theta(1-x_t)^2} \end{Bmatrix}\theta^{3/2}\delta^3 \\ +\sqrt{\frac{\theta}{\pi}}\begin{Bmatrix} \left[(1+x_t)-\frac{2\theta}{3}(1+x_t)^3\right]e^{-\theta(1+x_t)^2} \\ +\left[(1-x_t)-\frac{2\theta}{3}(1-x_t)^3\right]e^{-\theta(1-x_t)^2} \end{Bmatrix}\theta^2\delta^4 \\ +O[\theta^{5/2}\delta^5] \end{pmatrix}$$

$$R_{1,1} = \sqrt{\frac{\pi}{32\theta}} \begin{pmatrix} erf[\sqrt{2\theta}(1+x_t)] \\ +erf[\sqrt{2\theta}(1-x_t)] \\ +\sqrt{\frac{4}{\pi}}\begin{bmatrix} e^{-2\theta(1+x_t)^2} \\ -e^{-2\theta(1-x_t)^2} \end{bmatrix}\sqrt{2\theta}\delta \\ -\sqrt{\frac{8\theta}{\pi}}\begin{bmatrix} (1+x_t)e^{-2\theta(1+x_t)^2} \\ +(1-x_t)e^{-2\theta(1-x_t)^2} \end{bmatrix}2\theta\delta^2 \\ -\sqrt{\frac{4}{9\pi}}\begin{Bmatrix} [1-4\theta(1+x_t)^2]e^{-2\theta(1+x_t)^2} \\ -[1-4\theta(1-x_t)^2]e^{-2\theta(1-x_t)^2} \end{Bmatrix}(2\theta)^{3/2}\delta^3 \\ +\sqrt{\frac{2\theta}{\pi}}\begin{Bmatrix} \left[(1+x_t)-\frac{4\theta}{3}(1+x_t)^3\right]e^{-2\theta(1+x_t)^2} \\ +\left[(1-x_t)-\frac{4\theta}{3}(1-x_t)^3\right]e^{-2\theta(1-x_t)^2} \end{Bmatrix}4\theta^2\delta^4 \\ +O[\theta^{5/2}\delta^5] \end{pmatrix}$$

$$R_{1,2} = \sqrt{\frac{\pi}{32\theta}} \begin{Bmatrix} erf[\sqrt{2\theta}(1+x_t)] \\ +erf[\sqrt{2\theta}(1-x_t)] \end{Bmatrix} [1-2\theta\delta + 2\theta^2\delta^2 + O(\theta^3\delta^6)]$$

$$R_{2,2} = \sqrt{\frac{\pi}{32\theta}} \begin{pmatrix} erf[\sqrt{2\theta}(1+x_t)] \\ +erf[\sqrt{2\theta}(1-x_t)] \\ -\sqrt{\frac{4}{\pi}}\begin{bmatrix} e^{-2\theta(1+x_t)^2} \\ -e^{-2\theta(1-x_t)^2} \end{bmatrix}\sqrt{2\theta}\delta \\ -\sqrt{\frac{8\theta}{\pi}}\begin{bmatrix} (1+x_t)e^{-2\theta(1+x_t)^2} \\ +(1-x_t)e^{-2\theta(1-x_t)^2} \end{bmatrix}2\theta\delta^2 \\ +\sqrt{\frac{4}{9\pi}}\begin{Bmatrix} [1-4\theta(1+x_t)^2]e^{-2\theta(1+x_t)^2} \\ -[1-4\theta(1-x_t)^2]e^{-2\theta(1-x_t)^2} \end{Bmatrix}(2\theta)^{3/2}\delta^3 \\ +\sqrt{\frac{2\theta}{\pi}}\begin{Bmatrix} \left[(1+x_t)-\frac{4\theta}{3}(1+x_t)^3\right]e^{-2\theta(1+x_t)^2} \\ +\left[(1-x_t)-\frac{4\theta}{3}(1-x_t)^3\right]e^{-2\theta(1-x_t)^2} \end{Bmatrix}4\theta^2\delta^4 \\ +O[\theta^{5/2}\delta^5] \end{pmatrix}.$$

Each remaining element of **R** is determined by symmetry of **R** and was confirmed by SMS.

From Eq. 4.1 and Matrix Identity A1,

$$IMSPE = 1 - tr(\mathbf{L}^{-1}\mathbf{R}) = 1 - \sum_{j,k=0}^{N} L_{j,k}^{-1} R_{j,k} = 1 - \sum_{j=0}^{N} L_{j,j}^{-1} R_{j,j} - 2\sum_{j \neq k=0}^{N} L_{j,k}^{-1} R_{j,k},$$

and this can be written explicitly, via Taylor series through order $\theta\delta^2$, as follows, in which, following the order of terms in the last sum, the diagonal terms precede off-diagonal terms; and color is used to assist the reader in identifying cancelling terms, as well as terms to be collected, etc., as specified in the text:

$$IMSPE = 1 - (-1 + 2\theta\delta^2)$$

$$-\left(\frac{1}{8\theta\delta^2} + \frac{1}{4} + \frac{\theta\delta^2}{6}\right)\sqrt{\frac{\pi}{32\theta}} \begin{pmatrix} \textcolor{green}{erf[\sqrt{2\theta}(1+x_t)]} \\ \textcolor{green}{+erf[\sqrt{2\theta}(1-x_t)]} \\ \textcolor{red}{+\sqrt{\frac{4}{\pi}}\begin{bmatrix} e^{-2\theta(1+x_t)^2} \\ -e^{-2\theta(1-x_t)^2} \end{bmatrix}\sqrt{2\theta}\delta} \\ -\sqrt{\frac{8\theta}{\pi}}\begin{bmatrix} (1+x_t)e^{-2\theta(1+x_t)^2} \\ +(1-x_t)e^{-2\theta(1-x_t)^2} \end{bmatrix}2\theta\delta^2 \\ -\sqrt{\frac{4}{9\pi}}\left\{ \begin{matrix} [\mathbf{1-4\theta(1+x_t)^2}]e^{-\theta(1+x_t)^2} \\ -[\mathbf{1-4\theta(1-x_t)^2}]e^{-2\theta(1-x_t)^2} \end{matrix} \right\}(2\theta)^{3/2}\delta^3 \\ +\sqrt{\frac{2\theta}{\pi}}\left\{ \begin{matrix} [(1+x_t) - \frac{4\theta}{3}(1+x_t)^3]e^{-2\theta(1+x_t)^2} \\ +[(1-x_t) - \frac{4\theta}{3}(1-x_t)^3]e^{-2\theta(1-x_t)^2} \end{matrix} \right\}4\theta^2\delta^4 \\ +O(\theta^{5/2}\delta^5) \end{pmatrix}$$

$$-\left(\frac{1}{8\theta\delta^2} + \frac{1}{4} + \frac{\theta\delta^2}{6}\right)\sqrt{\frac{\pi}{32\theta}} \begin{pmatrix} \textcolor{green}{erf[\sqrt{2\theta}(1+x_t)]} \\ \textcolor{green}{+erf[\sqrt{2\theta}(1-x_t)]} \\ \textcolor{red}{-\sqrt{\frac{4}{\pi}}\begin{bmatrix} e^{-2\theta(1+x_t)^2} \\ -e^{-2\theta(1-x_t)^2} \end{bmatrix}\sqrt{2\theta}\delta} \\ -\sqrt{\frac{8\theta}{\pi}}\begin{bmatrix} (1+x_t)e^{-2\theta(1+x_t)^2} \\ +(1-x_t)e^{-2\theta(1-x_t)^2} \end{bmatrix}2\theta\delta^2 \\ +\sqrt{\frac{4}{9\pi}}\left\{ \begin{matrix} [\mathbf{1-4\theta(1+x_t)^2}]e^{-\theta(1+x_t)^2} \\ -[\mathbf{1-4\theta(1-x_t)^2}]e^{-2\theta(1-x_t)^2} \end{matrix} \right\}(2\theta)^{3/2}\delta^3 \\ +\sqrt{\frac{2\theta}{\pi}}\left\{ \begin{matrix} [(1+x_t) - \frac{4\theta}{3}(1+x_t)^3]e^{-2\theta(1+x_t)^2} \\ +[(1-x_t) - \frac{4\theta}{3}(1-x_t)^3]e^{-2\theta(1-x_t)^2} \end{matrix} \right\}4\theta^2\delta^4 \\ +O(\theta^{5/2}\delta^5) \end{pmatrix}$$

$$-\sqrt{\frac{\pi}{16\theta}}\begin{pmatrix} erf[\sqrt{\theta}(1+x_t)] \\ +erf[\sqrt{\theta}(1-x_t)] \\ \textcolor{red}{+\sqrt{\frac{4}{\pi}}\begin{bmatrix} e^{-\theta(1+x_t)^2} \\ -e^{-\theta(1-x_t)^2} \end{bmatrix}\sqrt{\theta}\delta} \\ -\sqrt{\frac{4\theta}{\pi}}\begin{bmatrix} (1+x_t)e^{-\theta(1+x_t)^2} \\ +(1-x_t)e^{-\theta(1-x_t)^2} \end{bmatrix}\theta\delta^2 \\ \textcolor{red}{-\sqrt{\frac{4}{9\pi}}\left\{\begin{array}{l}[1-2\theta(1+x_t)^2]e^{-\theta(1+x_t)^2} \\ -[1-2\theta(1-x_t)^2]e^{-\theta(1-x_t)^2}\end{array}\right\}\theta^{3/2}\delta^3} \\ +\sqrt{\frac{\theta}{\pi}}\left\{\begin{array}{l}\left[(1+x_t)-\frac{2\theta}{3}(1+x_t)^3\right]e^{-\theta(1+x_t)^2} \\ +\left[(1-x_t)-\frac{2\theta}{3}(1-x_t)^3\right]e^{-\theta(1-x_t)^2}\end{array}\right\}\theta^2\delta^4 \\ +O(\theta^{5/2}\delta^5) \end{pmatrix}$$

$$-\sqrt{\frac{\pi}{16\theta}}\begin{pmatrix} erf[\sqrt{\theta}(1+x_t)] \\ +erf[\sqrt{\theta}(1-x_t)] \\ \textcolor{red}{-\sqrt{\frac{4}{\pi}}\begin{bmatrix} e^{-\theta(1+x_t)^2} \\ -e^{-\theta(1-x_t)^2} \end{bmatrix}\sqrt{\theta}\delta} \\ -\sqrt{\frac{4\theta}{\pi}}\begin{bmatrix} (1+x_t)e^{-\theta(1+x_t)^2} \\ +(1-x_t)e^{-\theta(1-x_t)^2} \end{bmatrix}\theta\delta^2 \\ \textcolor{red}{+\sqrt{\frac{4}{9\pi}}\left\{\begin{array}{l}[1-2\theta(1+x_t)^2]e^{-\theta(1+x_t)^2} \\ -[1-2\theta(1-x_t)^2]e^{-\theta(1-x_t)^2}\end{array}\right\}\theta^{3/2}\delta^3} \\ +\sqrt{\frac{\theta}{\pi}}\left\{\begin{array}{l}\left[(1+x_t)-\frac{2\theta}{3}(1+x_t)^3\right]e^{-\theta(1+x_t)^2} \\ +\left[(1-x_t)-\frac{2\theta}{3}(1-x_t)^3\right]e^{-\theta(1-x_t)^2}\end{array}\right\}\theta^2\delta^4 \\ +O(\theta^{5/2}\delta^5) \end{pmatrix}$$

$$+2\left(\frac{1}{8\theta\delta^2}+\frac{1}{4}+\frac{\theta\delta^2}{6}\right)\sqrt{\frac{\pi}{32\theta}}\left\{\begin{array}{l}erf[\sqrt{2\theta}(1+x_t)] \\ +erf[\sqrt{2\theta}(1-x_t)]\end{array}\right\}[\textcolor{green}{\mathbf{1}}-2\theta\delta^2+2\theta^2\delta^4+O(\theta^3\delta^6)].$$

Canceling the red terms shows that odd powers of $\sqrt{\theta}\delta$ vanish. Then, after cancellation of the green terms, including one in the final square bracket of the ultimate equation, and keeping only terms through power $\theta\delta^2$, the following remains:

$IMSPE = 2 - 2\theta\delta^2$

$$-2\left(\frac{1}{8\theta\delta^2}+\frac{1}{4}\right)\sqrt{\frac{\pi}{32\theta}}\begin{pmatrix} -\sqrt{\frac{8\theta}{\pi}}\begin{bmatrix}(1+x_t)e^{-2\theta(1+x_t)^2} \\ +(1-x_t)e^{-2\theta(1-x_t)^2}\end{bmatrix}2\theta\delta^2 \\ +\sqrt{\frac{2\theta}{\pi}}\left\{\begin{array}{l}\left[(1+x_t)-\frac{4\theta}{3}(1+x_t)^3\right]e^{-2\theta(1+x_t)^2} \\ +\left[(1-x_t)-\frac{4\theta}{3}(1-x_t)^3\right]e^{-2\theta(1-x_t)^2}\end{array}\right\}4\theta^2\delta^4 \\ +O(\theta^{5/2}\delta^5) \end{pmatrix}$$

$$-2\sqrt{\frac{\pi}{16\theta}}\left\{\begin{array}{c}erf[\sqrt{\theta}(1+x_t)]\\+erf[\sqrt{\theta}(1-x_t)]\\-\sqrt{\frac{4\theta}{\pi}}\left[\begin{array}{c}(1+x_t)e^{-\theta(1+x_t)^2}\\+(1-x_t)e^{-\theta(1-x_t)^2}\end{array}\right]\theta\delta^2\end{array}\right\}$$

$$+2\left(\frac{1}{8\theta\delta^2}+\frac{1}{4}\right)\sqrt{\frac{\pi}{32\theta}}\left\{\begin{array}{c}erf[\sqrt{2\theta}(1+x_t)]\\+erf[\sqrt{2\theta}(1-x_t)]\end{array}\right\}[-2\theta\delta^2+2\theta^2\delta^4+O(\theta^3\delta^6)].$$

After noting most of the $\frac{\pi}{\theta}$'s cancel, and collecting powers of two, we have the following:

$$IMSPE = 2 - 2\theta\delta^2 - 2\left(\frac{1}{8\theta\delta^2}+\frac{1}{4}\right)\left(\begin{array}{c}-\left[\begin{array}{c}(1+x_t)e^{-2\theta(1+x_t)^2}\\+(1-x_t)e^{-2\theta(1-x_t)^2}\end{array}\right]\theta\delta^2\\+\left\{\begin{array}{c}\left[(1+x_t)-\frac{4\theta}{3}(1+x_t)^3\right]e^{-2\theta(1+x_t)^2}\\+\left[(1-x_t)-\frac{4\theta}{3}(1-x_t)^3\right]e^{-2\theta(1-x_t)^2}\end{array}\right\}\theta^2\delta^4\end{array}\right)$$

$$-\sqrt{\frac{\pi}{4\theta}}\left\{\begin{array}{c}erf[\sqrt{\theta}(1+x_t)]\\+erf[\sqrt{\theta}(1-x_t)]\end{array}\right\}$$

$$+\left[\begin{array}{c}(1+x_t)e^{-\theta(1+x_t)^2}\\+(1-x_t)e^{-\theta(1-x_t)^2}\end{array}\right]\theta\delta^2$$

$$-\left(\frac{1}{8\theta\delta^2}+\frac{1}{4}\right)\sqrt{\frac{\pi}{2\theta}}\left\{\begin{array}{c}erf[\sqrt{2\theta}(1+x_t)]\\+erf[\sqrt{2\theta}(1-x_t)]\end{array}\right\}[\theta\delta^2-\theta^2\delta^4+O(\theta^3\delta^6)].$$

Finally, collecting terms with common degree in $\theta\delta^2$ gives the following expansion, which has no terms with negative or odd powers of $\sqrt{\theta}\delta$:

$$IMSPE = 2 + \frac{1}{4}\left[\begin{array}{c}(1+x_t)e^{-2\theta(1+x_t)^2}\\+(1-x_t)e^{-2\theta(1-x_t)^2}\end{array}\right] - \sqrt{\frac{\pi}{4\theta}}\left\{\begin{array}{c}erf[\sqrt{\theta}(1+x_t)]\\+erf[\sqrt{\theta}(1-x_t)]\end{array}\right\} - \sqrt{\frac{\pi}{128\theta}}\left\{\begin{array}{c}erf[\sqrt{2\theta}(1+x_t)]\\+erf[\sqrt{2\theta}(1-x_t)]\end{array}\right\}$$

$$+\begin{pmatrix} -\frac{1}{4}\left\{\begin{array}{l}\left[(1+x_t)-\frac{4\theta}{3}(1+x_t)^3\right]e^{-2\theta(1+x_t)^2} \\ +\left[(1-x_t)-\frac{4\theta}{3}(1-x_t)^3\right]e^{-2\theta(1-x_t)^2}\end{array}\right\}^{-2} \\ +\frac{1}{2}\left[\begin{array}{l}(1+x_t)e^{-2\theta(1+x_t)^2} \\ +(1-x_t)e^{-2\theta(1-x_t)^2}\end{array}\right] \\ +\left[\begin{array}{l}(1+x_t)e^{-\theta(1+x_t)^2} \\ +(1-x_t)e^{-\theta(1-x_t)^2}\end{array}\right] \\ +\sqrt{\frac{\pi}{128\theta}}\left\{\begin{array}{l}erf\left[\sqrt{2\theta}(1+x_t)\right] \\ +erf\left[\sqrt{2\theta}(1-x_t)\right]\end{array}\right\} \\ -\sqrt{\frac{\pi}{32\theta}}\left\{\begin{array}{l}erf\left[\sqrt{2\theta}(1+x_t)\right] \\ +erf\left[\sqrt{2\theta}(1-x_t)\right]\end{array}\right\} \end{pmatrix}\theta\delta^2 + O(\theta^2\delta^4).$$

$$IMSPE = \begin{pmatrix} 2 \\ +\frac{1}{4}\left[\begin{array}{l}(1+x_t)e^{-2\theta(1+x_t)^2} \\ +(1-x_t)e^{-2\theta(1-x_t)^2}\end{array}\right] \\ -\sqrt{\frac{\pi}{4\theta}}\left\{\begin{array}{l}erf\left[\sqrt{\theta}(1+x_t)\right] \\ +erf\left[\sqrt{\theta}(1-x_t)\right]\end{array}\right\} \\ -\sqrt{\frac{\pi}{128\theta}}\left\{\begin{array}{l}erf\left[\sqrt{2\theta}(1+x_t)\right] \\ +erf\left[\sqrt{2\theta}(1-x_t)\right]\end{array}\right\} \end{pmatrix} + \begin{pmatrix} -\frac{1}{4}\left\{\begin{array}{l}\left[(1+x_t)-\frac{4\theta}{3}(1+x_t)^3\right]e^{-2\theta(1+x_t)^2} \\ +\left[(1-x_t)-\frac{4\theta}{3}(1-x_t)^3\right]e^{-2\theta(1-x_t)^2}\end{array}\right\}^{-2} \\ +\frac{1}{2}\left[\begin{array}{l}(1+x_t)e^{-2\theta(1+x_t)^2} \\ +(1-x_t)e^{-2\theta(1-x_t)^2}\end{array}\right] \\ +\left[\begin{array}{l}(1+x_t)e^{-\theta(1+x_t)^2} \\ +(1-x_t)e^{-\theta(1-x_t)^2}\end{array}\right] \\ -\sqrt{\frac{\pi}{128\theta}}\left\{\begin{array}{l}erf\left[\sqrt{2\theta}(1+x_t)\right] \\ +erf\left[\sqrt{2\theta}(1-x_t)\right]\end{array}\right\} \end{pmatrix}\theta\delta^2 + O(\theta^2\delta^4).$$

Accumulating some terms in the second large parenthesis gives the following:

$$IMSPE = \begin{pmatrix} 2 \\ +\frac{1}{4}\left[\begin{array}{l}(1+x_t)e^{-2\theta(1+x_t)^2} \\ +(1-x_t)e^{-2\theta(1-x_t)^2}\end{array}\right] \\ -\sqrt{\frac{\pi}{4\theta}}\left\{\begin{array}{l}erf\left[\sqrt{\theta}(1+x_t)\right] \\ +erf\left[\sqrt{\theta}(1-x_t)\right]\end{array}\right\} \\ -\sqrt{\frac{\pi}{128\theta}}\left\{\begin{array}{l}erf\left[\sqrt{2\theta}(1+x_t)\right] \\ +erf\left[\sqrt{2\theta}(1-x_t)\right]\end{array}\right\} \end{pmatrix} + \begin{pmatrix} +\frac{\theta}{3}\left\{\begin{array}{l}\left[(1+x_t)^3\right]e^{-2\theta(1+x_t)^2} \\ +\left[(1-x_t)^3\right]e^{-2\theta(1-x_t)^2}\end{array}\right\}^{-2} \\ +\frac{1}{4}\left[\begin{array}{l}(1+x_t)e^{-2\theta(1+x_t)^2} \\ +(1-x_t)e^{-2\theta(1-x_t)^2}\end{array}\right] \\ +\left[\begin{array}{l}(1+x_t)e^{-\theta(1+x_t)^2} \\ +(1-x_t)e^{-\theta(1-x_t)^2}\end{array}\right] \\ -\sqrt{\frac{\pi}{128\theta}}\left\{\begin{array}{l}erf\left[\sqrt{2\theta}(1+x_t)\right] \\ +erf\left[\sqrt{2\theta}(1-x_t)\right]\end{array}\right\} \end{pmatrix}\theta\delta^2 + O(\theta^2\delta^4).$$

Identifying a common factor $(1+x_t)$ gives the following:

$$IMSPE = \begin{pmatrix} +\frac{1}{4}\begin{bmatrix} (1+x_t)e^{-2\theta(1+x_t)^2} \\ +(1-x_t)e^{-2\theta(1-x_t)^2} \end{bmatrix}^2 \\ -\sqrt{\frac{\pi}{4\theta}}\begin{Bmatrix} erf[\sqrt{\theta}\,(1+x_t)] \\ +erf[\sqrt{\theta}\,(1-x_t)] \end{Bmatrix} \\ -\sqrt{\frac{\pi}{128\theta}}\begin{Bmatrix} erf[\sqrt{2\theta}(1+x_t)] \\ +erf[\sqrt{2\theta}(1-x_t)] \end{Bmatrix} \end{pmatrix} + \begin{pmatrix} +\left[\frac{1}{4}+\frac{\theta}{3}(1+x_t)^2\right](1+x_t)e^{-2\theta(1+x_t)^2} \\ +\left[\frac{1}{4}+\frac{\theta}{3}(1-x_t)^2\right](1-x_t)e^{-2\theta(1-x_t)^2} \\ +\begin{bmatrix} (1+x_t)e^{-\theta(1+x_t)^2} \\ +(1-x_t)e^{-\theta(1-x_t)^2} \end{bmatrix} \\ -\sqrt{\frac{\pi}{128\theta}}\begin{Bmatrix} erf[\sqrt{2\theta}(1+x_t)] \\ +erf[\sqrt{2\theta}(1-x_t)] \end{Bmatrix} \end{pmatrix}^{-2} \theta\delta^2 + O(\theta^2\delta^4).$$

(R. 3.2)

**R.4 Expansions in powers of $\sqrt{\theta}\delta$, using SMS**

$IMSPE * \theta^{3/2} =$

$$\begin{aligned}
&-\frac{1}{16}\sqrt{2}\,\text{erf}\!\left(\sqrt{2}\,\sqrt{\theta}\,(1+xt)\right)\sqrt{\pi}\,\theta + \frac{1}{16}\sqrt{2}\,\text{erf}\!\left(\sqrt{2}\,\sqrt{\theta}\,(-1+xt)\right)\sqrt{\pi}\,\theta \\
&-\frac{1}{2}\theta\sqrt{\pi}\,\text{erf}\!\left(\sqrt{\theta}\,(1+xt)\right) + \frac{1}{2}\theta\sqrt{\pi}\,\text{erf}\!\left(\sqrt{\theta}\,(-1+xt)\right) \\
&+\frac{1}{4}e^{-2\theta(1+xt)^2}\theta^{3/2}\,xt - \frac{1}{4}e^{-2\theta(-1+xt)^2}\theta^{3/2}\,xt + \frac{1}{4}e^{-2\theta(1+xt)^2}\theta^{3/2} \\
&+\frac{1}{4}e^{-2\theta(-1+xt)^2}\theta^{3/2} + 2\theta^{3/2} + \left(e^{-\theta(1+xt)^2}\theta^{5/2}\,xt - e^{-\theta(-1+xt)^2}\theta^{5/2}\,xt\right. \\
&+\frac{1}{4}e^{-2\theta(1+xt)^2}\theta^{5/2}\,xt - \frac{1}{4}e^{-2\theta(-1+xt)^2}\theta^{5/2}\,xt + \frac{1}{3}e^{-2\theta(1+xt)^2}\theta^{7/2}\,xt^3 \\
&-\frac{1}{3}e^{-2\theta(-1+xt)^2}\theta^{7/2}\,xt^3 + e^{-2\theta(1+xt)^2}\theta^{7/2}\,xt^2 + e^{-2\theta(-1+xt)^2}\theta^{7/2}\,xt^2 \\
&+e^{-2\theta(1+xt)^2}\theta^{7/2}\,xt - e^{-2\theta(-1+xt)^2}\theta^{7/2}\,xt - \frac{1}{16}\sqrt{2}\,\text{erf}\!\left(\sqrt{2}\,\sqrt{\theta}\,(1+xt)\right)\sqrt{\pi}\,\theta^2 \\
&+\frac{1}{16}\sqrt{2}\,\text{erf}\!\left(\sqrt{2}\,\sqrt{\theta}\,(-1+xt)\right)\sqrt{\pi}\,\theta^2 + \frac{1}{3}e^{-2\theta(1+xt)^2}\theta^{7/2} \\
&+\frac{1}{3}e^{-2\theta(-1+xt)^2}\theta^{7/2} + \frac{1}{4}e^{-2\theta(1+xt)^2}\theta^{5/2} + \frac{1}{4}e^{-2\theta(-1+xt)^2}\theta^{5/2} \\
&+e^{-\theta(1+xt)^2}\theta^{5/2} + e^{-\theta(-1+xt)^2}\theta^{5/2} - 2\theta^{5/2}\right)\delta^2 + O(\delta^4)
\end{aligned}$$

(R.4.1)

A slight clipping of $\theta^2$, on the right-hand side of the eighth row of the ultimate equation, is of no concern. Typeset, $IMSPE * \theta^{3/2}$ is the following:

$$\text{IMSPE} * \theta^{3/2} = \begin{pmatrix} -\sqrt{\dfrac{\pi}{128}}\theta \begin{cases} \text{erf}[\sqrt{2\theta}(1+x_t)] \\ -\text{erf}[\sqrt{2\theta}(1-x_t)] \end{cases} \\ -\sqrt{\dfrac{\pi}{4}}\theta \begin{cases} \text{erf}[\sqrt{\theta}(1+x_t)] \\ -\text{erf}[\sqrt{\theta}(1-x_t)] \end{cases} \\ +\dfrac{1}{4}\theta^{3/2} x_t \begin{bmatrix} e^{-2\theta(1+x_t)^2} \\ -e^{-2\theta(1-x_t)^2} \end{bmatrix} \\ +\dfrac{1}{4}\theta^{3/2} \begin{bmatrix} e^{-2\theta(1+x_t)^2} \\ +e^{-2\theta(1-x_t)^2} \end{bmatrix} \\ +2\theta^{3/2} \end{pmatrix} + \begin{pmatrix} \theta^{5/2} x_t \begin{bmatrix} e^{-\theta(1+x_t)^2} \\ -e^{-\theta(1-x_t)^2} \end{bmatrix} \\ +\dfrac{1}{4}\theta^{5/2} x_t \begin{bmatrix} e^{-2\theta(1+x_t)^2} \\ -e^{-2\theta(1-x_t)^2} \end{bmatrix} \\ +\dfrac{1}{3}\theta^{7/2} x_t^3 \begin{bmatrix} e^{-2\theta(1+x_t)^2} \\ -e^{-2\theta(1-x_t)^2} \end{bmatrix} \\ +\theta^{7/2} x_t^2 \begin{bmatrix} e^{-2\theta(1-x_t)^2} \\ +e^{-2\theta(1-x_t)^2} \end{bmatrix} \\ +\theta^{7/2} x_t \begin{bmatrix} e^{-\theta(1-x_t)^2} \\ +e^{-\theta(1-x_t)^2} \end{bmatrix} \\ -\sqrt{\dfrac{\pi}{128}}\theta^2 \begin{cases} \text{erf}[\sqrt{2\theta}(1+x_t)] \\ +\text{erf}[\sqrt{2\theta}(1-x_t)] \end{cases} \\ +\dfrac{1}{3}\theta^{7/2} \begin{bmatrix} e^{-2\theta(1+x_t)^2} \\ +e^{-2\theta(1-x_t)^2} \end{bmatrix} \\ +\dfrac{1}{4}\theta^{5/2} \begin{bmatrix} e^{-2\theta(1+x_t)^2} \\ +e^{-2\theta(1-x_t)^2} \end{bmatrix} \\ +\theta^{5/2} \begin{bmatrix} e^{-\theta(1-x_t)^2} \\ +e^{-\theta(1-x_t)^2} \end{bmatrix} \\ -2\theta^{5/2} \end{pmatrix} \delta^2 + O(\theta^2 \delta^4).$$

Collecting terms, row-by-row, and dividing by $\theta^{3/2}$ gives the following:

$$\text{IMSPE} = \begin{pmatrix} -\sqrt{\dfrac{\pi}{128\theta}} \begin{cases} \text{erf}[\sqrt{2\theta}(1+x_t)] \\ -\text{erf}[\sqrt{2\theta}(-1+x_t)] \end{cases} \\ -\sqrt{\dfrac{\pi}{4\theta}} \begin{cases} \text{erf}[\{\sqrt{\theta}(1+x_t)\}] \\ -\text{erf}[\sqrt{\theta}(-1+x_t)] \end{cases} \\ +\dfrac{1}{4}\begin{bmatrix} (1+x_t)e^{-2\theta(1+x_t)^2} \\ -(1-x_t)e^{-2\theta(-1+x_t)^2} \end{bmatrix} \\ 2 \end{pmatrix} + \begin{pmatrix} x_t \begin{bmatrix} e^{-\theta(1+x_t)^2} \\ -e^{-\theta(-1+x_t)^2} \end{bmatrix} \\ +\dfrac{1}{4} x_t \begin{bmatrix} e^{-2\theta(1+x_t)^2} \\ -e^{-2\theta(-1+x_t)^2} \end{bmatrix} \\ +\dfrac{\theta}{3} x_t^3 \begin{bmatrix} e^{-2\theta(1+x_t)^2} \\ -e^{-2\theta(-1+x_t)^2} \end{bmatrix} \\ +\theta x_t^2 \begin{bmatrix} e^{-2\theta(1-x_t)^2} \\ +e^{-2\theta(-1+x_t)^2} \end{bmatrix} \\ +\theta x_t \begin{bmatrix} e^{-\theta(1-x_t)^2} \\ +e^{-\theta(-1+x_t)^2} \end{bmatrix} \\ -\sqrt{\dfrac{\pi}{128\theta}} \begin{cases} \text{erf}[\sqrt{2\theta}(1+x_t)] \\ -\text{erf}[\sqrt{2\theta}(-1+x_t)] \end{cases} \\ +\dfrac{\theta}{3} \begin{bmatrix} e^{-2\theta(1+x_t)^2} \\ +e^{-2\theta(-1+x_t)^2} \end{bmatrix} \\ +\dfrac{1}{4} \begin{bmatrix} e^{-2\theta(1+x_t)^2} \\ +e^{-2\theta(-1+x_t)^2} \end{bmatrix} \\ +\begin{bmatrix} e^{-\theta(1-x_t)^2} \\ +e^{-\theta(-1+x_t)^2} \end{bmatrix} \\ -2 \end{pmatrix} \theta \delta^2 + O(\theta^2 \delta^4).$$

Rearranging rows gives the following:

$$IMSPE = \begin{pmatrix} 2 \\ +\frac{1}{4}\begin{bmatrix} (1+x_t)e^{-2\theta(1+x_t)^2} \\ +(1-x_t)e^{-2\theta(-1+x_t)^2} \end{bmatrix} \\ -\sqrt{\frac{\pi}{4\theta}} \begin{Bmatrix} erf[\sqrt{\theta}(1+x_t)] \\ +erf[\sqrt{\theta}(1-x_t)] \end{Bmatrix} \\ -\sqrt{\frac{\pi}{128\theta}} \begin{Bmatrix} erf[\sqrt{2\theta}(1+x_t)] \\ +erf[\sqrt{2\theta}(1-x_t)] \end{Bmatrix} \end{pmatrix} + \begin{pmatrix} -2 \\ \begin{bmatrix} (1+x_t)e^{-\theta(1+x_t)^2} \\ +(1-x_t)e^{-\theta(-1+x_t)^2} \end{bmatrix} \\ +\frac{1}{4}\begin{bmatrix} (1+x_t)e^{-2\theta(1+x_t)^2} \\ +(1-x_t)e^{-2\theta(-1+x_t)^2} \end{bmatrix} \\ +\frac{\theta}{3}\begin{bmatrix} (1+x_t)^3 e^{-2\theta(1+x_t)^2} \\ +(1-x_t)^3 e^{-2\theta(-1+x_t)^2} \end{bmatrix} \\ -\sqrt{\frac{\pi}{128\theta}} \begin{Bmatrix} erf[\sqrt{2\theta}(1+x_t)] \\ -erf[\sqrt{2\theta}(-1+x_t)] \end{Bmatrix} \end{pmatrix} \theta\delta^2 + O(\theta^2\delta^4).$$

Further rearrangement gives the following, which is identical to the hand algebra in Eq. R.3.2:

$$IMSPE = \begin{pmatrix} 2 \\ +\frac{1}{4}\begin{bmatrix} (1+x_t)e^{-2\theta(1+x_t)^2} \\ +(1-x_t)e^{-2\theta(-1+x_t)^2} \end{bmatrix} \\ -\sqrt{\frac{\pi}{4\theta}} \begin{Bmatrix} erf[\sqrt{\theta}(1+x_t)] \\ +erf[\sqrt{\theta}(1-x_t)] \end{Bmatrix} \\ -\sqrt{\frac{\pi}{128\theta}} \begin{Bmatrix} erf[\sqrt{2\theta}(1+x_t)] \\ +erf[\sqrt{2\theta}(1-x_t)] \end{Bmatrix} \end{pmatrix} + \begin{pmatrix} -2 \\ +\left[\frac{1}{4}+\frac{\theta}{3}(1+x_t)^2\right](1+x_t)e^{-2\theta(1+x_t)^2} \\ +\left[\frac{1}{4}+\frac{\theta}{3}(1-x_t)^2\right](1-x_t)e^{-2\theta(1-x_t)^2} \\ +\begin{bmatrix} (1+x_t)e^{-\theta(1+x_t)^2} \\ +(1-x_t)e^{-\theta(1-x_t)^2} \end{bmatrix} \\ -\sqrt{\frac{\pi}{128\theta}} \begin{Bmatrix} erf[\sqrt{2\theta}(1+x_t)] \\ +erf[\sqrt{2\theta}(1-x_t)] \end{Bmatrix} \end{pmatrix} \theta\delta^2 + O(\theta^2\delta^4).$$

(R. 4.2)

### R.5 Expansions in powers of $\sqrt{\theta}\delta$, using symmetry operators

A new alternative approach follows, where operators $S_w$, $D_w$, and $Z_w$ evaluate the objects to their right sides in products, by changing the sign of $w$, doubling $w$, or setting $w$ to zero, respectively. Also, we define the coefficients in the power-series-in-$\delta^2$ expansion of $R_{0,1}$, via $R_{0,1} = R_{0,1}^{(0)} + R_{0,1}^{(2)}\theta\delta^2 + R_{0,1}^{(4)}\theta^2\delta^4 + O(\delta^6)$, where $R_{0,1}^{(0)}$, $R_{0,1}^{(2)}$, $\cdots$ are independent of $\delta$ but may depend on $\theta$. We also simplify the notation further by using the generic symbol $\mathcal{R}$ for the $R_{0,1}$'s, via the following definitions $\mathcal{R} \equiv R_{0,1}$, $\mathcal{R}^{(0)} \equiv R_{0,1}^{(0)}$, $\mathcal{R}^{(2)} \equiv R_{0,1}^{(2)}$, $\cdots$. We have

$$L = \begin{bmatrix} 0 & | & 1 & 1 \\ -- & | & -- & -- \\ \cdot & | & 1 & V_{1,2} \\ \cdot & | & \cdot & 1 \end{bmatrix},$$

$$L^{-1} = \begin{bmatrix} -1+\frac{1-V_{1,2}}{2} & | & \frac{1}{2} & \frac{1}{2} \\ ----- & | & -- & -- \\ \cdot & | & \frac{1}{2(1-V_{1,2})} & \frac{-1}{2(1-V_{1,2})} \\ \cdot & | & \cdot & \frac{1}{2(1-V_{1,2})} \end{bmatrix},$$

$$R = \begin{bmatrix} 1 & | & \mathcal{R} & S_\delta \mathcal{R} \\ -- & | & --- & ---------- \\ \cdot & | & D_\theta \mathcal{R} & e^{-2\theta\delta^2} Z_\delta D_\theta \mathcal{R} \\ \cdot & | & \cdot & S_\delta D_\theta \mathcal{R} \end{bmatrix}, \text{ and}$$

$$IMSPE = 1 - tr(L^{-1}R) = 2 - \frac{1-V_{1,2}}{2} - \frac{1}{2(1-V_{1,2})}(1+S_\delta)D_\theta\mathcal{R} - (1+S_\delta)\mathcal{R} + \frac{e^{-2\theta\delta^2}}{1-V_{1,2}} Z_\delta D_\theta \mathcal{R}. \quad (R.5.1)$$

Note that the power-series-in-$\sqrt{\theta}\delta$ expansions of $V_{1,2}$ and $\mathcal{R}$ include only non-negative powers, the factor $(1 + S_\delta)$ annihilates odd powers and doubles even powers, and $Z_\delta$ annihilates any non-zero powers, so the expansion of $IMSPE$ includes only non-negative even powers, giving the following, after plugging in the definitions $V_{1,2} = e^{-4\theta\delta^2}$ and $\mathcal{R} = \mathcal{R}^{(0)} + \mathcal{R}^{(2)}\theta\delta^2 + \mathcal{R}^{(4)}\theta^2\delta^4$, noting $Z_\delta D_\theta \mathcal{R} = D_\theta \mathcal{R}^{(0)}$ and $\theta D_\theta = D_\theta \theta/2,$ and rearranging terms:

$$IMSPE = 2 - \frac{1-e^{-4\theta\delta^2}}{2} - 2\begin{pmatrix} \mathcal{R}^{(0)} \\ +\mathcal{R}^{(2)}\theta\delta^2 \end{pmatrix} - \frac{1}{1-e^{-4\theta\delta^2}} D_\theta \begin{pmatrix} \mathcal{R}^{(0)} \\ +\mathcal{R}^{(2)}\theta\delta^2 \\ +\mathcal{R}^{(4)}\theta^2\delta^4 \end{pmatrix} + \frac{e^{-2\theta\delta^2}}{1-e^{-4\theta\delta^2}} D_\theta \mathcal{R}^{(0)} + O(\theta^2\delta^4)$$

$$= 2 - \frac{1-e^{-4\theta\delta^2}}{2} - 2\begin{pmatrix} \mathcal{R}^{(0)} \\ +\mathcal{R}^{(2)}\theta\delta^2 \end{pmatrix} - \frac{1-e^{-2\theta\delta^2}}{1-e^{-4\theta\delta^2}} D_\theta \mathcal{R}^{(0)} - \frac{1}{1-e^{-4\theta\delta^2}} D_\theta \begin{pmatrix} \mathcal{R}^{(2)}\theta\delta^2 \\ +\mathcal{R}^{(4)}\theta^2\delta^4 \end{pmatrix} + O(\theta^2\delta^4)$$

$$= 2 - \frac{1-e^{-4\theta\delta^2}}{2} - 2\begin{pmatrix} \mathcal{R}^{(0)} \\ +\mathcal{R}^{(2)}\theta\delta^2 \end{pmatrix} - \frac{1}{1+e^{-2\theta\delta^2}} D_\theta \mathcal{R}^{(0)} - \frac{1}{1-e^{-4\theta\delta^2}} D_\theta \begin{pmatrix} \mathcal{R}^{(2)}\theta\delta^2 \\ +\mathcal{R}^{(4)}\theta^2\delta^4 \end{pmatrix} + O(\theta^2\delta^4)$$

$$= 2 - 2\theta\delta^2 - 2\begin{pmatrix} \mathcal{R}^{(0)} \\ +\mathcal{R}^{(2)}\theta\delta^2 \end{pmatrix} - \frac{1}{2(1-\theta\delta^2)} D_\theta \mathcal{R}^{(0)} - \frac{1}{4\theta\delta^2(1-2\theta\delta^2)} D_\theta \begin{pmatrix} \mathcal{R}^{(2)}\theta\delta^2 \\ +\mathcal{R}^{(4)}\theta^2\delta^4 \end{pmatrix} + O(\theta^2\delta^4)$$

$$= 2 - 2\theta\delta^2 - 2\begin{pmatrix} \mathcal{R}^{(0)} \\ +\mathcal{R}^{(2)}\theta\delta^2 \end{pmatrix} - \frac{1+\theta\delta^2}{2} D_\theta \mathcal{R}^{(0)} - \frac{1+2\theta\delta^2}{4} D_\theta \begin{pmatrix} 2\mathcal{R}^{(2)} \\ +2\mathcal{R}^{(4)}\theta\delta^2 \end{pmatrix} + O(\theta^2\delta^4)$$

$$= 2 - 2\mathcal{R}^{(0)} - \frac{D_\theta \mathcal{R}^{(0)}}{2} - \frac{D_\theta \mathcal{R}^{(2)}}{2} + \left(-2 - 2\mathcal{R}^{(2)} - \frac{D_\theta \mathcal{R}^{(0)}}{4} - \frac{D_\theta \mathcal{R}^{(2)}}{2} - \frac{D_\theta \mathcal{R}^{(4)}}{2}\right)\theta\delta^2 + O(\theta^2\delta^4)$$

$$= \left[2 - 2\left(1 + \frac{D_\theta}{4}\right)\mathcal{R}^{(0)} - \frac{D_\theta}{2}\mathcal{R}^{(2)}\right] + \left[-2 - \frac{D_\theta}{4}\mathcal{R}^{(0)} - 2\left(1 + \frac{D_\theta}{4}\right)\mathcal{R}^{(2)} - \frac{D_\theta \mathcal{R}^{(4)}}{2}\right]\theta\delta^2 + O(\theta^2\delta^4).$$

(R.5.2)

From Eq. R.3.1 we have the following for $\mathcal{R}^{(0)}$, $\mathcal{R}^{(2)}$, and $\mathcal{R}^{(4)}$:

$$\mathcal{R}^{(0)} = \sqrt{\frac{\pi}{16\theta}} \left\{ \begin{array}{c} erf[\sqrt{\theta}(1+x_t)] \\ +erf[\sqrt{\theta}(1-x_t)] \end{array} \right\} = \sqrt{\frac{\pi}{16\theta}} (1 + S_{x_t}) erf[\sqrt{\theta}(1+x_t)],$$

$$\mathcal{R}^{(2)} = -\sqrt{\frac{\pi}{16\theta}} \sqrt{\frac{4\theta}{\pi}} \left[ \begin{array}{c} (1+x_t)e^{-\theta(1+x_t)^2} \\ +(1-x_t)e^{-\theta(1-x_t)^2} \end{array} \right] \sqrt{\theta} = -\sqrt{\frac{\pi}{16\theta}} \sqrt{\frac{4\theta}{\pi}} (1 + S_{x_t})(1+x_t)e^{-\theta(1+x_t)^2}, \text{ and}$$

$$\mathcal{R}^{(4)} = \sqrt{\frac{\pi}{16\theta}} \sqrt{\frac{\theta}{\pi}} \left\{ \begin{array}{c} \left[(1+x_t) - \frac{2\theta}{3}(1+x_t)^3\right] e^{-\theta(1+x_t)^2} \\ +\left[(1-x_t) - \frac{2\theta}{3}(1-x_t)^3\right] e^{-\theta(1-x_t)^2} \end{array} \right\}$$

$$= \sqrt{\frac{\pi}{16\theta}} \sqrt{\frac{\theta}{\pi}} (1 + S_{x_t}) \left[ (1 + x_t) - \frac{2\theta}{3} (1 + x_t)^3 \right] e^{-\theta(1+x_t)^2}.$$

Substituting these into Eq. R.5.2 gives the following:

$$IMSPE = \begin{pmatrix} -2\left(1 + \frac{D_\theta}{4}\right) \sqrt{\frac{\pi}{16\theta}} \begin{Bmatrix} erf[\sqrt{\theta}(1 + x_t)] \\ +erf[\sqrt{\theta}(1 - x_t)] \end{Bmatrix} \\ +\frac{D_\theta}{2} \sqrt{\frac{\pi}{16\theta}} \begin{Bmatrix} \sqrt{\frac{4\theta}{\pi}} \begin{bmatrix} (1 + x_t) e^{-\theta(1+x_t)^2} \\ +(1 - x_t) e^{-\theta(1-x_t)^2} \end{bmatrix} \end{Bmatrix} \end{pmatrix}^2$$

$$+ \begin{pmatrix} -\frac{D_\theta}{4} \sqrt{\frac{\pi}{16\theta}} \begin{Bmatrix} erf[\sqrt{\theta}(1 + x_t)] \\ +erf[\sqrt{\theta}(1 - x_t)] \end{Bmatrix} \\ +2\left(1 + \frac{D_\theta}{4}\right) \sqrt{\frac{\pi}{16\theta}} \sqrt{\frac{4\theta}{\pi}} \begin{bmatrix} (1 + x_t) e^{-\theta(1+x_t)^2} \\ +(1 - x_t) e^{-\theta(1-x_t)^2} \end{bmatrix} \\ -\frac{D_\theta}{2} \sqrt{\frac{\pi}{16\theta}} \sqrt{\frac{\theta}{\pi}} \begin{Bmatrix} \left[ (1 + x_t) - \frac{2\theta}{3}(1 + x_t)^3 \right] e^{-\theta(1+x_t)^2} \\ + \left[ (1 - x_t) - \frac{2\theta}{3}(1 - x_t)^3 \right] e^{-\theta(1-x_t)^2} \end{Bmatrix} \end{pmatrix}^{-2} \theta \delta^2 + O(\theta^2 \delta^4). \quad (R.5.3)$$

$$= \left\{ -2\left(1 + \frac{D_\theta}{4}\right) \sqrt{\frac{\pi}{16\theta}} (1 + S_{x_t}) erf[\sqrt{\theta}(1 + x_t)] \atop +\frac{D_\theta}{2} \sqrt{\frac{\pi}{16\theta}} \sqrt{\frac{4\theta}{\pi}} (1 + S_{x_t})(1 + x_t) e^{-\theta(1+x_t)^2} \right\}^2$$

$$+ \left\{ \begin{array}{c} -\frac{D_\theta}{4} \sqrt{\frac{\pi}{16\theta}} (1 + S_{x_t}) erf[\sqrt{\theta}(1 + x_t)] \\ +2\left(1 + \frac{D_\theta}{4}\right) \sqrt{\frac{\pi}{16\theta}} \sqrt{\frac{4\theta}{\pi}} (1 + S_{x_t})(1 + x_t) e^{-\theta(1+x_t)^2} \\ -\frac{D_\theta}{2} \sqrt{\frac{\pi}{16\theta}} \sqrt{\frac{\theta}{\pi}} (1 + S_{x_t}) \left[ (1 + x_t) - \frac{2\theta}{3}(1 + x_t)^3 \right] e^{-\theta(1+x_t)^2} \end{array} \right\}^{-2} \theta \delta^2 + O(\theta^2 \delta^4).$$

(R.5.3a)

Expanding Eq. R.5.3, paying particular attention to the fact that each $D_\theta$ operates on all the $\theta$'s to its right in a multiplicative sequence, including, if present in its sequence, the $\theta$ in the factor $\theta \delta^2$, we obtain the following:

$$IMSPE = \begin{pmatrix} -\sqrt{\dfrac{\pi}{4\theta}}\left\{\begin{array}{l}erf[\sqrt{\theta}(1+x_t)]\\+erf[\sqrt{\theta}(1-x_t)]\end{array}\right\} \\ -\sqrt{\dfrac{\pi}{128\theta}}\left\{\begin{array}{l}erf[\sqrt{2\theta}(1+x_t)]\\+erf[\sqrt{2\theta}(1-x_t)]\end{array}\right\} \\ +\dfrac{1}{4}\left[\begin{array}{l}(1+x_t)e^{-2\theta(1+x_t)^2}\\+(1-x_t)e^{-2\theta(1-x_t)^2}\end{array}\right] \end{pmatrix}^{2} + \begin{pmatrix} -\sqrt{\dfrac{\pi}{128\theta}}\left\{\begin{array}{l}erf[\sqrt{2\theta}(1+x_t)]\\+erf[\sqrt{2\theta}(1-x_t)]\end{array}\right\} \\ +\left[\begin{array}{l}(1+x_t)e^{-\theta(1+x_t)^2}\\+(1-x_t)e^{-\theta(1-x_t)^2}\end{array}\right] \\ +\dfrac{1}{2}\left[\begin{array}{l}(1+x_t)e^{-2\theta(1+x_t)^2}\\+(1-x_t)e^{-2\theta(1-x_t)^2}\end{array}\right] \\ -\dfrac{1}{4}\left\{\begin{array}{l}\left[(1+x_t)-\dfrac{4\theta}{3}(1+x_t)^3\right]e^{-\theta(1+x_t)^2}\\+\left[(1-x_t)-\dfrac{4\theta}{3}(1-x_t)^3\right]e^{-\theta(1-x_t)^2}\end{array}\right\} \end{pmatrix}^{-2}\theta\delta^2 + O(\theta^2\delta^4).$$

(R.5.4)

$$= \left\{-\dfrac{1}{2}\left(1+\dfrac{D_\theta}{4}\right)\sqrt{\dfrac{\pi}{\theta}}(1+S_{x_t})erf[\sqrt{\theta}(1+x_t)] \atop +\dfrac{1}{4}(1+S_{x_t})(1+x_t)e^{-\theta(1+x_t)^2}\right\}^{2}$$

$$+\left\{\begin{array}{l}-\dfrac{D_\theta}{16}\sqrt{\dfrac{\pi}{\theta}}(1+S_{x_t})erf[\sqrt{\theta}(1+x_t)]\\+\left(1+\dfrac{D_\theta}{4}\right)(1+S_{x_t})(1+x_t)e^{-\theta(1+x_t)^2}\\-\dfrac{D_\theta}{8}(1+S_{x_t})\left[(1+x_t)-\dfrac{2\theta}{3}(1+x_t)^3\right]e^{-\theta(1+x_t)^2}\end{array}\right\}^{-2}\theta\delta^2 + O(\theta^2\delta^4).$$

(R.5.4a)

$$= \left\{-\dfrac{\sqrt{\pi}}{2}\left(1+\dfrac{D_\theta}{4}\right)\sqrt{\dfrac{1}{\theta}}(1+S_{x_t})erf[\sqrt{\theta}(1+x_t)] \atop +\dfrac{1}{4}(1+S_{x_t})(1+x_t)e^{-\theta(1+x_t)^2}\right\}^{2}$$

$$+\left\{\begin{array}{l}-\dfrac{\sqrt{\pi}}{16}D_\theta\sqrt{\dfrac{1}{\theta}}(1+S_{x_t})erf[\sqrt{\theta}(1+x_t)]\\+(1+S_{x_t})(1+x_t)e^{-\theta(1+x_t)^2}\\+\dfrac{D_\theta}{4}(1+S_{x_t})(1+x_t)e^{-\theta(1+x_t)^2}\\-\dfrac{D_\theta}{8}(1+S_{x_t})\left[(1+x_t)-\dfrac{2\theta}{3}(1+x_t)^3\right]e^{-\theta(1+x_t)^2}\end{array}\right\}^{-2}\theta\delta^2 + O(\theta^2\delta^4).$$

(R.5.4b)

Next: Pulling out common $(1+S_{x_t})$'s, and noting $D_\theta\sqrt{\dfrac{1}{\theta}} = \sqrt{\dfrac{1}{2\theta}}D_\theta$, so $\dfrac{D_\theta}{4}\sqrt{\dfrac{1}{\theta}} = \sqrt{\dfrac{1}{32\theta}}D_\theta$, and $\left(1+\dfrac{D_\theta}{4}\right)\sqrt{\dfrac{1}{\theta}} = \left(\sqrt{\dfrac{1}{\theta}}+\sqrt{\dfrac{1}{32\theta}}\right)D_\theta = \left(1+\sqrt{\dfrac{1}{32}}\right)\sqrt{\dfrac{1}{\theta}}D_\theta$, etc.

$$IMSPE_{1,2,2} = (1 + S_{x_t}) \left\{ \begin{matrix} -\frac{\sqrt{\pi}}{2}\left(1 + \sqrt{\frac{1}{32}}\right)\sqrt{\frac{1}{\theta}} D_\theta erf[\sqrt{\theta}(1 + x_t)] \\ +\frac{1}{4}(1 + x_t)e^{-\theta(1+x_t)^2} \end{matrix} \right\}^1$$

$$+ (1 + S_{x_t}) \left\{ \begin{matrix} -\frac{\sqrt{\pi}}{16}\sqrt{\frac{1}{2\theta}} D_\theta erf[\sqrt{\theta}(1 + x_t)] \\ +(1 + x_t)e^{-\theta(1+x_t)^2} \\ +\frac{D_\theta}{4}(1 + x_t)e^{-\theta(1+x_t)^2} \\ -\frac{D_\theta}{8}\left[(1 + x_t) - \frac{2\theta}{3}(1 + x_t)^3\right]e^{-\theta(1+x_t)^2} \end{matrix} \right\}^{-1} \theta\delta^2 + O(\theta^2\delta^4). \quad (R.5.4c)$$

$$= (1 + S_{x_t}) \left\{ \begin{matrix} -\left(1 + \sqrt{\frac{1}{32}}\right)\sqrt{\frac{\pi}{4\theta}} D_\theta erf[\sqrt{\theta}(1 + x_t)] \\ +\frac{1}{4}(1 + x_t)e^{-\theta(1+x_t)^2} \end{matrix} \right\}^1$$

$$+ (1 + S_{x_t}) \left\{ \begin{matrix} -\frac{1}{16}\sqrt{\frac{\pi}{2\theta}} D_\theta erf[\sqrt{\theta}(1 + x_t)] \\ +(1 + x_t)e^{-\theta(1+x_t)^2} \\ +\frac{D_\theta}{4}(1 + x_t)e^{-\theta(1+x_t)^2} \\ -\frac{D_\theta}{8}\left[(1 + x_t) - \frac{2\theta}{3}(1 + x_t)^3\right]e^{-\theta(1+x_t)^2} \end{matrix} \right\}^{-1} \theta\delta^2 + O(\theta^2\delta^4). \quad (R.5.4d)$$

$$= (1 + S_{x_t}) \left( \left\{ \begin{matrix} -\frac{16+\sqrt{8}}{16}\sqrt{\frac{\pi}{4\theta}} D_\theta erf[\sqrt{\theta}(1 + x_t)] \\ +\frac{1}{4}(1 + x_t)e^{-\theta(1+x_t)^2} \end{matrix} \right\}^1 + \left\{ \begin{matrix} -\frac{1}{16}\sqrt{\frac{\pi}{4\theta}} D_\theta erf[\sqrt{\theta}(1 + x_t)] \\ +(1 + x_t)e^{-\theta(1+x_t)^2} \\ +\frac{1}{8}D_\theta(1 + x_t)e^{-\theta(1+x_t)^2} \\ +\frac{\theta D_\theta}{12}(1 + x_t)^3 e^{-\theta(1+x_t)^2} \end{matrix} \right\}^{-1} \theta\delta^2 \right) + O(\theta^2\delta^4).$$

(R.5.4e)

$$= (1 + S_{x_t}) \left\{ \begin{matrix} & [1 - \theta\delta^2] \\ -\frac{16+\sqrt{8}}{16}\sqrt{\frac{\pi}{4\theta}} D_\theta erf[\sqrt{\theta}(1 + x_t)]\left[1 + \frac{1}{16+\sqrt{8}}\theta\delta^2\right] \\ +\frac{1}{4}(1 + x_t)e^{-\theta(1+x_t)^2} & [1 + 4\theta\delta^2] \\ +\left[1 + \frac{2\theta}{3}(1 + x_t)^2\right]\frac{1}{8}D_\theta(1 + x_t)e^{-\theta(1+x_t)^2}\theta\delta^2 \end{matrix} \right\} + O(\theta^2\delta^4). \quad (R.5.4f)$$

$$IMSPE_{1,1,2} = 2\left(-\sqrt{\frac{\pi}{16\theta}}\left\{\begin{matrix} erf[\sqrt{\theta}(1 + x_1)] \\ +erf[\sqrt{\theta}(1 - x_1)] \end{matrix}\right\}^1\right) = (1 + S_{x_1})\left(-\sqrt{\frac{\pi}{4\theta}} erf[\sqrt{\theta}(1 + x_1)]\right).$$

$IMSPE_{1,0,2} = 1$.

Upon further rearrangement, this is identical to the expressions worked out by hand algebra or SMS algebra of Eqs. R.3.2 and R.4.2, respectively.

$$IMSPE = \left( \begin{array}{c} \frac{1}{4}\left[ (1+x_t)e^{-2\theta(1+x_t)^2} \\ +(1-x_t)e^{-2\theta(-1+x_t)^2} \right]^2 \\ -\sqrt{\frac{\pi}{4\theta}}\left\{ \begin{array}{c} erf[\sqrt{\theta}(1+x_t)] \\ +erf[\sqrt{\theta}(1-x_t)] \end{array} \right\} \\ -\sqrt{\frac{\pi}{128\theta}}\left\{ \begin{array}{c} erf[\sqrt{2\theta}(1+x_t)] \\ +erf[\sqrt{2\theta}(1-x_t)] \end{array} \right\} \end{array} \right) + \left( \begin{array}{c} \left[ \frac{1}{4} + \frac{\theta}{3}(1+x_t)^2 \right](1+x_t)e^{-2\theta(1+x_t)^2} \\ +\left[ \frac{1}{4} + \frac{\theta}{3}(1-x_t)^2 \right](1-x_t)e^{-2\theta(1-x_t)^2} \\ +\left[ (1+x_t)e^{-\theta(1+x_t)^2} \\ +(1-x_t)e^{-\theta(1-x_t)^2} \right] \\ -\sqrt{\frac{\pi}{128\theta}}\left\{ \begin{array}{c} erf[\sqrt{2\theta}(1+x_t)] \\ +erf[\sqrt{2\theta}(1-x_t)] \end{array} \right\} \end{array} \right)^{-2} \theta\delta^2 + O(\theta^2\delta^4).$$

(R.5.5)

## Appendix S. Support for Section 6, when $[d, n, v] = [1, 2, 3/2]$

For $[d, n, v] = [1,2,3/2]$, with design points denoted $x_1$ and $x_2$, we have the following, from Table 4.1:

$$\mathbf{L} = \begin{bmatrix} 0 & | & 1 & 1 \\ -- & | & -- & ------ \\ \cdot & | & 1 & (1+\sqrt{3Q})e^{-\sqrt{3Q}} \\ \cdot & | & \cdot & 1 \end{bmatrix}, \text{ where } Q \equiv \theta(x_1-x_2)^2, \text{ from Sec. 4. Via Matrix Identity A2,}$$

$$\mathbf{L}^{-1} = \frac{1}{2} * \begin{bmatrix} -1-(1+\sqrt{3Q})e^{-\sqrt{3Q}} & | & 1 & 1 \\ ----------- & | & -------- & -------- \\ \cdot & | & \frac{1}{1-(1+\sqrt{3Q})e^{-\sqrt{3Q}}} & \frac{-1}{1-(1+\sqrt{3Q})e^{-\sqrt{3Q}}} \\ \cdot & | & \cdot & \frac{1}{1-(1+\sqrt{3Q})e^{-\sqrt{3Q}}} \end{bmatrix}. \text{ Then, by Eq. H.1,}$$

$$\mathbf{R} = \left( \begin{array}{c|cc} 1 & \frac{1}{2\sqrt{3\theta}}\left\{ \begin{array}{c} 2\left[ \begin{array}{c} 1-e^{-\sqrt{3\theta}(1+x_1)} \\ +1-e^{-\sqrt{3\theta}(1-x_1)} \end{array} \right] \\ -\sqrt{3\theta}\left[ \begin{array}{c} (1+x_1)e^{-\sqrt{3\theta}(1+x_1)} \\ +(1-x_1)e^{-\sqrt{3\theta}(1-x_1)} \end{array} \right] \end{array} \right\} & \frac{1}{2\sqrt{3\theta}}\left\{ \begin{array}{c} 2\left[ \begin{array}{c} 1-e^{-\sqrt{3\theta}(1+x_2)} \\ +1-e^{-\sqrt{3\theta}(1-x_2)} \end{array} \right] \\ -\sqrt{3\theta}\left[ \begin{array}{c} (1+x_2)e^{-\sqrt{3\theta}(1+x_2)} \\ +(1-x_2)e^{-\sqrt{3\theta}(1-x_2)} \end{array} \right] \end{array} \right\} \\ -- & ---------------- & ---------------- \\ \cdot & MRSE, via\ Eq.I.1 & MRSE, via\ Eq.I.1 \\ \cdot & \cdot & MRSE, via\ Eq.I.1 \end{array} \right),$$

where $MRSE$ stands for a machine-readable symbolic expression. Then, using Eq. 4.1,

$$IMSPE = 1 - \frac{1}{2}tr\begin{bmatrix}\begin{bmatrix}-1-\left(1+\sqrt{3Q}\right)e^{-\sqrt{3Q}} & | & 1 & 1 \\ --------- & | & -------- & -------- \\ \cdot & | & \frac{1}{1-(1+\sqrt{3Q})e^{-\sqrt{3Q}}} & \frac{-1}{1-(1+\sqrt{3Q})e^{-\sqrt{3Q}}} \\ \cdot & | & \cdot & \frac{1}{1-(1+\sqrt{3Q})e^{-\sqrt{3Q}}}\end{bmatrix} \\ * \begin{pmatrix} 1 & | & \frac{1}{2\sqrt{3\theta}}\begin{Bmatrix}2\begin{bmatrix}1-e^{-\sqrt{3\theta}(1+x_1)} \\ +1-e^{-\sqrt{3\theta}(1-x_1)}\end{bmatrix} \\ -\sqrt{3\theta}\begin{bmatrix}(1+x_1)e^{-\sqrt{3\theta}(1+x_1)} \\ +(1-x_1)e^{-\sqrt{3\theta}(1-x_1)}\end{bmatrix}\end{Bmatrix} & \frac{1}{2\sqrt{3\theta}}\begin{Bmatrix}2\begin{bmatrix}1-e^{-\sqrt{3\theta}(1+x_2)} \\ +1-e^{-\sqrt{3\theta}(1-x_2)}\end{bmatrix} \\ -\sqrt{3\theta}\begin{bmatrix}(1+x_2)e^{-\sqrt{3\theta}(1+x_2)} \\ +(1-x_2)e^{-\sqrt{3\theta}(1-x_2)}\end{bmatrix}\end{Bmatrix} \\ -- & | & ------------------ & ------------------ \\ \cdot & | & MRSE, via\ Eq.I.1 & MRSE, via\ Eq.I.1 \\ \cdot & | & \cdot & MRSE, via\ Eq.I.1\end{pmatrix}\end{bmatrix}.$$

(S.1)

## Appendix T.  Support for Sub-Section 6, when $[d, n, \nu] = [1, 2, 5/2]$

For $[d, n, \nu] = [1,2,5/2]$, with design points denoted $x_1$ and $x_2$, we have the following:

$$\boldsymbol{L} = \begin{bmatrix}0 & | & 1 & 1 \\ -- & | & -- & ----------- \\ \cdot & | & 1 & \left(1+\sqrt{5Q}+\frac{5Q}{3}\right)e^{-\sqrt{5Q}} \\ \cdot & | & \cdot & 1\end{bmatrix}, \text{ where from Sec. 4, } Q \equiv \theta(x_1-x_2)^2. \text{ Via Matrix}$$

Identity A2 and, then, via Eq. J.2,

$$\boldsymbol{L^{-1}} = \frac{1}{2} * \begin{bmatrix}-1-\left(1+\sqrt{5Q}+\frac{5Q}{3}\right)e^{-\sqrt{5Q}} & | & 1 & 1 \\ --------------- & | & --------- & --------- \\ \cdot & | & \frac{1}{1-\left(1+\sqrt{5Q}+\frac{5Q}{3}\right)e^{-\sqrt{5Q}}} & \frac{-1}{1-\left(1+\sqrt{5Q}+\frac{5Q}{3}\right)e^{-\sqrt{5Q}}} \\ \cdot & | & \cdot & \frac{1}{1-\left(1+\sqrt{5Q}+\frac{5Q}{3}\right)e^{-\sqrt{5Q}}}\end{bmatrix}.$$

$$\boldsymbol{R} = \begin{bmatrix}1 & | & \frac{1}{6\sqrt{5\theta}}\begin{pmatrix}8\begin{Bmatrix}\begin{bmatrix}1-e^{-\sqrt{5\theta}(1+x_1)}\end{bmatrix} \\ +\begin{bmatrix}1-e^{-\sqrt{5\theta}(1-x_1)}\end{bmatrix}\end{Bmatrix} \\ -5\sqrt{5\theta}\begin{Bmatrix}(1+x_1)e^{-\sqrt{5\theta}(1+x_1)} \\ +(1-x_1)e^{-\sqrt{5\theta}(1-x_1)}\end{Bmatrix} \\ -5\theta\begin{Bmatrix}(1+x_1)^2 e^{-\sqrt{5\theta}(1+x_1)} \\ +(1-x_1)^2 e^{-\sqrt{5\theta}(1-x_1)}\end{Bmatrix}\end{pmatrix} & \frac{1}{6\sqrt{5\theta}}\begin{pmatrix}8\begin{Bmatrix}\begin{bmatrix}1-e^{-\sqrt{5\theta}(1+x_2)}\end{bmatrix} \\ +\begin{bmatrix}1-e^{-\sqrt{5\theta}(1-x_2)}\end{bmatrix}\end{Bmatrix} \\ -5\sqrt{5\theta}\begin{Bmatrix}(1+x_2)e^{-\sqrt{5\theta}(1+x_2)} \\ +(1-x_2)e^{-\sqrt{5\theta}(1-x_2)}\end{Bmatrix} \\ -5\theta\begin{Bmatrix}(1+x_2)^2 e^{-\sqrt{5\theta}(1+x_2)} \\ +(1-x_2)^2 e^{-\sqrt{5\theta}(1-x_2)}\end{Bmatrix}\end{pmatrix} \\ -- & | & ------------------ & ------------------ \\ \cdot & | & MRSE, via\ Eq.K.1 & MRSE, via\ Eq.K.1 \\ \cdot & | & \cdot & MRSE, via\ Eq.K.1\end{bmatrix},$$

where $MRSE$ stands for a machine-readable symbolic expression. Then, using Eq. 4.1,

$$IMSPE = 1 - \frac{1}{2} tr \left\{ \begin{bmatrix} \begin{bmatrix} -1 - \left(1 + \sqrt{5Q} + \frac{5Q}{3}\right) e^{-\sqrt{5Q}} & | & 1 & 1 \\ \text{------} & | & \text{------} & \text{------} \\ \cdot & | & \frac{1}{1-\left(1+\sqrt{5Q}+\frac{5Q}{3}\right)e^{-\sqrt{5Q}}} & \frac{-1}{1-\left(1+\sqrt{5Q}+\frac{5Q}{3}\right)e^{-\sqrt{5Q}}} \\ \cdot & | & \cdot & \frac{1}{1-\left(1+\sqrt{5Q}+\frac{5Q}{3}\right)e^{-\sqrt{5Q}}} \end{bmatrix} \\ * \begin{bmatrix} 1 & | & \frac{1}{6\sqrt{5\theta}} \begin{pmatrix} 8 \left\{ \begin{bmatrix} 1 - e^{-\sqrt{5\theta}\,(1+x_1)} \end{bmatrix} \\ + \begin{bmatrix} 1 - e^{-\sqrt{5\theta}\,(1-x_1)} \end{bmatrix} \right\} \\ -5\sqrt{5\theta} \left\{ \begin{matrix} (1+x_1)e^{-\sqrt{5\theta}\,(1+x_1)} \\ +(1-x_1)e^{-\sqrt{5\theta}\,(1-x_1)} \end{matrix} \right\} \\ -5\theta \left\{ \begin{matrix} (1+x_1)^2 e^{-\sqrt{5\theta}\,(1+x_1)} \\ +(1-x_1)^2 e^{-\sqrt{5\theta}\,(1-x_1)} \end{matrix} \right\} \end{pmatrix} & \frac{1}{6\sqrt{5\theta}} \begin{pmatrix} 8 \left\{ \begin{bmatrix} 1 - e^{-\sqrt{5\theta}\,(1+x_2)} \end{bmatrix} \\ + \begin{bmatrix} 1 - e^{-\sqrt{5\theta}\,(1-x_2)} \end{bmatrix} \right\} \\ -5\sqrt{5\theta} \left\{ \begin{matrix} (1+x_2)e^{-\sqrt{5\theta}\,(1+x_2)} \\ +(1-x_2)e^{-\sqrt{5\theta}\,(1-x_2)} \end{matrix} \right\} \\ -5\theta \left\{ \begin{matrix} (1+x_2)^2 e^{-\sqrt{5\theta}\,(1+x_2)} \\ +(1-x_2)^2 e^{-\sqrt{5\theta}\,(1-x_2)} \end{matrix} \right\} \end{pmatrix} \\ \text{--} & | & \text{------} & \text{------} \\ \cdot & | & MRSE, via\ Eq.K.1 & MRSE, via\ Eq.K.1 \\ \cdot & | & \cdot & MRSE, via\ Eq.K.1 \end{bmatrix} \end{bmatrix} \right\}.$$

(T.1)